\documentclass{article}
\usepackage{custom_tex}
\graphicspath{{./fig/}}

\begin{document}

\title{Batch Predictive Inference}

\date{\today}

\author{Yonghoon Lee}
\author{Eric Tchetgen Tchetgen}
\author{Edgar Dobriban
\footnote{E-mail addresses: 
\texttt{yhoony31@wharton.upenn.edu},
\texttt{ett@wharton.upenn.edu},
\texttt{dobriban@wharton.upenn.edu}.
ETT and ED have jointly advised YL on the project.
}}

\affil{Department of Statistics and Data Science, The Wharton School, University of Pennsylvania}

\maketitle

\begin{abstract}
Constructing prediction sets with coverage guarantees for unobserved outcomes is a core problem in modern statistics.
Methods for predictive inference have been developed for a wide range of settings, but usually only consider test data points one at a time.
Here we study the problem of distribution-free predictive inference for a functions of batch of multiple test points, 
aiming
to construct prediction sets
for functions---such as the mean or median---of any number of unobserved test datapoints.
This setting 
includes constructing simultaneous prediction sets with a high probability of coverage, 
and selecting datapoints satisfying a specified condition (e.g., being large) while controlling the number of false claims.
Here, 
for the general task of predictive inference on a function of a batch of test points,
we introduce a methodology called \textit{batch predictive inference (batch PI)}, 
and provide a distribution-free coverage guarantee under exchangeability of the calibration and test data. 
Batch PI requires the quantiles of a \emph{rank ordering function} defined on certain subsets of ranks.
While computing these quantiles is NP-hard in general, 
we show that it can be done efficiently in many cases of interest, most notably for batch score functions with a compositional structure---which includes examples of interest such as the mean---via a dynamic programming algorithm that we develop.
Batch PI
has advantages over baseline approaches (such as partitioning the calibration data or directly extending conformal prediction) in many settings, 
as it can deliver informative prediction sets even using small calibration sample sizes. 
We illustrate that our procedures provide informative inference across the use cases mentioned above, through experiments on both simulated data and a  drug-target interaction dataset.
\end{abstract}

\tableofcontents
\medskip

\sloppy

\section{Introduction}

Consider 
a supervised learning setting  
where we have a 
dataset $(X_1,Y_1), \ldots, (X_n,Y_n)$ drawn from $P_{X,Y} = P_X \times P_{Y \mid X}$ on $\X \times \Y$ and a batch of new test inputs $X_{n+1}, \ldots, X_{n+m}$ from $P_X$.
Our task is to predict and make inference for the unobserved outcomes $Y_{n+1}, \ldots, Y_{n+m}$. This setting includes both regression and classification.
Beyond point predictions, it is of significant interest to construct prediction sets for various functions 
of the unobserved outcomes $Y_{n+1}, \ldots, Y_{n+m}$. 
For example, given a regression function
$\hat{\mu} : \X \rightarrow \Y$
 trained using a subset of the data, 
 one might aim to construct a prediction set for $Y_{n+1}$ of the form $\Ch(X_{n+1}) = \hat{\mu}(X_{n+1}) \pm (\text{constant})$, which satisfies the \emph{marginal coverage guarantee}
$\mathbb{P}\{Y_{n+1} \in \Ch(X_{n+1})\} \ge 1-\alpha$,
for a predetermined level $\alpha \in (0,1)$.

Distribution-free inference aims to achieve such inferential targets without imposing distributional assumptions on the sampling distribution $P_{X,Y}$, and dates back at least to the pioneering works of \cite{Wilks1941}, \cite{Wald1943}, \cite{scheffe1945non} in the 1940s, and \cite{tukey1947non,tukey1948nonparametric}. 
For example, conformal prediction~\citep[e.g.,][etc.]{saunders1999transduction,vovk1999machine, vovk2005algorithmic} provides a general framework for achieving marginal coverage 
under exchangeability.
Many recent works have explored the possibility of improving or generalizing this framework to achieve stronger targets, reduce computational costs, or enable inference with non-exchangeable data, etc, see \Cref{relw}. 
However, 
method development for joint inference on functions of multiple test points has been limited.

In this work, we develop methodology for distribution-free joint inference on multiple test points. At a high level, this problem is connected to two major areas of statistical research:

\begin{enumerate}
    \item \textit{Simultaneous inference on multiple quantities.} In many real-world problems, there are 
    multiple quantities of interest for inference---e.g., multiple applicants for a job \citep{cohen2020efficient,barigozzi2016screening}, patients undergoing screening or a particular treatment \citep{nielsen1999principles,colombo2007screening}, drug candidates in high-throughput screening \citep{mayr2009novel,macarron2011impact}, 
    multiple endpoints in medical trials \citep{budig2024simultaneous},
    weather or other variables in weather forecasting \citep{neeven2018conformal, messoudi2022ellipsoidal, sampson2024conformal}. 
    For testing problems, a series of methods have been developed for multiplicity adjustment to obtain valid multiple testing procedures \cite[see e.g.,][etc]{lehmann2005testing, rupert2012simultaneous}. 
    However, for predictive inference problems, methodological development remains limited.
    We will show that existing approaches often struggle to provide useful valid inferential guarantees in this setting.

    \item \textit{Inference on a finite population.}
    In applications such as survey studies and randomized trials \citep{kalton2020introduction,hariton2018randomised}, researchers are often interested in analyzing a finite population rather than a hypothetical infinite population \citep[see e.g.,][etc]{abadie2014finite}---for example, the distribution of treatment effects across the group of individuals who received the treatment. A school administrator may want to anticipate the effect of a new teaching method specifically on the students in a program, rather than on a hypothetical broader student population, see e.g., \cite{kautz2017comparing}. 
    Similarly, in the analysis of network data, researchers are often interested in understanding how a message or intervention spreads through a specific social network \cite{newman2018networks}, and network models that include exchangeable feature observations have been studied~\citep{mao2021consistent}.
    
\end{enumerate}

More specifically, this problem includes several inferential tasks as special cases:

\begin{enumerate}
    \item {\bf Inference on the mean of a test dataset; including on counterfactuals.}
    Consider predicting the mean of the test outcomes via a prediction set $\Ch$ such that
    \[\mathbb{P}\Big\{\frac{1}{m}\sum\nolimits_{j=1}^m Y_{n+j} \in \Ch(X_{n+1},\ldots,X_{n+m})\Big\} \ge 1-\alpha.\]
    This problem has a range of use cases and we illustrate it in a problem of inference on the mean difference between counterfactual outcomes under different treatments. Specifically, consider a randomized trial where $A \in \{0,1\}$ denotes the treatment assignment, and $Y^{a=0}$ and $Y^{a=1}$ represent the counterfactual outcomes under control and treatment, respectively. For each individual $i = 1, \ldots, n$, we observe the triplet $(X_i, A_i, (1 - A_i) Y_i^{a=0} + A_i Y_i^{a=1})$---that is, we observe only the counterfactual corresponding to the assigned treatment.
Our goal is to construct a prediction set $\ch(\D)$ for the mean of the unobserved counterfactual outcomes among the treated subgroup:
$$
\mathbb{P}\Big\{\frac{1}{N^1}\sum_{i : A_i = 1} Y_i^{a=0} \in \ch(\D)\Big\} \ge 1 - \alpha,
$$
where $N^1 = |\{i : A_i = 1\}|$, and $\D$ here denotes the full set of the observed data.
     When the number of test datapoints is small, methods based on concentration inequalities can generally be conservative for the above problems, producing wide intervals.
     In contrast, as we demonstrate empirically, our methods can still be informative. It is also worth mentioning that our method also works for the median and other quantiles; and in particular for the median counterfactual. \\
    \item {\bf Prediction sets for multiple unobserved outcomes.}
    Consider constructing an algorithm $\Ch$ 
    that likely obtains at least $1-\delta$ empirical coverage over the test set, i.e., 
     \[\mathbb{P}\Big\{\frac{1}{m}\sum\nolimits_{j=1}^m \One{Y_{n+j} \in \Ch(X_{n+j})} \ge 1-\delta\Big\} \ge 1-\alpha.\]
     Compared to applying conformal prediction separately to individual test points to obtain 
     $\mathbb{P}\{Y_{n+j} \in \Ch(X_{n+j})\} \ge 1-\alpha'$, $j=1, \ldots, m$, for some $\alpha'$, the above coverage guarantee directly states that most prediction sets cover the true outcome with a  well-calibrated and pre-specified high-probability. For instance, this allows us to construct prediction sets for a machine learning classifier, such that for a test set of interest, most labels are covered with a given probability. \\

    \item {\bf Selection of datapoints with error  control.}  
    Consider selecting test datapoints in the test set whose responses satisfy a specific condition, such as $Y_{n+j} > c$ for a predetermined threshold $c$. 
    As $(Y_{n+j})_{1 \le j \le m}$ are unobserved, a potential approach is to construct a selection criterion based on the training and calibration data, e.g., of the form $\hat{\mu}(X_{n+j}) > \hat{T}$. 
    One possible inferential target
    is the control of the probability of making more than $k$ errors  at level $\alpha$, i.e., 
    \[\mathbb{P}\Big\{\sum\nolimits_{j=1}^m \One{\hat{\mu}(X_{n+j}) > \hat{T}, Y_{n+j} \le c} > k\Big\} < \alpha,\]\
    where $k$ is a predetermined target error bound.
    This is analogous to the notion of family-wise error rate (FWER) control in multiple hypothesis testing. 
    As an example, we use this method to select promising drug-target pairs.
\end{enumerate}

We provide more details on the above examples in Section~\ref{sec:applications}. 
The examples turn out to be special cases of the following general problem:
 Given the calibration data $\Dn = \{(X_i,Y_i)\}_{1 \le i \le n}$ and a function $g : \mathcal{P}(\X \times \Y)\rightarrow \R$\footnote{For a set $A$, we write $\mathcal{P}(A)$ to denote its power set.} that takes the set of test observations as the input, construct a prediction set $\ch(\Dn)$ that satisfies 
\[\PP{g\left(\{(X_{n+1},Y_{n+1}), \ldots, (X_{n+m},Y_{n+m})\}\right) \in \ch(\Dn)} \ge 1-\alpha.\]

For instance, the high-probability coverage property for multiple unobserved outcomes can be achieved by taking $g$ to be a specific quantile of the non-conformity scores of the test data.
More generally, 
we propose a \emph{batch predictive inference} methodology applicable to a wide range of target functions $g$. 
We then explain use cases, 
including those described above.\\

\noindent {\bf Notation.}
We write $\R$ to denote the set of real 
numbers 
and $\R_{\ge 0}$ to denote the set of nonnegative reals. The set of positive integers is denoted by $\N$. For a positive integer $n \in \N$, we write $[n]$ to denote the set $\{1,2,\ldots,n\}$ and
for any $a, b\in[n]$ with $a\le b$
write $X_{a:b}$ to denote the vector $(X_a,X_2,\ldots,X_b)^\top$. 
We will denote the all ones vector of size $m$ as $1_m$.
For a function $f : A \rightarrow B$, We write $\mathrm{Im}(f)$ to denote the image of a function $f$, and $f \big|_{C}$ to denote the restriction of $f$ to $C \subset A$. 
For a real number $x$, we write $\lfloor x \rfloor$, $\lceil x \rceil$, and $\text{round}(x)$ to denote the floor, ceiling, and rounding of $x$ (with 1/2 rounding up) to the nearest integer, respectively. We let $a_+ = \max\{a,0\}$ for a real number $a \in \R$. 
We denote the number of ways to choose $r$ items with replacement from $n$ items as $_{n}\mathrm{H}_r$. Let $\R^m_{\uparrow} = \{x\in\R^m: x_1\le x_2 \le \ldots \le x_m\}$ be the set of monotone non-increasing vectors. For two vectors $u=(u_1,\ldots,u_d)^\top, v=(v_1,\ldots,v_d)^\top \in \R^d$, 
we write $u \preceq v$ if $u_i \le v_i$ for all $i=1,2,\ldots,d$. 

We write $\sum_{i=1}^k p_i \delta_{v_i}$ to denote the discrete distribution with support $\{v_1,v_2,\ldots,v_k\}$ and the probability masses $(p_1,p_2,\ldots,p_k)$. For a distribution $P$, we define two types of quantile functions $Q_\tau(P) = \inf\{t \in \R : \Pp{X \sim P}{X \leq t} \ge \tau\}$ and $Q_\tau'(P) = \sup\{t \in \R : \Pp{X \sim P}{X \geq t} \geq 1-\tau\}$\footnote{It holds that $Q_\tau'(P) = -Q_{1-\tau}(-P)$, where $-P$ denotes the distribution of $-X$ when $X \sim P$.}.
For an event $E$, we write $\One{E}$ to denote its corresponding indicator variable. 
All objects (sets and functions) considered will be measurable with respect to appropriate sigma-algebras (typically the Borel sigma-algebra generated by open sets), which will not be mentioned further.
For a set $D$, $\mathcal{P}(D)$ denotes its power set; or the Borel sigma algebra on $D$ if that is well-defined.
 We write $\mathcal{N}(\mu,\sigma^2;[a,b])$ to denote the truncated normal distribution with mean $\mu$, variance $\sigma^2$, and truncation set $[a,b]$. 

\subsection{Main contributions}

Our contributions can be summarized as below:

\begin{enumerate}
    \item {\bf Batch predictive inference (batch PI):} 
    We develop the batch predictive inference (batch PI): methodology for distribution-free inference on a function of multiple unobserved test outcomes. 
    Our targets include a broad range of functions satisfying a certain monotonicity property, such as the mean or quantiles. 
    Furthermore, we extend this approach to achieve simultaneous inference on multiple quantiles of test scores. 
    Batch PI can provide useful inference 
    when the calibration dataset size is comparable to---or even smaller than---the test size, a scenario in which we show that baseline approaches fail.\\

    \item {\bf Efficient algorithms for the batch PI procedure:}
     We show that the batch PI procedure is generally NP-hard to compute, but it can be simplified for many target functions of practical interest, such as the mean and quantiles. For quantiles, and more generally for ``sparse"  functions depending only on a few quantiles, we establish how the computational burden can be reduced substantially, making the approach feasible in routine applications. 
     For functions satisfying a certain compositional structure (e.g., the mean), 
     we present a polynomial-time dynamic programming algorithm for batch PI.\\
    
    \item {\bf Use cases in statistical inference problems:}
    We develop use cases of the batch PI methodology in  various statistical inferential problems:
    (1) constructing simultaneous prediction sets for multiple individual outcomes,  (2) selecting individuals with error control, and (3) inference on counterfactual variables.
    The last use case relies on a more general methodology that we develop for the setting of coverage under covariate shift.\\

    \item {\bf Empirical evaluation:}
    We empirically examine the performance of batch PI-based methods in simulations 
    and via an illustration on a drug-target interaction dataset. 
    The empirical results support that our procedure achieves the theoretical guarantees, and provides practically useful predictive inference.
    
\end{enumerate}

\subsection{Problem setting}
We observe data points $\Dn = \{(X_1,Y_1),(X_2,Y_2), \ldots, (X_n,Y_n)\} \subset \X \times \Y$, where $\X$ is a feature space and $\Y$ is an outcome space. 
Here and below, sets refer to multisets, and allow repetitions of elements.
We denote $\mathcal{D}_n$ as a calibration dataset, in the sense that it will be used for inference, e.g., to construct a prediction set.
We then observe a test dataset consisting of $m \geq 1$ test features $X_{n+1}, \ldots, X_{n+m}$, for which the corresponding outcomes $Y_{n+1}, \ldots, Y_{n+m}$ are not observed.
We denote each data point as $Z_i = (X_i,Y_i)$, for $i\in[n+m]$. 

Given a real-valued function $g : \mathcal{P}(\X \times \Y) \rightarrow \R$ 
of interest taking as input a subset of $\X \times \Y$, our goal is to construct a \emph{prediction set} for the unobserved value 
$g(\{Z_{n+1}, \ldots, Z_{n+m}\})$; which depends on the unobserved outcomes $Y_{n+1}, \ldots, Y_{n+m}$.
Specifically, we aim to construct a procedure $\ch : (\X \times \Y)^n \rightarrow \mathcal{P}(\R)$ such that
\begin{equation}\label{eqn:pred_set}
    \PP{g(\{Z_{n+1}, \ldots, Z_{n+m}\}) \in \ch(\Dn)} \ge 1-\alpha
\end{equation}
holds for a predefined level $\alpha \in (0,1)$, regardless of the sampling distribution.
We are interested in a general setting where $m$ is not necessarily significantly smaller than the calibration set size $n$ (in contrast to cases with trivial solutions, as we will describe later), and may even be larger. We will argue that this setting covers a wide range of important scenarios.

We now need some notations: For any vector $v\in\R^m$, 
let $v_{\uparrow}=(v_{(1)}, \ldots, v_{(m)})$ be the vector $v$ sorted in a non-decreasing order.
For $z = (z_1,\ldots,z_m) \in (\X \times \Y)^m$ and a ``score" function $s : \X \times \Y \rightarrow \R$, define 
$s(z) = (s(z_1), s(z_2), \ldots, s(z_m))$ 
by element-wise application of $s$. 
We denote $S_i = s(Z_i)$ for all $i\in[m+n]$.

 We require the following structural monotonicity condition for the target function $g$.

\begin{condition}[Monotonicity of the target function]\label{asm:target}
There is a \emph{batch score
function}\footnote{Let $\inf s = \inf\{s(x,y): (x,y)\in \X \times \Y\}$
and $\sup s = \sup\{s(x,y): (x,y)\in \X \times \Y\}$.
When $s$ is unbounded, 
we need the function
$h$ to be defined for all values
$s_1\le \ldots \le s_m$ such that 
$s_i \in (\inf s,\sup s)$ for all $i\in \{2, \ldots, m-1\}$
and either $s_1 = \inf s$ or $s_m = \sup s$.
We define $h(-\infty,s_2,  \ldots, s_m) = -\infty$
if $s_1 = \inf s = -\infty$, 
and
$h(s_1,  \ldots,s_{m-1},\infty) = \infty$
if $s_m = \sup s = \infty$.
}
$h : \R^m_{\uparrow} \rightarrow \R$
and a (non-batch, per-datapoint) \emph{score function}
$s : \X \times \Y \rightarrow \R$,
such that
\begin{equation}\label{gh}
g(\{z_1,\ldots,z_m\}) = h(s(z)_{\uparrow})
\end{equation}
holds for all $z\in (\X \times \Y)^m$.
Moreover, the function $h$ is monotone non-decreasing with respect to each coordinate, i.e.,
\begin{equation}\label{eqn:h_monotone}
    \textnormal{for any } v, \tilde{v} \in \R^m \textnormal{ with } v \preceq \tilde{v}, \textnormal{ we have } h(v_{\uparrow}) \le h(\tilde{v}_{\uparrow}).
\end{equation}
\end{condition}

Condition~\ref{asm:target} covers a broad range of targets, from the mean $h(s_1,\ldots,s_m) = \frac{s_1+\ldots+s_m}{m}$ and the $q$-th quantile $h(s_1,\ldots,s_m) = s_{(\lceil(q m)\rceil)}$, $q\in (0,1)$, to more general targets such as the truncated mean or the proportion of scores exceeding a certain threshold.
In many settings, $h$ represents a fixed function of interest, while $s$ is typically constructed using a separate dataset. 
For instance, in regression tasks, we can consider nonconformity scores such as $s:(x,y) \mapsto |y - \hat{\mu}(x)|$, where $\hat{\mu}$ is fitted on a separate dataset. As a simpler example, one can consider $s(y) = y$ and $m = 2$, with $h(s_1, s_2) = \frac{s_1 + s_2}{2}$, where the goal becomes inference on the average of two test outcomes, $(Y_{n+1} + Y_{n+2})/2$.

As a remark, if the cardinality of  $\X \times \Y $ is at most that of $\R$---e.g., $\X \subset \R^d$ for some positive integer $d \geq 1$ and $\Y = \R$---then \eqref{gh} holds, and
only the monotonicity property \eqref{eqn:h_monotone} 
imposes a condition.\footnote{To see that, observe that in this case, there is an injective map $f: \X \times \Y \to \R$. Let $\mathcal{I}\subset \R$ be the image of $f$. Then, for any $v\in \R^m_{\uparrow}\cap \mathcal{I}^m$, we can define $h(v) = g(\{f^{-1}(v_1), \ldots, f^{-1}(v_m)\})$, and for $v\in \R^m_{\uparrow}\setminus\mathcal{I}^m$, we can define $h(v)$  arbitrarily. Since $f$ is injective, $h$ is well-defined, satisfying \eqref{gh} by definition.}

\subsection{Related work}
\label{relw}

The idea of distribution-free prediction sets dates back at least to the pioneering works of \cite{Wilks1941}, \cite{Wald1943}, \cite{scheffe1945non}, and \cite{tukey1947non,tukey1948nonparametric}.
Distribution-free inference has been extensively studied in recent works \citep[see, e.g.,][]{saunders1999transduction,vovk1999machine,papadopoulos2002inductive,vovk2005algorithmic,Vovk2013, lei2013distribution,lei2014distribution,lei2018distribution,angelopoulos2023conformal,guan2023localized, romano2020classification,liang2023conformal,dobriban2023symmpi}. 
Predictive inference methods
\citep[e.g.,][etc]{geisser2017predictive}  
have been developed under various assumptions
\citep[see, e.g.,][]{bates2021distribution,park2021pac,park2022pac,sesia2022conformal,qiu2023prediction,li2022pac,kaur2022idecode,si2024pac,lee2024simultaneous}.
Overviews of the field are provided by \cite{vovk2005algorithmic, shafer2008tutorial}, and \cite{angelopoulos2023conformal}. For exchangeable data, conformal prediction and split conformal prediction \citep{vovk2005algorithmic, papadopoulos2002inductive} provide a general framework for distribution-free predictive inference.

Distribution-free predictive inference for multiple test points has been extensively studied in the context of outlier detection and selection \citep{10.1214/22-AOS2244, jin2023selection, jin2023model, gui2024conformal}. These works apply multiple testing methods to conformal p-values for inference on multiple test outcomes. \cite{vovk2013transductive} discuss transductive conformal methods for constructing a prediction region for the vector of test outcomes, where transductive means that the predictor (inducing the non-conformity score) used to construct the prediction sets can depend on the test dataset. 
\cite{lee2024simultaneous} introduces a method for constructing simultaneous prediction sets for multiple outcomes under covariate shift with a conditional guarantee.

\cite{gazin2024transductive} 
study a closely related problem setting to our paper, 
transductive conformal inference with adaptive scores.
In this scenario, they derive the joint distribution of multiple test conformal p-values in the case of no ties between non-conformity scores, which is equivalent to the joint distribution of their ranks, and which we use in the proof of our Theorem \ref{thm:batch_PI}.
\cite{gazin2024transductive} also give intriguing equivalent characterizations of this distribution, for instance in terms of P\'olya urns (see also \cite{gazin2024asymptoticsconformalinference} for a functional CLT for the coverage). 
Further, they apply these results 
to several problems, including controlling the false coverage rate of the prediction sets for multiple test points.
For this problem, \cite{MARQUESF2025} derived the distribution of the coverage.
This problem is also considered in one of our use cases in this work, and we will provide further discussion in Section~\ref{sec:pred_pac}.

In Section~\ref{sec:cov_shift}, we discuss how our procedure can be applied to situations involving covariate shift. This is relevant in light of the recent literature, which has shown significant interest in extending the conformal prediction framework to handle non-exchangeable data. For instance, \cite{tibshirani2019conformal} proposes weighted conformal prediction for predictive inference under covariate shift, and their method is further developed in works such as \cite{lei2021conformal, candes2023conformalized}, and~\cite{guan2023localized}. \cite{qiu2023prediction} and \cite{yang2022} introduce adaptive prediction methods with unknown covariate shift. \cite{10.1214/23-AOS2276} introduces a robust conformal prediction approach for non-exchangeable data. Other works have explored applying the conformal prediction framework to structured datasets. For example, \cite{dunn2022distribution, lee2023distribution}, and \cite{duchi2024predictive} provide conformal-type methods for data with a hierarchical structure, while \cite{dobriban2023symmpi} provides a method for data with group symmetries.

\section{Main results}
Here and below, we will suppose that the calibration and test data $(X_1,Y_1), \ldots, (X_{n+m},Y_{n+m})$ are exchangeable, unless explicitly specified otherwise.
If $m=1$, i.e., we have only one test point, then the coverage guarantee~\eqref{eqn:pred_set} can be achieved simply by standard distribution-free prediction methods, such as full and split conformal prediction \citep{vovk2005algorithmic, papadopoulos2002inductive},
for any function $g$.
For example, if we set $g(z)$ as the nonconformity score, i.e., $g({z}) = |y-\hat{\mu}(x)|$, for all $z = (x,y)$,
then the condition~\eqref{eqn:pred_set} is equivalent to the standard marginal coverage guarantee $\mathbb{P}\big\{s(X_{n+1},Y_{n+1}) \in \ch(\Dn)\big\} \ge 1-\alpha$,
and split conformal prediction attains this guarantee with the following prediction set~\citep{saunders1999transduction,vovk1999machine,vovk2005algorithmic, papadopoulos2002inductive}.
\[\ch(\Dn) = \Bigg(-\infty, Q_{1-\alpha}\left(\sum_{i=1}^n \frac{1}{n+1}\delta_{s(X_i,Y_i)} + \frac{1}{n+1}\delta_{\infty}\right)\Bigg].\]

However, for multiple test points, it turns out that constructing a useful distribution-free prediction set that satisfies~\eqref{eqn:pred_set} is a nontrivial task. 
One can imagine a number of direct approaches, such as directly extending split conformal or full conformal prediction; however, it turns out that their usefulness is limited to a small range of settings, as we discuss next. 
The reader may directly skip to Section \ref{bpim} to read the description of our proposed method.
\subsection{Baseline approaches}
In this Section, we consider several possible reasonable alternative approaches to our Batch PI approach introduced in the next Section, and we discuss their limitations.
\label{baseline}
\subsubsection{Partitioning the calibration data}
\label{part}
A potential approach  
to achieve \eqref{eqn:pred_set}
is to partition the calibration data, to obtain multiple groups of observations that are exchangeable with the test set. 
Specifically, suppose $n = mq+r$ where $q$ is a non-negative integer and $0 \le r \le m-1$. Let
\[\tilde{Z}_k = \{Z_{(k-1)m+1},Z_{(k-1)m+2},\ldots,Z_{km}\} \text{
for $k \in [q]$
and } \tilde{Z}_\textnormal{test} = \{Z_{n+1}, \ldots, Z_{n+m}\}.\]
Then it is clear that $g(\tilde{Z}_1), \ldots, g(\tilde{Z}_q), g(\tilde{Z}_\textnormal{test})$ are exchangeable, 
and thus we can apply 
split conformal prediction to obtain the following prediction set for $g(\tilde{Z}_\textnormal{test})$:
\begin{equation}\label{eqn:baseline_partition}
    \scalebox{1}{$\ch(\Dn) = \left[Q_{\beta}'\left(\sum_{k=1}^q \frac{1}{q+1}\delta_{g(\tilde{Z}_k)} + \frac{1}{q+1}\delta_{-\infty}\right), Q_{1-\gamma}\left(\sum_{k=1}^q \frac{1}{q+1}\delta_{g(\tilde{Z}_k)} + \frac{1}{q+1}\delta_{\infty}\right)\right],$}
\end{equation}
where $\beta, \gamma \in [0,1]$ satisfies $\beta+\gamma = \alpha$. 
For example one can set $\beta=\gamma=\alpha/2$ for the construction of a two-sided prediction interval, while $\beta=0,\gamma=\alpha$ yields a one-sided interval.
The above method achieves the coverage guarantee~\eqref{eqn:pred_set},
but the usefulness is limited to the case where $n \gg m$.
For example, if $n < m(1/\alpha-1)$ so that $q+1 < 1/\alpha$ holds, 
then it leads to a trivial prediction set.

\subsubsection{Extending split conformal prediction}\label{sec:split}

Instead of constructing exchangeable groups, one can directly leverage individual-level exchangeability.
Let 
\begin{equation}\label{eqn:uS_bS}
\begin{split}
\bS_i &= S_i\One{1 \le i \le n}+(\sup s)\One{n+1 \le i \le n+m}, \\
\uS_i &= S_i\One{1 \le i \le n}+(\inf s)\One{n+1 \le i \le n+m},
\end{split}
\end{equation}
where $S_i = s(Z_i)$.
 For $s_1 \le s_2 \le \ldots \le s_m$, we define $h(s_1, s_2, \ldots, s_m)$ as $\sup h$ if $s_m = \sup s$ and $h$ is not well-defined, e.g., $\sup s = +\infty$ and $h(s_1,\ldots,s_m) = \sum_{j=1}^m s_i$. Similarly, we define $h(s_1, s_2, \ldots, s_m)$ as $\inf h$ 
 if $s_1 = \inf s$ and $h$ is not well-defined; while noting that only one of the two cases can occur below. 
 Then, the adapted split conformal prediction set $\ch(\Dn)$ is defined as:

\begin{align}\label{eqn:baseline_CP}
\scalebox{1}{$\ch(\Dn) = \left[Q_{\beta}'\left(\sum_{\substack{1 \le i_1 < \ldots\\ < i_m \le n+m}} \frac{\delta_{h(\uS_{i_1}, \ldots, \uS_{i_m})}}{\binom{n+m}{m}} \right), Q_{1-\gamma}\left(\sum_{\substack{1 \le i_1 < \ldots\\ < i_m \le n+m}} \frac{\delta_{h(\bS_{i_1}, \ldots, \bS_{i_m})}}{\binom{n+m}{m}} \right)\right],$}
\end{align}

where $\beta, \gamma \ge 0$ are predefined levels satisfying $\beta+\gamma=\alpha$. It can be shown that this is a valid distribution-free prediction set,
based on 
arguments similar to those used in the proof for split conformal prediction.
Specifically, under Condition~\ref{asm:target}, the prediction set $\Ch$ from~\eqref{eqn:baseline_CP} satisfies the coverage guarantee~\eqref{eqn:pred_set}.

However, this approach still faces limitations unless $n \gg m$. For instance, consider the scenario where $n = m$ . 
Then half of the $(\bar{S}_i)_{1 \le i \le n+m}$ values are $\sup s$, likely leading to a near-trivial upper bound in~\eqref{eqn:baseline_CP}.

\subsubsection{Split conformal prediction with Bonferroni correction}\label{sec:cf_bonferroni}
Alternatively, one may consider bounding individual scores and then combining them using a Bonferroni-type approach. Specifically, let $\hat{q}_{\beta/m}'$ and $\hat{q}_{1-\gamma/m}$ denote the lower and upper score bounds obtained from split conformal prediction, using the adjusted level $\beta/m$ and $\gamma/m$ (where $\beta+\gamma=\alpha$):
\begin{align*}
    \hat{q}_{\beta/m}' = Q_{\beta/m}'\left(\sum_{i=1}^n \tfrac{1}{n+1}\delta_{S_i} + \tfrac{1}{n+1}\delta_{\inf s}\right),
    \hat{q}_{1-\gamma/m} = Q_{1-\gamma/m}\left(\sum_{i=1}^n \tfrac{1}{n+1}\delta_{S_i} + \tfrac{1}{n+1}\delta_{\sup s}\right).
\end{align*}
Then the following prediction set attains the coverage guarantee at level $1-\alpha$:
\[\left[h(\hat{q}_{\beta/m}',\hat{q}_{\beta/m}',\cdots,\hat{q}_{\beta/m}'),\;h(\hat{q}_{1-\gamma/m},\hat{q}_{1-\gamma/m},\cdots,\hat{q}_{1-\gamma/m})\right].\]
The proof follows directly from the union bound and the monotonicity of $h$. 
This method suffers from issues similar to the previous ones: unless $n > m/\alpha$ (i.e., $n \gg m$), we have $\hat{q}_{\beta/m}' = \inf s$ and $\hat{q}_{1-\gamma/m} = \sup s$, which lead to a trivial prediction set.

\subsubsection{Extending full conformal prediction}
\label{full}
To avoid the issue of having a large mass at $\infty$ or $-\infty$, one may try to construct a full conformal-type prediction set instead of relying on split conformal-type constructions. For example, we can first construct a joint prediction set
$\Ch(X_{n+1}, \ldots, X_{n+m})$
for $(y_{(n+1):(n+m)})$ as
\begin{equation}\label{eqn:baseline_full_CP}
\scalebox{0.95}{$\bigg\{\tilde{y}=(y_{(n+1):(n+m)}) : h(S_{(n+1):(n+m)}^{\tilde{y}}) 
\le Q_{1-\alpha}\Big(\sum\limits_{1 \le i_1 < \ldots < i_m \le n+m} \frac{1}{\binom{n+m}{m}} \delta_{h(S_{i_1}^{\tilde{y}}, \ldots, S_{i_m}^{\tilde{y}})}\Big)\bigg\}$},
\end{equation}
where $S_i^{\tilde{y}} = s^{\tilde{y}}(X_i,Y_i)$ and $s^{\tilde{y}}$ is the nonconformity score constructed from $(X_1,Y_1),\ldots,$ $(X_n,Y_n)$ and $(X_{n+1}, y_{n+1}), \ldots, (X_{n+m},y_{n+m})$---this step is essentially equivalent to the transductive conformal prediction~\cite{vovk2013transductive}. Then the prediction set for $g(\{Z_{n+1}, \ldots, Z_{n+m}\})$ can be constructed as
\[\scalebox{0.95}{$\ch(\Dn) = \left\{g(\{(X_{n+1},y_{n+1}), \ldots, (X_{n+m},y_{n+m})\})) : (y_{(n+1):(n+m)}) \in \Ch(X_{n+1}, \ldots, X_{n+m})\right\}$}.\]

However, this full-conformal type procedure suffers greatly from a heavy computational load. 
Computing the prediction set~\eqref{eqn:baseline_full_CP} requires repeating the computation of scores and quantiles for all tuples $(y_{(n+1):(n+m)})$ in $\mathbb{R}^m$.
Even if we carry out these steps on a grid, the number of steps increases exponentially with the size of the test set, making this procedure computationally infeasible in most practical scenarios.

To summarize, none of these direct approaches are practically viable in the setting we are interested in---in terms of the usefulness of the prediction set or the computational burden---and therefore will not be given further consideration.

\subsection{Proposed method: batch PI}
\label{bpim}

In this section, we introduce our batch PI procedure,
which
can be less conservative and more computationally efficient than the baseline methods described above.
To introduce our method, 
it is helpful to review the idea of split conformal prediction. 
Suppose we have only one test input $X_{n+1}$. 
The first step is to construct a nonconformity score function $s:\X \times \Y \rightarrow \R$; based on data that is independent of the calibration and test datasets. 
Let us write $S_i = s(X_i,Y_i)$ for $i \in [n+1]$. 
The split conformal prediction set is given by
\begin{equation}\label{eqn:split_CP}
    \Ch(x) = \left\{y \in \Y : s(x,y) \le \textnormal{$\lceil(1-\alpha)(n+1)\rceil$-th smallest value of $S_1,S_2,\ldots,S_n$}\right\}.
\end{equation}
It is known that 
if $(X_1,Y_1), \ldots, (X_n,Y_n), (X_{n+1},Y_{n+1})$ are exchangeable, 
the prediction set $\Ch$ from~\eqref{eqn:split_CP} satisfies the following coverage guarantee~\citep{vovk2005algorithmic}:
\(\mathbb{P}\big\{Y_{n+1} \in \Ch(X_{n+1})\big\} \ge 1-\alpha.\)

The key intuition is as follows:
Let $S_{(1)}, \ldots, S_{(n)}$ be the order statistics of $S_1,\ldots, S_n$,
breaking ties uniformly at random. 
Then, the rank $R \in [n+1]$ such that $S_{(R)}$ is the smallest upper bound among the observed scores for the unobserved score $S_{n+1}$ follows a uniform distribution over $[n+1]$;
where we define $S_{(n+1)} = +\infty$.
Then, because $Y_{n+1} \in \Ch(X_{n+1})$ is implied by 
$R \le \lceil(1-\alpha)(n+1)\rceil$,
the coverage probability is at least $1-\alpha$.

We now consider the setting of multiple test points (test size $m\ge1$).
Since we will need to consider not just one rank, but rather the ranks of all the test data points among the $n$ calibration data points,  
we define the set $H$ of monotone non-decreasing sequences of length $m$, of positive integers between one and $n+1$ as
\begin{align}\label{h}
    H &= \left\{r_{1:m} : 1 \le r_1 \le \ldots \le r_m \le n+1\right\}.
\end{align}
Note that $|H|  = _{n+1}\mathrm{H}_m = \binom{n+m}{m}$. 
This set will represent the ranks of the test data points among the calibration data points\footnote{
Denoting the order statistics of the test scores $S_{n+1}, \cdots, S_{n+m}$ as $S_{(1)}^\text{test}, \cdots, S_{(m)}^\text{test}$, our strategy is to bound each order statistic---which is unobserved---by one of the observed scores. Let $S_{(1)}, \cdots S_{(n)}$ be the order statistics of the calibration scores which we have access to. 
Now, for each $j=1,2,\cdots,m$, we consider the smallest  (observed) $S_{(r_j)}$ which bounds (the unobserved) $S_{(j)}^\text{test}$. We can then bound our target as well. For example, if we are interested in the mean, we leverage
\(({S_{(1)}^\text{test}+ \cdots + S_{(m)}^\text{test}})/{m} \leq ({S_{(r_1)}+\cdots+S_{(r_m)}})/{m}.\)
This motivates the definition of $H$ as the set of ranks the test scores.}.

Moreover, we also need a way to order these ranks.
In the standard conformal case where $m=1$, the ranks are ordered as $1\le \ldots\le n+1$, but for our case,
there is no default ordering.
Hence to allow for the maximum flexibility, we introduce a general
\emph{rank-ordering function}
$\tilde{h} : H \rightarrow \R$ that we will use to prioritize the ranks. We will later discuss at length the choice of this function.

Given the
rank-ordering function
$\tilde{h} : H \rightarrow \R$, 
as well as lower and upper error levels $\beta, \gamma \in [0,1]$ satisfying $\beta+\gamma = \alpha$,
we consider the following two quantiles of the distribution of the rank-ordering function given a uniform distribution over the set $H$,
\begin{equation}\label{eqn:q_L_U}
    q_L = Q_\beta'\bigg(\sum_{r_{1:m} \in H}  \frac{1}{\binom{n+m}{m}}\delta_{\tilde{h}(r_{1:m})}\bigg), \quad 
    q_U = Q_{1-\gamma}\bigg(\sum_{r_{1:m} \in H} \frac{1}{\binom{n+m}{m}}\delta_{\tilde{h}(r_{1:m})}\bigg).
\end{equation}
By definition,
if $R_{1:m}$ is distributed uniformly over $H$, then $\mathbb{P}\big\{\tilde{h}(R_{1:m}) \in [q_L, q_U]\big\}\ge 1-\alpha$.
However, since we are interested in covering the values of the function $h$ (or equivalently $g$),
we also need a way to define an appropriate range of $h$ values.
We do this by first considering the pre-image of $[q_L,q_U]$  under $\tilde h$, and then considering its image under $h$.
It turns out that we also need to consider certain corner cases (e.g., when the rank is $n+1$), 
and so with $S_{(0)} = \inf s$ and $S_{(n+1)} = \sup s$\footnote{The notations $S_{(0)}$ and $S_{(n+1)}$ are introduced solely for notational convenience in the expressions for $B_L$ and $B_U$, and they do not correspond to actual order statistics.}, 
 we define
\begin{equation}\label{eqn:B_L_U}
\begin{split}
    B_L&= \min\left\{h(S_{(r_1-1)}, \ldots, S_{(r_m-1)}) : r_{1:m} \in H,\, \tilde{h}(r_{1:m}) \ge q_L\right\},\\
    B_U&= \max\left\{h(S_{(r_1)}, \ldots, S_{(r_m)}) : r_{1:m} \in H,\, \tilde{h}(r_{1:m}) \le q_U\right\}.
\end{split}
\end{equation}
Then we construct the
\textit{batch predictive inference (batch PI)}
prediction set as
\begin{equation}\label{eqn:batch_PI}
    \ch(\Dn) = \big[ B_L, B_U \big].
\end{equation}
See Algorithm~\ref{bpi}.
For completeness, we also provide a one-sided version of the batch PI prediction set algorithm, which simplifies slightly, see Algorithm~\ref{bpi1}.
The validity of batch PI is proved in Theorem~\ref{thm:batch_PI}.

\begin{algorithm}
\caption{Batch Predictive Inference (batch PI)}
\label{bpi}
{\bf Input:} {Calibration data $\mathcal{D}_n= \{(X_1,Y_1),(X_2,Y_2), \ldots, (X_n,Y_n)\}$. 
Score function $s: \X \times \Y \rightarrow \R$. Test set size $m$. 
Batch score function $h : \R^m_{\uparrow} \rightarrow \R$.
Rank-ordering function $\tilde{h}: \N^m \rightarrow \R$. Target coverage level $1-\alpha\in[0,1]$. Lower and upper error levels $\beta, \gamma \in [0,1]$ satisfying $\beta+\gamma = \alpha$}

{\bf Step 1:} With $H = \left\{ r_{1:m}:=(r_1,\ldots,r_m)^\top : 1 \le r_1 \le \ldots \le r_m \le n+1 \right\}$, compute the sample quantiles induced by the rank-ordering function $\tilde h$:
\[q_L = Q_\beta'\bigg(\sum_{r_{1:m} \in H}  \frac{1}{\binom{n+m}{m}}\delta_{\tilde{h}(r_{1:m})}\bigg), \quad 
    q_U = Q_{1-\gamma}\bigg(\sum_{r_{1:m} \in H} \frac{1}{\binom{n+m}{m}} \delta_{\tilde{h}(r_{1:m})}\bigg).\]

{\bf Step 2:} Compute the scores $S_i = s(X_i,Y_i)$ for $i=1,2,\ldots,n$.

{\bf Step 3:} Compute the bounds, with $S_{(0)} = \inf s$, and $S_{(n+1)} = \sup s$:
\begin{align*}
    B_L&= \min\left\{h(S_{(r_1-1)}, \ldots, S_{(r_m-1)}) : r_{1:m} \in H,\, \tilde{h}(r_{1:m}) \ge q_L\right\},\\
    B_U&= \max\left\{h(S_{(r_1)}, \ldots, S_{(r_m)}) : r_{1:m} \in H,\, \tilde{h}(r_{1:m}) \le q_U\right\}.
\end{align*}

{\bf Return:} Prediction set $\ch(\mathcal{D}_n)=\big[ B_L, B_U \big]$
\end{algorithm}

\begin{theorem}[Validity of batch PI]\label{thm:batch_PI}
Suppose that Condition~\ref{asm:target} holds, and that the data points $Z_1,\ldots,Z_n, Z_{n+1}, \ldots, Z_{n+m}$ are exchangeable. Then the batch PI prediction set from~\eqref{eqn:batch_PI} satisfies
$\PP{g(\{Z_{n+1}, \ldots, Z_{n+m}\}) \in \ch(\Dn)} \ge 1-\alpha$.
\end{theorem}

The proof is deferred to the Appendix, and here we offer some intuition. 
Suppose the scores $S_1,\ldots,S_{n+m}$ are distinct almost surely\footnote{Almost sure distinctness is not an assumption of the theorem; it is assumed here solely for simplicity in the intuitive proof sketch.}, and define $R_{n+1}, \ldots, R_{n+m}$ as
\[R_{n+j} = \min\{r \in \{1,2,\ldots,n\} : S_{(r)} \ge S_{n+j}\}, \textnormal{ for } j \in [m],\]
where we let $R_{n+j} = n+1$ if $S_{(n)} < S_{n+j}$. Let $(R_{(n+j)})_{j \in [m]}$ be their order statistics. 
Through the exchangeability condition, 
it follows that 
$(R_{(n+1)}, \ldots, R_{(n+m)}) \sim \textnormal{Unif}(H)$.
Thus, for any subset $I$ of $H$ with $|I| \ge (1-\alpha)|H|,$  $\PP{(R_{(n+1)}, \ldots, R_{(n+m)}) \in I}$ 
$ \ge 1-\alpha$.

Denoting the $j$-th order statistics in $S_{n+1},\cdots,S_{n+m}$ as $S_{(j)}^\text{test}$ for $j \in [m]$, we thus have
by construction that
\[\PP{h(S_{(1)}^\text{test}, \ldots, S_{(m)}^\text{test}) \in [B_L, B_U]} \ge \PP{h(S_{R_{(n+1)}}, \ldots, S_{R_{(n+m)}}) \in [B_L, B_U]} \ge 1-\gamma,\]
as desired.
While the fully rigorous proof 
follows a similar argument, 
it requires more elaborate reasoning.

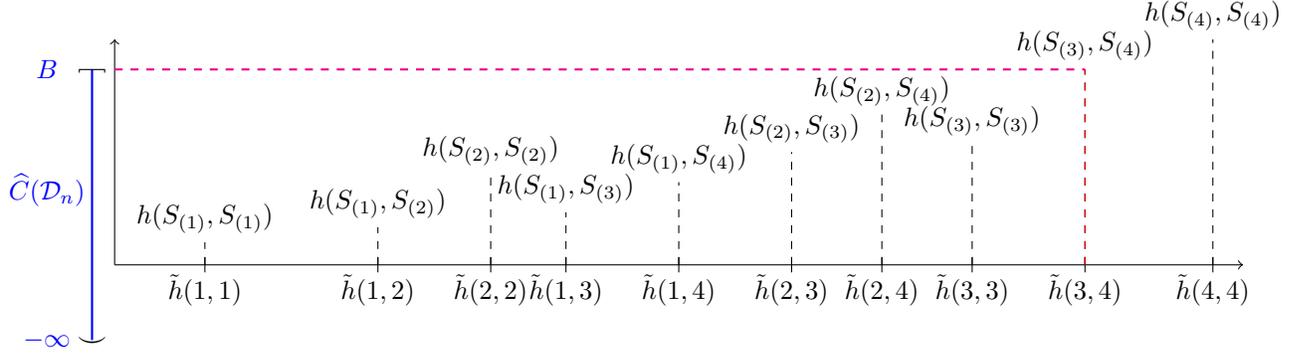
\begin{figure}[ht]
  \centering
\begin{tikzpicture}[scale=0.8]
   \pgfmathsetmacro{\xa}{1.2}
    \pgfmathsetmacro{\xb}{3.5}
    \pgfmathsetmacro{\xc}{4.8}
    \pgfmathsetmacro{\xd}{6.0}
    \pgfmathsetmacro{\xe}{7.5}
    \pgfmathsetmacro{\xf}{9.0}
    \pgfmathsetmacro{\xg}{10.2}
    \pgfmathsetmacro{\xh}{11.4}
    \pgfmathsetmacro{\xi}{12.9}
    \pgfmathsetmacro{\xj}{14.6}

    \pgfmathsetmacro{\ya}{0.3}
    \pgfmathsetmacro{\yb}{0.5}
    \pgfmathsetmacro{\yc}{1.3}
    \pgfmathsetmacro{\yd}{0.7}
    \pgfmathsetmacro{\ye}{1.1}
    \pgfmathsetmacro{\yf}{1.5}
    \pgfmathsetmacro{\yg}{2.2}
    \pgfmathsetmacro{\yh}{1.6}
    \pgfmathsetmacro{\yi}{2.6}
    \pgfmathsetmacro{\yj}{3.1}

\node (a) at (\xa,0) [below] {$\tilde{h}(1,1)$};
\node (b) at (\xb,0) [below] {$\tilde{h}(1,2)$};
\node (c) at (\xc,0) [below] {$\tilde{h}(2,2)$};
\node (d) at (\xd,0) [below] {$\tilde{h}(1,3)$};
\node (e) at (\xe,0) [below] {$\tilde{h}(1,4)$};
\node (f) at (\xf,0) [below] {$\tilde{h}(2,3)$};
\node (g) at (\xg,0) [below] {$\tilde{h}(2,4)$};
\node (h) at (\xh,0) [below] {$\tilde{h}(3,3)$};
\node (i) at (\xi,0) [below] {$\tilde{h}(3,4)$};
\node (j) at (\xj,0) [below] {$\tilde{h}(4,4)$};

    \draw[->] (0,0) -- (15,0);
    \draw[->] (0,0) -- (0,3);

    \foreach \x in {\xa,\xb,\xc,\xd,\xe,\xf,\xg,\xh,\xi,\xj} {
        \draw (\x,0.1) -- (\x,-0.1);
    }

\node at (\xa,\ya) [above] {$h(S_{(1)}, S_{(1)})$};
\node at (\xb,\yb) [above] {$h(S_{(1)}, S_{(2)})$};
\node at (\xc,\yc) [above] {$h(S_{(2)}, S_{(2)})$};
\node at (\xd,\yd) [above] {$h(S_{(1)}, S_{(3)})$};
\node at (\xe,\ye) [above] {$h(S_{(1)}, S_{(4)})$};
\node at (\xf,\yf) [above] {$h(S_{(2)}, S_{(3)})$};
\node at (\xg,\yg) [above] {$h(S_{(2)}, S_{(4)})$};
\node at (\xh,\yh) [above] {$h(S_{(3)}, S_{(3)})$};
\node at (\xi,\yi) [above] {$h(S_{(3)}, S_{(4)})$};
\node at (\xj,\yj) [above] {$h(S_{(4)}, S_{(4)})$};

    \draw[dashed] (\xa,0) -- (\xa,\ya);
    \draw[dashed] (\xb,0) -- (\xb,\yb);
    \draw[dashed] (\xc,0) -- (\xc,\yc);
    \draw[dashed] (\xd,0) -- (\xd,\yd);
    \draw[dashed] (\xe,0) -- (\xe,\ye);
    \draw[dashed] (\xf,0) -- (\xf,\yf);
    \draw[dashed] (\xg,0) -- (\xg,\yg);
    \draw[dashed] (\xh,0) -- (\xh,\yh);
    \draw[dashed] (\xi,0) -- (\xi,\yi);
    \draw[dashed] (\xj,0) -- (\xj,\yj);

    \draw[red, dashed] (\xi,0) -- (\xi,\yi);
    \draw[magenta,dashed, thick] (0,\yi) -- (\xi,\yi);

    \pgfmathsetmacro{\delt}{-0.9}
    \node at (\delt,\yi) [blue, font=\scriptsize] {$B$};
    \node at (\delt,-1) [blue, font=\scriptsize] {$-\infty$};
    \node at (\delt,1) [blue, font=\scriptsize] {$\ch(\Dn)$};
    
    \pgfmathsetmacro{\del}{-0.3}
    \draw[blue, thick] (\del,-1) -- (\del,\yi);
    \node at (\del,-1) [rotate=90, font=\scriptsize] {$($};
    \node at (\del,\yi) [rotate=90, font=\scriptsize] {$]$};
\end{tikzpicture}
  \caption{\footnotesize An illustration of the batch PI method with $n=3$ calibration data points, $m=2$ test data points, and coverage $1-\alpha=0.9$. 
  Here we show hypothetical (arbitrarily chosen) values for $\tilde h$ and $h$.
  The values of $h$ shown satisfy the monotonicity constraint from Assumption \ref{asm:target}, which 
  for pairs $1\le i\le j\le 4$ and $1\le k\le l\le 4$
  reduces to $h(S_{(i)}, S_{(j)}) \le h(S_{(k)}, S_{(l)})$ if $i\le j$ and $k\le l$.
  The value $q$ is defined as the $(1-\alpha)$-th quantile of the $\tilde h$ values. The value $B$ is defined as the maximum of the $h$ values to the ``left" of $q$. Then the batch PI prediction set is $\ch(\mathcal{D}_n)=\big(-\infty, B \big]$, and is shown on the left.}
  \label{fig:vis}
\end{figure}

While batch PI offers valid coverage, computing it requires finding the quantiles $q_L,q_U$,
as well as the interval endpoints $B_L,B_U$.
Specifically, the procedure includes the following computations:
\begin{enumerate}
\item $q_L$ and $q_U$ involves the computation of $\tilde{h}(r_1,\ldots,r_m)$ for $\binom{n+m}{m}$ elements in $H$.
\item $B_L$ and $B_U$ involves the computation of $h(S_{(r_1)},\ldots,S_{(r_m)})$ for $\lceil(1-\alpha)\binom{n+m}{m}\rceil$ rank vectors. 
\end{enumerate}
Since $\binom{n+m}{m} \sim (1+n/m)^m$, the computational cost of an enumeration-based approach for batch PI can be extremely high when there are many calibration and test datapoints.

To confirm that this computation is indeed hard in general, we 
take the perspective of standard computational complexity theory \citep[e.g.,][]{garey1979computers}, 
where the difficulty of problems is characterized according to the number of steps it takes to execute them on a standard model of computation called the Turing machine.
Tractable problems usually have a polynomial running time,
while there is a potentially broader class of problems---called NP---whose solutions can be verified in polynomial time.
There is a large set of difficult combinatorial problems---called NP-hard problems---that are at least as hard as any problem in NP.
By showing that solving the prediction set problem can be used to solve the so-called vertex cover problem \citep[e.g.,][]{garey1979computers}, 
we show that computing batch PI is in general NP-hard.

\begin{proposition}[NP-hardness of Batch PI]\label{prop:np_hard}
Computing $B_L$ and $B_U$ in~\eqref{eqn:B_L_U} is NP-hard (as a function of $n$) for general functions $h,\tilde{h}$, even when $n=m$.   
\end{proposition}

However, 
we will show in the remainder of the paper that
the computation can often be simplified at a feasible computational cost for target functions $h$ of practical interest: functions of a small number of quantiles (including single quantiles) and functions with a compositional structure.

\begin{remark}[Choice of the rank ordering function]\label{chr} 

For the choice of the rank ordering function $\tilde{h}$, we have the following considerations.
To ensure validity, this function cannot depend on the calibration scores $S_1,\ldots, S_n$.
However, 
to obtain a short and informative prediction sets, the values 
$h(S_{(r_1)}, \ldots, S_{(r_m)})$ when varying 
$(r_1,\ldots,r_m)$
should be similarly ordered as the values $\tilde{h}(r_1,\ldots,r_m)$. To see this, observe that the upper bound $B_U$ in~\eqref{eqn:B_L_U} is, roughly speaking, defined as the ``maximum of the $h$ values among those with small $\tilde{h}$ values". We describe below two heuristic strategies to achieve this goal, and provide a more detailed discussion in~\Cref{sec:ex_rank_ordering}.\\

{\bf Strategy 1: Rank-ordering functionally identical to the batch score.}
In many settings, a simple choice would be to set $\tilde{h}=h \big|_{H}$, namely the restriction of the batch score function to the set of ranks (if that restriction is well defined). 
For instance, if we are interested in the mean of test scores, i.e., $h(s_1,\ldots,s_m) = \frac{1}{m}\sum_{j=1}^m s_j$, then one choice would be to set $\tilde{h}(r_1,\ldots,r_m) = \frac{1}{m}\sum_{j=1}^m r_j$. 
This ensures that the mean of the scores corresponding to a ``smaller" rank vector tends to be smaller than that corresponding to a ``larger" rank vector. \\

{\bf Strategy 2: Rank ordering based on independent split.}
Another approach 
is to use a split
$\tilde{Z}_1, \ldots, \tilde{Z}_n$
of the data to construct $\tilde{S}_1= s(\tilde{Z}_1), \ldots, \tilde{S}_n=s(\tilde{Z}_n)$
with the same distribution as 
 $S_1,\ldots, S_n$ from the remaining split (which will be used in the batch PI procedure).
Then we can consider the rank-ordering function defined as $\tilde{h}(r_1,\ldots,r_m) = h(\tilde{S}_{(r_1)}, \ldots, \tilde{S}_{(r_m)})$. 
\end{remark}

\subsection{Computationally tractable examples of batch PI}
We now turn to discussing how the batch PI procedure simplifies to become computationally tractable in examples of interest.

\subsubsection{Inference on a quantile}\label{sec:examples}

Given $\delta\in(0,1)$,
consider forming a prediction set for
 the $(1-\delta)$-th sample quantile of the unobserved scores $S_{n+1}, \ldots, S_{n+m}$,
\[S_{(\zeta)}^\textnormal{test} = \textnormal{$\zeta$-th smallest value in $(S_{n+1}, S_{n+2}, \ldots, S_{n+m})$, where $\zeta = \lceil (1-\delta) m\rceil$}.\]
This problem has many motivations, for instance in predicting tail events. 
Consider for instance predicting the 95th percentile of the stock returns among several stocks. This becomes a problem of 
predictive
inference on the quantiles.
Similarly, if we are interested in the median of the hours of sunshine or rain levels over the next few days (or locations, etc), this is a problem of 
predictive
inference on the quantiles.

Formally, inference on $S_{(\zeta)}^\textnormal{test}$
 corresponds to the batch score
function $h:(s_1,s_2,\ldots,s_m) \mapsto s_{\zeta}$ in Condition~\ref{asm:target}. 
Observe that for this special case, we have full access to the ordering of $h$ values without knowing the exact score values, i.e., we know $S_{(i_1)} \le S_{(i_2)}$ when $i_1 \le i_2$, even if the actual values of $S_{(i_1)}$ and $S_{(i_2)}$ are unknown.
Therefore, 
denoting 
by $r_{(\zeta)}$ the $\zeta$-th smallest element in $r=(r_1,\ldots,r_m)$, 
we can set $\tilde{h}(r_1,\ldots,r_m) = r_{(\zeta)}$. 
This choice of $\tilde{h}$ recovers the exact ordering of $h$ values, i.e.,
\[h(S_{(r_1)},\ldots,S_{(r_m)}) \le h(S_{(r'_1)},\ldots,S_{(r'_m)})
\;\text{ if and only if }\;
\tilde{h}(r_1,\ldots,r_m) \le \tilde{h}(r'_1,\ldots,r'_m).\]
Thus, as per our discussion from Remark \ref{chr}, this choice of $\tilde h$ is ``optimal" in a sense.
Then, by observing\footnote{Here \( p_{n,m,\zeta} \) is the \emph{probability mass function of the \(\zeta\)-th order statistic} from a random sample of size \( m \) drawn \emph{without replacement} from a finite population of size \( n + m \)
\citep[e.g.,][p. 243]{wilks1962mathematical}.
}
\[p_{n,m,\zeta}(k) := \frac{|\{r \in H : r_{(\zeta)} = k\}|}{|H|} = \frac{_{k}\mathrm{H}_{\zeta-1} \cdot _{n-k+2}\mathrm{H}_{m-\zeta}}{_{n+1}\mathrm{H}_m} = \frac{\tbinom{k+\zeta-2}{\zeta-1}\tbinom{n+m-k-\zeta+1}{m-\zeta}}{\binom{n+m}{{m}}}\]
for $k \in [n+1]$, 
we have the following explicit expressions:
\begin{equation}\label{eqn:q_L_U_quantile}
    q_L = Q_{\beta}'\bigg(\sum\nolimits_{k=1}^{n+1} p_{n,m,\zeta}(k)\cdot \delta_{k}\bigg),\quad q_U = Q_{1-\gamma}\bigg(\sum\nolimits_{k=1}^{n+1} p_{n,m,\zeta}(k)\cdot \delta_{k}\bigg).
\end{equation}
Next, observe that $B_U$ in~\eqref{eqn:B_L_U} for this setting can be simplified as
$B_U = S_{(q_U)}$, and similarly $B_{L}= S_{(q_L-1)}$. 
Therefore, 
batch PI reduces to 
the following $(1-\alpha)$-prediction set for $S_{(\zeta)}^\textnormal{test}$:
\begin{equation}\label{eqn:ch_quantile}
    \ch^{\textnormal{bPI-q}}(\Dn) = \left[S_{(q_L-1)}, S_{(q_U)}\right].
\end{equation}

\begin{corollary}[Batch PI for quantiles]\label{cor:quantile}
If the data points $Z_1,\ldots,Z_n, Z_{n+1}, \ldots, Z_{n+m}$ are exchangeable,
    the prediction set $\ch^{\textnormal{bPI-q}}(\Dn)$ from~\eqref{eqn:q_L_U_quantile} and~\eqref{eqn:ch_quantile} satisfies
    \(\mathbb{P}\{S_{(\zeta)}^\textnormal{test} \in \ch^{\textnormal{bPI-q}}(\Dn)\} \geq 1-\alpha.\)
    Furthermore, if the scores $(S_i)_{i \in [n+m]}$ are all distinct almost surely, the following holds:
    \[\mathbb{P}\{S_{(\zeta)}^\textnormal{test} \in \ch^{\textnormal{bPI-q}}(\Dn)\}\leq  1-\alpha + \eps_{n,m,\zeta}, \text{ where } \eps_{n,m,\zeta} = \max_{k \in [n+1]} p_{n,m,\zeta}(k) = O(\tfrac{1}{n}).\]
\end{corollary}

Above, we additionally obtain an upper bound on the coverage for inference on quantiles. 
This equals $1 - \alpha + \frac{1}{n + 1}$ when $m = \zeta = 1$---i.e., the above result recovers the guarantee for the standard conformal prediction when the test size is one. 
For this procedure, the computational cost arises only from computing $q_L$ and $q_U$, and is relatively low, since these are quantiles of discrete distributions with support size $n+1$.

Moreover, in this case, we can also show an optimality result. 
Consider prediction sets of the form $\{y : s(x,y) \leq S_{(r)}\}$, where  $r \in [n+1]$. 
Indeed, in the simpler case of standard conformal prediction, it is known that all prediction sets of the form 
$\Ch(x) = \{y \in \Y : s(x,y) \le f(S_1, \ldots, S_n)\}$ 
that have distribution-free coverage and where 
$f$ is a symmetric function are of this form 
\citep{robbins1944distribution}.
Thus, the focus on such prediction sets
is not restrictive. 
Now, based on the arguments in Section~\ref{bpim}, the coverage rate of a prediction set of this form is equal to $\PP{R_{(n+m_\delta)} \leq r}$, and the batch PI method finds the smallest $r$ such that this probability is at least $1 - \alpha$, based on the exact distribution of $R_{(n + m_\delta)}$.
This also leads to a tight upper bound, and implies that it dominates any other prediction set of the form $\{y : s(x, y) \leq S_{(r)}\}$ that achieves valid coverage.

\begin{proposition}[Optimality of batch PI for the quantile]\label{prop:optimality_batchPI}
Consider constructing prediction sets $\Ch(x) = \{y \in \Y : s(x,y) \le S_{(r)}\}$ for some fixed rank $r \in [n+1]$ for the quantile $S_{(\zeta)}^\textnormal{test}$ of the test datapoints, 
where $S_i = s(X_i, Y_i)$ are the nonconformity scores computed on the calibration data.
Among all such procedures satisfying the 
distribution-free guarantee
 \(\mathbb{P}\{S_{(\zeta)}^\textnormal{test} \in \ch^{\textnormal{bPI-q}}(\Dn)\} \geq 1-\alpha\) under exchangeability,
\emph{the batch PI procedure yields the shortest prediction sets}. 
\end{proposition}
\vspace{-8mm}

\begin{algorithm}
\caption{\footnotesize Batch PI for Inference on a Quantile}
\label{quantile_inference}
{\bf Input:} {Calibration data $\mathcal{D}_n= \{(X_1,Y_1),(X_2,Y_2), \ldots, (X_n,Y_n)\}$.
Score function $s: \mathcal{X} \times \mathcal{Y} \rightarrow \mathbb{R}$.
Test set size $m$.
Target quantile level $1-\delta \in (0,1)$.
Target coverage level $1-\alpha \in [0,1]$.
Lower and upper error levels $\beta, \gamma \in [0,1]$ satisfying $\beta + \gamma = \alpha$.}

\textbf{Step 1:} With $\zeta = \lceil (1-\delta) m \rceil$,
compute the sample quantiles:
\[
    \scalebox{0.95}{$q_L = Q_{\beta}'\left(\sum\limits_{k=1}^{n+1} \frac{\binom{k+\zeta-2}{\zeta-1}\binom{n+m-k-\zeta+1}{m-\zeta}}{\binom{n+m}{m}}\cdot \delta_{k}\right), \quad q_U = Q_{1-\gamma}\left(\sum\limits_{k=1}^{n+1} \frac{\binom{k+\zeta-2}{\zeta-1}\binom{n+m-k-\zeta+1}{m-\zeta}}{\binom{n+m}{m}}\cdot \delta_{k}\right)$}.
\]

\textbf{Step 2:} Compute the scores $S_i = s(X_i,Y_i)$ for $i=1,2,\ldots,n$; denote $S_{(n+1)} = +\infty$.

{\bf Return:} Prediction set $\ch^{\textnormal{bPI-q}}(\mathcal{D}_n)=\left[ S_{(q_L-1)}, S_{(q_U)} \right]$
\end{algorithm}

In Section~\ref{sec:sparse}, we extend the above method to describe the simplification of the batch PI procedure for general 
\emph{sparse functions} 
$h$, 
where $h(s_1,\ldots,s_m)$ depends only on a small number of the $s_j$s.

\subsubsection{Inference on the mean and general compositionally structured functions}
\label{comp-struct}

In this section, we show how to compute the batch predictive inference prediction sets efficiently in a general setting 
where the rank ordering and batch score functions have a certain
compositional structure,
a setting that includes the important case of the mean.
Recall that for a given 
rank-ordering function
$\tilde{h} : \N^m \rightarrow \R$, 
the computation of $q_L,q_U$ 
from \eqref{eqn:q_L_U} requires,
for all \( k \in \text{range}( \tilde{h} ) \), that we compute the number of \(r_{1:m} \in H\), such that
$\tilde{h}(r_{1:m}) = k$.

To introduce our algorithm and ideas, let us first
consider the simpler case where the function $\tilde{h}$ is the sum, 
\( \tilde{h}(r_{1:m}) = \sum_{j=1}^{m} r_j \),
for all $r_{1:m}\in H$.
This is equivalent to the mean, up to scale.
In that case, the problem becomes to find the number---denoted \( C_{m,n,k} \)---of 
the positive integer solutions $r_{1:m}= (r_1, \ldots, r_m)$
to the equation $r_1 + r_2 + \ldots + r_m = k$
with \( 1 \le r_1 \le \ldots \le r_m \le n \).
These are known as 
the number of partitions of $k$ with at most $m$ parts, 
each of size at most $n$ \citep{stanley2011enumerative}, and efficient recursive algorithms are known for computing them.
Once we have \( C_{m,n,k} \), we can simplify
$q_U$ to $q_U = Q_{1-\gamma}\Big(
\sum_{k \in \text{range}( \tilde{h} ) } \frac{C_{m,n,k}}
{\binom{n+m}{m}}\delta_{k}\Big).$

For pedagogical purposes, we first present the idea for computing these numbers for the mean.
Consider any $a\in [n]$.
For a solution $r_{1:m}$, 
if \( r_m = a \), then \( r_1 + \ldots + r_{m-1} = k - a \). 
By definition, there are $C_{m-1, n, k-a}$ such solutions.
Thus, by considering all possible values of $a$ for $r_m\in [n]$, we obtain the recursion
$
C_{m,n,k} = \sum_{a=1}^{n} C_{m-1, n, k-a}.
$

More generally, consider finding
the number of 
\( 1 \le r_1 \le \ldots \le r_m \le n \) such that
$\tilde{h}(r_{1:m}) = k$.
Suppose that for all \(r \ge 1\), there is a strictly increasing  function  \(\tilde\Gamma(\cdot; r):\{0,1,\ldots\}\to \{0,1,\ldots\} \) 
such that
for any \(\kappa \ge 1\),
\beq\label{htc}
\tilde h(r_{1:\kappa}) = \tilde\Gamma(\tilde h(r_{1:(\kappa-1)}); r_\kappa).
\eeq
Here the function $\tilde\Gamma$ could be made to depend on $\kappa$, i.e., having $\tilde h(r_{1:\kappa}) = \tilde\Gamma_{\kappa}(\tilde h(r_{1:(\kappa-1)}); r_\kappa),
$ but we omit this for simplicity.
For instance, for 
our previous example of the sum,
$\tilde h(s_{1:\kappa}) = \sum_{j\in[\kappa]} s_{j}$, we can take
$
\tilde\Gamma_{}(a; r)=a+r,
$
for all positive integers $\kappa, r,a$.
Then the same reasoning by partitioning on the possible values of $r_m$ yields that
$C_{m,n,k} = \sum_{a=1}^{n} C_{m-1, a, \tilde\Gamma^{-1}(k;a)}$,
where $\tilde\Gamma^{-1}(\cdot;a)$ denotes the inverse of the function $x\mapsto\tilde\Gamma(x;a)$. 
Here, the understanding is that if the equation $\tilde\Gamma(x;a)=k$ does not have a solution in $x$, then $C_{m-1, a, \tilde\Gamma^{-1}(k;a)}=0$.

This recursion immediately leads to a  dynamic programming algorithm similar to the one for the mean.
For all algorithms mentioned in this section, see~\Cref{alg}.
The initial conditions $C_{1,n,k}$ are either one or zero, depending on whether or not the equations 
$\tilde\Gamma(0;s)= k$
have a solution $1 \le s \le n$.  \\

The running time of this algorithm is \( O(mkn^2) \) flops, due to a triple loop (each going up to $m,k,n$, respectively) and as the innermost computation takes $O(n)$ steps.
Thus, since the range of 
$\tilde{h}$
ranges between $m$ and $(n+1)m$, computing 
$q_U$ 
by computing $C_{m, n, k}$
for all \( k \in \text{range}( \tilde{h} ) \)
has complexity \( O(m^2kn^3) \).\footnote{Alternatively, for even faster computation with moderate sample sizes, one can estimate the quantiles $q_L$ and $q_U$ using sample quantiles. 
Specifically, drawing a sample from $H$ is equivalent to drawing $m$ samples from a uniform distribution over $[n+1]$ with replacement, allowing us to construct samples of $\tilde{h}(r_{1:m}), r_{1:m} \sim \text{Unif}(H)$. This approach leads to an accurate estimate of $q_L$ and $q_U$ if a sufficient number of samples is drawn.}\\

The computation of the interval endpoints $B_L, B_U$ from \eqref{eqn:B_L_U} can be performed efficiently in a similar way  (see \Cref{alg}). 

\begin{remark}\label{rmk:concentration}
If the calibration and test set sizes are very large, 
the above algorithms can still have a high cost.
However, in certain cases of interest, especially for the central case of the mean, 
a straightforward procedure for inference 
is based on concentration inequalities. 
For instance, 
if  $Y \in [a,b]$ almost surely, then by McDiarmid's inequality,
the prediction set
\[\scalebox{1}{$\ch(\Dn) = 
\left(\frac{1}{n}\sum_{i=1}^n Y_i \pm (b-a)\sqrt{\frac{1}{2}\left(\frac{1}{n}+\frac{1}{m}\right)\log \frac{2}{\alpha}}
\right) \cap [a,b]$}\]
has $(1-\alpha)$ coverage 
for the mean of test outcomes, under the i.i.d. assumption.
Thus, very large sample sizes $n,m$ can be handled with concentration inequalities, while for moderate sample sizes, our algorithms remain computationally efficient---under the weaker assumption of exchangeability---whereas the concentration-based method may result in trivial prediction sets. In Section~\ref{sec:exp_counterfactual}, we provide experimental results comparing the performance of the batch PI-based method and the concentration-based method.


\end{remark}

\subsection{Inference under covariate shift}\label{sec:cov_shift}

Our methods presented so far are valid when the test and calibration data 
are drawn from the same population, but this might not always hold in applications. This phenomenon has been referred to as \textit{dataset shift} \protect\citep[see, e.g.,][]{quinonero2009dataset,shimodaira2000improving,Sugiyama2012}.
An important form of dataset shift is \textit{covariate shift}: 
a changed feature distribution, and an unchanged distribution of the outcome given features. 
The shift may arise due to a change in the sampling probabilities of various sub-populations, or due to a patient's features changing over time, 
while 
the distribution of the outcome given the features stays fixed \protect\citep{quinonero2009dataset}.
There is a growing body of work on distribution-free predictive inference under covariate shift, see e.g., 
\cite{tibshirani2019conformal,qiu2023prediction, yang2022, park2021pac, cauchois2024robust, lei2021conformal}.
However, to our knowledge, methods for batch predictive inference have not been developed yet in this setting.

Here, 
we develop methods for batch predictive inference under covariate shift. 
This refers to 
the following 
distribution of the data points:
\begin{equation}\label{eqn:dist_cov_shift}
\begin{split}
    &(X_1,Y_1), (X_2,Y_2)\ldots,(X_n,Y_n) \iidsim P_X \times P_{Y \mid X},\\
    &(X_{n+1},Y_{n+1}), (X_{n+2},Y_{n+2})\ldots,(X_{n+m},Y_{n+m}) \iidsim Q_X \times P_{Y \mid X},
\end{split}
\end{equation}
where $P_X$ and $Q_X$ represent two distinct distributions on $\mathcal{X}$, and $P_{Y \mid X}$ denotes the conditional distribution of $Y$ given $X$, which is consistent across both the calibration and test datasets. 
Our objective is to construct a prediction set for a function of the test points under this setting, with coverage at least $1-\alpha$:
\begin{equation}\label{eqn:guarantee_shift}
    \Pp{Z_{1:n} \iidsim P_X \times P_{Y \mid X}, Z_{(n+1):(n+m)} \iidsim Q_X \times P_{Y \mid X}}{g(\{Z_{n+1}, \ldots, Z_{n+m}\}) \in \ch(\Dn)} \ge 1-\alpha.
\end{equation}
We consider the setting of a known 
likelihood ratio 
$dP/dQ$, which is required 
for nontrivial distribution-free prediction sets even in the setting of one test datapoint~\citep{qiu2023prediction, yang2022}. 
We develop a method leveraging rejection sampling to obtain an exchangeable dataset, 
and then applying the batch PI procedure; similarly to \cite{park2021pac,qiu2023prediction} for standard conformal prediction. We present more details in~\Cref{sec:cov_shift_appendix}.

\section{Use cases}\label{sec:applications}

In this section, 
we discuss use cases of batch PI:
(1)
inference on
counterfactual variables;
(2)
simultaneous predictive inference with PAC-coverage;
and
(3)
selection of individuals with error control--—the latter two are based on 
inference on one quantile.
All three will be
illustrated empirically  in Section~\ref{sec:simulations}.

\subsection{Inference on counterfactual variables}
 
We consider a randomized trial setting where the underlying data structure is 
\[(X_i,A_i, Y_i^{a=0}, Y_i^{a=1})_{1 \le i \le n} \iidsim P_X \times P_{A \mid X} \times P_{Y^{a=0} \mid X} \times P_{Y^{a=1} \mid X},\] where $X$ denotes the feature, $A \in \{0,1\}$ denotes the treatment, and $Y^{a=0}$ and $Y^{a=1}$ denote the counterfactual outcomes under $A=0$ and $A=1$, respectively. We only observe $(X_i,A_i, Y_i)_{1 \le i \le n}$, where we assume the consistency condition $Y_i = (1-A_i)Y_i^{a=0} + A_i Y_i^{a=1}$.

We consider the task of inference on the counterfactual outcomes $\{Y_i^{a=0} : A_i = 1\}$ in the treated group. 
Under the consistency assumption, the problem is equivalent to inference on missing outcomes/test points under covariate shift, 
with $\{(X_i, Y_i^{a=0}) : A_i = 0\}$ as the calibration set and $\{X_i : A_i = 1\}$ as the test inputs.

Therefore, based on the discussion in Section~\ref{sec:cov_shift}, 
we obtain procedures for the  following tasks:
\begin{enumerate}
    \item \textit{Inference on the mean of counterfactuals}:
    Construct $\ch(\Dn)$ such that\\
    $\PP{\frac{1}{N^1}\sum_{i : A_i = 1} Y_i^{a=0} \in \ch(\Dn)}$ $ \ge 1-\alpha$,
    where $N^1 = |\{i : A_i = 1\}|$.
    \item \textit{Inference on the median of counterfactuals}:
    Construct $\ch(\Dn)$ such that\\
    $\PP{\textnormal{Median}(\{Y_i^{a=0} : A_i = 1\}) \in \ch(\Dn)} \ge 1-\alpha$.\footnote{For inference on the median, and more generally on a quantile, we also obtain an upper bound on the coverage based on Corollary~\ref{cor:quantile}.}
    \item \textit{Inference on multiple quantiles of counterfactuals}:   
    Construct $L, U \in \R^l$ such that\\
    $\PP{L \preceq (Y_{(\zeta_1)}^{a=0}, \ldots, Y_{(\zeta_l)}^{a=0}) \preceq U} \ge 1-\alpha$,
    where $Y_{(\zeta)}^{a=0}$ is the $\zeta$-th smallest value of $\{Y_i^{a=0} : A_i = 1\}$.
\end{enumerate}

\subsection{Simultaneous predictive inference of multiple unobserved responses}\label{sec:pred_pac}

Consider
constructing prediction sets $\Ch(X_{n+1}), \Ch(X_{n+2}), \ldots, \Ch(X_{n+m})$ for $Y_{n+1}, Y_{n+2}, \ldots, Y_{n+m}$ respectively, 
such that most of the unobserved outcomes are covered by their corresponding prediction sets. 
A simple approach is to construct standard split conformal prediction sets, 
leading to 
marginal coverage for each prediction set, i.e., 
$\mathbb{P}\{Y_{n+j} \in \Ch(X_{n+j})\} \ge 1-\alpha, \text{ for all } j \in [m]$.

However, this
does not characterize the simultaneous---joint---behavior of the prediction sets. 
For instance, it does not directly guarantee how many of the test outcomes will be covered.
Since each marginal coverage guarantee is with respect to the distribution of $(X_1,Y_1), \ldots, (X_n,Y_n), (X_{n+j}, Y_{n+j})$, the $m$ coverage events $\{\{Y_{n+j} \in \Ch(X_{n+j})\}, j \in [m]\}$ have a joint distribution with a potentially complex dependence structure. 
Nonetheless, 
the distribution of the coverage 
$\frac{1}{m}\sum_{j=1}^m \mathbbm{1}\{Y_{n+j} \in \Ch(X_{n+j}) \}$
was discussed in \cite{MARQUESF2025,huang2024uqgnn}, and this enables constructing prediction sets with various guarantees.
Our goal is to construct prediction sets with the following
probably approximately correct (PAC)-type\footnote{This can also be 
viewed as an analogue of the family-wise error rate, and more generally of the $k$-family-wise error rate from multiple hypothesis testing \citep{lehmann2005generalizations}.
For a positive integer $k$, 
set
$\delta=k/m$, so that
$k=\delta m$. Then this guarantee is equivalent to 
$\PP{\sum_{j=1}^m \One{Y_{n+j} \notin \Ch(X_{n+j}) } \ge k} \le \alpha$.
Now, since $\sum_{j=1}^m \One{Y_{n+j} \notin \Ch(X_{n+j}) }$ 
is the number of errors, 
this can be viewed as a direct
analogue of the $k$-family-wise error rate \citep{lehmann2005generalizations}.
In particular, if $k=1$ (i.e., $\delta=1/m$), it can be viewed as an analogue of the  
family-wise error rate.}
\citep{Park2020}
guarantees:
\begin{equation}\label{eqn:guarantee_pac}
\scalebox{0.9}{$
    \PP{\frac{1}{m}\sum_{j=1}^m \One{Y_{n+j} \in \Ch(X_{n+j}) } \ge 1-\delta} \ge 1-\alpha,
$}
\end{equation}
where $\alpha, \delta \in (0,1)$ are predefined levels. 
This directly controls the proportion of test outcomes covered by the prediction sets.
For illustration purposes, 
we will show that the batch PI procedure can be applied to achieve the above guarantee.

Let $s:\X \times \Y \rightarrow \R^+$ be a nonconformity score, constructed independently of the calibration data. 
Define $m_\delta = \lceil(1-\delta)m\rceil$, and 
the following prediction set, 
which is a direct application of the procedure for inference on a single quantile:
\begin{equation}\label{eqn:appl_pac}
    \Ch(x) = \left\{y \in \Y : s(x,y) \le S_{(r_{\delta,\alpha})} \right\}, \text{ where } r_{\delta,\alpha} = Q_{1-\alpha}\bigg(\sum_{k=1}^{n+1} \tfrac{\binom{k+m_\delta-2}{m_\delta-1}\binom{n+m-k-m_\delta+1}{m-m_\delta}}{\binom{n+m}{m}}\cdot \delta_k\bigg).
\end{equation}
As a consequence of Corollary~\ref{cor:quantile}, we establish the following guarantee for the procedure described above.

\begin{corollary}\label{cor:appl_pac}
  If $(X_1,Y_1), \ldots, (X_{n+m},Y_{n+m})$ are exchangeable,
    then the prediction set $\Ch$ from~\eqref{eqn:appl_pac} satisfies
    \(1-\alpha \leq \PP{\frac{1}{m}\sum_{j=1}^m \One{Y_{n+j} \in \Ch(X_{n+j}) } \ge 1-\delta}\)
    \(\leq 1-\alpha+\eps_{n,m,m_\delta},\)
    where the upper bound holds under the assumption that all the scores $(s(X_i,X_i))_{i \in [n+m]}$ are almost surely distinct, and $\eps_{n,m,m_\delta}$ is defined in Corollary~\ref{cor:quantile}.
\end{corollary}

\begin{remark}[Comparison with Markov inequality-based approach]\label{rmk:pac_baseline}
The PAC-type guarantee~\eqref{eqn:guarantee_pac} can also be achieved by applying standard split conformal prediction to each test points separately, at an adjusted level $\delta \cdot \alpha$---i.e., the procedure $\ch_n^\text{Markov} = \{y \in \Y : s(x,y) \leq S_{(\lceil(1-\delta\alpha)(n+1)\rceil)}\}$. To see this, denote $C_j = \mathbbm{1}\{Y_{n+j} \in \Ch(X_{n+j})\}$. 
Then, by Markov's inequality, we have
\[\mathbb{P}\bigg\{\frac{1}{m}\sum_{j=1}^m C_j < 1-\delta\bigg\} = \mathbb{P}\bigg\{\frac{1}{m}\sum_{j=1}^m (1-C_j) > \delta\bigg\} \leq \frac{1}{\delta}\cdot\mathbb{E}\bigg[\frac{1}{m}\sum_{j=1}^m (1-C_j)\bigg] \leq \alpha,\]
provided that $\EE{C_j} \geq 1-\delta\alpha$ holds for all $j \in [m]$.

However, this method does not yield a tight bound.
In fact, 
from Proposition \ref{prop:optimality_batchPI}, 
it follows 
that the batch PI-based prediction set $\ch_n(x)$ in~\eqref{eqn:appl_pac} is always a subset of $\ch_n^\text{Markov}(x)$, for any $x \in \X$. 
We also provide relevant experimental results in Section~\ref{sec:sim_1}, where 
we show that our method outperforms 
the Markov-adjustment based method.
    
\end{remark}

\begin{remark}[Comparison with the PAC guarantee for calibration-dataset-conditional coverage]
    Consider the setting where the data points are i.i.d. 
    Let $C_j = \One{Y_{n+j} \in \Ch(X_{n+j})}$ denote the coverage indicator for the $j$th test point, $j\in [m]$, 
    and let $\mathcal{D}_\text{cal}$ denote the calibration set. 
    Then we have 
    \[C_1,\cdots,C_m \mid \mathcal{D}_\text{cal} \iidsim \textnormal{Bernoulli}(p_C), \text{ where } p_C = \PPst{Y \in \Ch(X)}{\mathcal{D}_\text{cal}},\]
    and thus $\bar{C} = \frac{1}{m}\sum_{j=1}^m C_j$ converges to $p_C$ almost surely as $m\to\infty$, conditional on $\mathcal{D}_\text{cal}$.
    It follows that
    \begin{align*}
        \PP{\bar{C} \ge 1-\delta} = \EE{\EEst{\One{\bar{C} \ge 1-\delta}}{\mathcal{D}_\text{cal}}} \xrightarrow{m \rightarrow \infty} \EE{\One{p_C \ge 1-\delta}} = \PP{p_C \ge 1-\delta},
    \end{align*}
    by applying the dominated convergence theorem twice. 
    Therefore, as $m\to\infty$,
    the prediction set~\eqref{eqn:appl_pac} 
    converges to achieving the PAC guarantee for the calibration conditional-coverage property \citep{vovk2013conditional,Park2020}
    $ \PP{p_C \ge 1-\delta} \ge 1-\alpha$.
    The advantage of
    the prediction set~\eqref{eqn:appl_pac} is that it controls the coverage rate also for small test sizes $m$.
\end{remark}

\begin{remark}
This problem was also studied previously in \cite{gazin2024transductive}, where the authors further aim to provide uniform control over the false coverage rate. This can be expressed in our notation as:
\( \PP{\forall\, \alpha \in (0,1),\;\frac{1}{m}\sum_{j=1}^m \One{Y_{n+j} \in \Ch^{(\alpha)}(X_{n+j}) } \ge 1-\gamma_{\alpha,\delta}}\)
\(\ge 1-\delta,\)
where $(\gamma_{\alpha,\delta})_{0 < \alpha < 1}$ is a family of (random) bounds.
They provide a concentration inequality-based approach to achieve this stronger notion of coverage, 
with a score of the form $S_i = |Y_i - \hat{\mu}(X_i, \mathcal{D}_\text{train}, X_{1:n+m})|$---i.e., the score is constructed using the training data, as well as the calibration and test covariates. 
They also briefly mention the weaker target (equivalent to~\eqref{eqn:guarantee_pac}) in the appendix,
providing a method based on an implicit formula---which turns out to be equivalent to the method in~\eqref{eqn:appl_pac} after reorganization.

\end{remark}

\subsection{Selection of test datapoints}\label{sec:selection}

Next, we consider
selecting the individuals in the test set whose outcome values satisfy a certain condition---for instance, selecting individuals whose outcome values exceed a threshold, i.e., $Y_i > c$ for some $c \in \R$. This setting was investigated by~\cite{jin2023selection} and~\cite{jin2023model}, where they discuss applications to candidate screening, drug discovery, etc. 
Denoting the ``null"  events as
$E_{j} =\{ Y_{n+j} \le c\}, j=1,2,\ldots,m$,
we can view this problem as controlling an error measure depending on the number of true events declared to be false. 
Previous work \citep{jin2023selection,jin2023model} has developed methods for controlling a quantity analogous to the false discovery rate~\citep{benjamini1995controlling}.
Here, we introduce a different procedure, which applies batch PI, directly controlling the number of false claims on the test set.

We assume that $Y$ is bounded below---without loss of generality, suppose $Y \ge 0$ almost surely. 
Generally, for unbounded $Y$, we can apply a monotone transformation to obtain a bounded outcome $\tilde{Y}$---e.g., $\tilde{Y} = \tanh(Y)$---and then apply the procedure below. 
Let $\hat{\mu} : \X \rightarrow \R_{\ge 0}$ be an estimated mean function, constructed on a separate independent dataset.
Let $s(x,y) = \hat{\mu}(x) \One{y \le c}$ for all $x,y$, and define $S_i = s(X_i,Y_i)$ for $i=1,2,\ldots,n$. 
We write $S_{(1)}, \ldots, S_{(n)}$ to denote the order statistics of $S_1,\ldots,S_n$. 
Next, for a target number of errors $\eta \in \{0\} \cup [m]$, let
\[\hat{T} = S_{(q_\eta)}, \text{ where }q_\eta = Q_{1-\alpha}\left(\sum_{k=1}^{n+1} \frac{\binom{k+m-\eta-2}{m-\eta-1}\binom{n+\eta-k+1}{\eta}}{\binom{n+m}{m}}\cdot \delta_{k}\right),\]
following the formula in~\eqref{eqn:q_L_U_quantile} with $\zeta = m-\eta$ and $\gamma=\alpha$.
Then we consider the following selection rule:
\begin{equation}\label{eqn:selection_rule}
    \textnormal{declare $E_j$ to be false if $\hat{\mu}(X_{n+j}) > \hat{T}$}.
\end{equation}
This satisfies the following property:
\begin{corollary}\label{cor:selection}
    Suppose $\hat{\mu}(X) \ge 0$ holds almost surely. Then the selection procedure~\eqref{eqn:selection_rule} controls the number of false claims by $\eta$ with probability at least $1-\alpha$, i.e.,
    \begin{equation}\label{eqn:fd_control}
        \PP{\sum_{j=1}^m \One{\hat{\mu}(X_{n+j}) > \hat{T}, Y_{n+j} \le c} \le \eta} \ge 1-\alpha.
    \end{equation}
\end{corollary}

If $\eta=0$, then~\eqref{eqn:fd_control} is equivalent to controlling the probablity of making at least one false claims with probability at most $\alpha$, which is analogous to the control of the family-wise error rate (FWER) in multiple hypothesis testing.
More generally,~\eqref{eqn:fd_control} is
analogous to the control of the $k$-family-wise error rate ($k$-FWER) \citep{lehmann2005generalizations} in multiple hypothesis testing.

As a remark, if we are generally interested in selecting individuals whose outcome satisfies a condition $\mathcal{C}$ using an estimator $\hat{f}(\cdot)$ (which is nonnegative), we can apply the same procedure with the score function $s(x,y) = \hat{f}(x)\One{y \text{ satisfies }\mathcal{C}}$, and then select the individuals whose $\hat{f}$ value exceeds $\hat{T}$.

\subsubsection{Comparison with p-value-based methods}\label{sec:fwer_baseline}
For the selection problem with the guarantee~\eqref{eqn:fd_control}, one might consider first constructing p-values and then applying a standard multiple testing procedure that controls the $k$-FWER~\citep{lehmann2005generalizations}. 
Specifically, we
prove the following (see Appendix~\ref{sec:proofs} for the proof):

\begin{proposition}\label{prop:k_fwer}
    For the events $E_1, \ldots, E_m$, suppose there exist random variables $p_1, \ldots, p_m$ such that $\PP{p_j \leq \alpha \text{ and } E_j \text{ holds}} \leq \alpha$ for all $\alpha \in (0,1)$ and for all $j \in [m]$. Then the selection rule that selects $E_j$ such that $p_j \leq \frac{(k+1)\alpha}{m}$ controls the $k$-FWER at level $\alpha$, i.e.,
    \begin{equation}\label{eqn:fwer_p_val}
        \PP{\sum_{j=1}^m \One{p_j \leq \frac{(k+1) \alpha}{m}, Y_{n+j} \le c} \le k} \ge 1-\alpha.
    \end{equation} 
\end{proposition}

The proof is deferred to the Appendix. Note that when $k = 0$, the procedure reduces to the simple Bonferroni method,
which enjoys FWER control. 
As the choice of the random variable $p_j$,
\cite{jin2023selection} proposes to use the following conformal p-value:
\begin{equation}\label{eqn:cf_p_val_1}
    p_j = \frac{\sum_{i=1}^n\One{c - \hat{\mu}(X_{n+j}) > Y_i - \hat{\mu}(X_i)} + 1}{n+1}.
\end{equation}
\cite{jin2024confidence} introduces a more powerful conformal p-value, defined as\footnote{\cite{jin2023selection} and \cite{jin2024confidence} discuss a more general form of these conformal p-values, of which the p-values~\eqref{eqn:cf_p_val_1} and~\eqref{eqn:cf_p_val_2} are special cases.}
\begin{equation}\label{eqn:cf_p_val_2}
    p_j = \frac{\sum_{i=1}^n\One{\hat{\mu}(X_{n+j}) < \hat{\mu}(X_i), Y_i \leq c} + 1}{n+1}.
\end{equation}
However,
the multiple testing procedure of~\cite{lehmann2005generalizations}, 
can be conservative
when combined with these p-values. 
We provide a comparison between these methods and the batch PI-based method through experiments in Section~\ref{sec:exp_fwer}.

\begin{remark}
    In the proof of Corollary~\ref{cor:selection}, we show that for a rejection threshold $\hat{T}$ and the rejection rule $\hat{\mu}(X) > \hat{T}$, the $\eta$-FWER is equal to the probability $\PP{S_{(m-\eta)}^\text{test} \leq \hat{T}}$. The batch PI procedure finds the optimal threshold $\hat{T}$ based on the exact distribution of the rank of $S_{(m-\eta)}$, and thus the resulting selection rule dominates any selection rule of the form $\hat{\mu}(X) > \tilde{T}$ with $\tilde{T}$ determined by calibration scores, including the conformal $p$-value-based methods. We omit the details here for brevity, but one can verify that the $p$-value in~\eqref{eqn:cf_p_val_2} yields a selection rule of this form, and that the $p$-value in~\eqref{eqn:cf_p_val_1} is deterministically larger than (i.e., dominated by) the $p$-value in~\eqref{eqn:cf_p_val_2}.
\end{remark}

\section{Simulations}\label{sec:simulations}

In this section, we illustrate the performance of batch PI-based procedures across different experiments\footnote{Code to reproduce the experiments is available at \url{https://github.com/yhoon31/batch-PI}.}.

\subsection{Simultaneous predictive inference of multiple unobserved outcomes}\label{sec:sim_1}

We generate the data according to the distribution
\[X \sim N_p(\mu_x,5 \cdot I_p), Y \mid X \sim \mathcal{N}(\beta_1^\top X + (\beta_2^\top X)^2, |\beta_3^\top X|^2),\]
where we set the dimension as $p=20$, and the mean vectors $\mu_x$ and $\beta_1,\beta_2,\beta_3$ are randomly generated by drawing each component from uniform distributions over the unit interval. 
First, we generate a training dataset of size $n_\text{train}=200$, 
and then fit a random forest regression estimator to estimate the mean function $\hat{\mu}(\cdot)$. 

\begin{figure}[ht]
  \centering
  \includegraphics[width=0.8\linewidth]{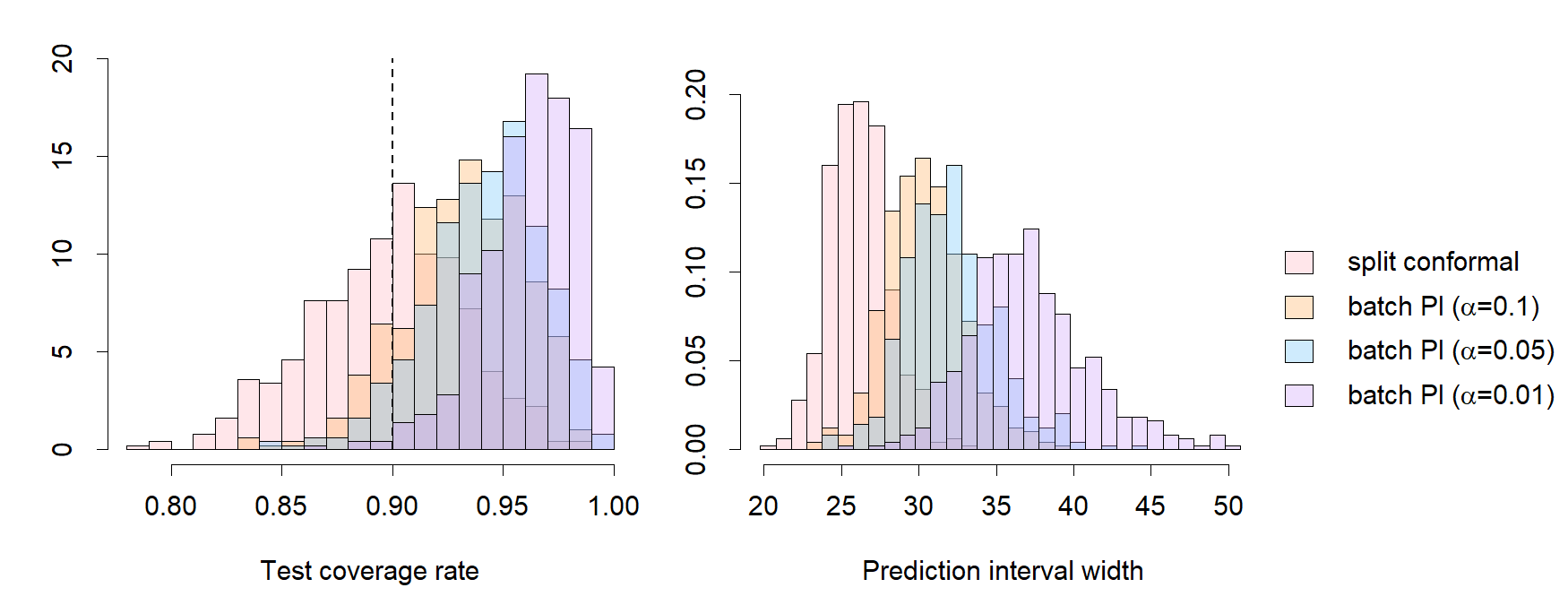}
  \caption{\footnotesize Test coverage rates and prediction interval widths of split conformal and batch PI prediction sets. 
  }
  \label{fig:hist_pac}
\end{figure}

Next, we repeat the following steps 500 times: We generate a calibration set of size $n=200$ and a test set of size $m=100$. We then apply the batch PI procedure described in Section~\ref{sec:pred_pac} at level $\delta=0.1$ and $\alpha=0.1,0.05,0.01$. For comparison, we also run split conformal prediction at level $0.1$.
The two methods provide the following guarantees, respectively:

\begin{align}
\text{Split conformal prediction: } \EE{\hat{r}} \ge 0.9, \quad \text{batch PI: } \PP{\hat{r} \ge 0.9} \ge 1-\alpha,
\end{align}

where $\hat{r} = \frac{1}{m}\sum_{j=1}^m \One{Y_{n+j} \in \Ch(X_{n+j})}$ denotes the coverage rate over the test set.
We sample $\hat{r}$ 
500 times
for both methods,
and compare the
estimated means and the probability of $\hat{r}$ exceeding $0.9$. The results are summarized in Table~\ref{table:result_pac} and Figure~\ref{fig:hist_pac}.

\begin{table}[ht]
\begin{center} 
\begin{tabular}{lll}
\hline
& $\EE{\text{coverage}}$ & $\PP{\text{coverage} \ge 0.9}$\\
\hline
split conformal & 0.9022 (0.0016)  & 0.6100 (0.0218)\\
batch PI ($\alpha=0.1$) & 0.9366 (0.0012)  & 0.9280 (0.0116)\\
batch PI ($\alpha=0.05$) & 0.9468 (0.0012)  & 0.9660 (0.0081)\\
batch PI ($\alpha=0.01$) & 0.9663 (0.0010)  & 0.9940 (0.0035)\\
\hline
\end{tabular} 
\end{center}
\caption{\footnotesize Mean of test coverage, probability of test coverage being larger than $0.9$, and the mean prediction interval width of the split conformal and batch PI prediction sets, with standard errors.}
\label{table:result_pac}
\end{table}

Table~\ref{table:result_pac} shows that both methods achieve their target guarantees tightly. 
As further supported by Figure~\ref{fig:hist_pac}, the batch PI-based method achieves stronger control over the test coverage rate by permitting slightly wider prediction sets. 
Specifically, in all three settings ($\alpha=0.1,0.05,0.01$), 
the test coverage rate of batch PI
exceeds 0.9 in a fraction $(1-\alpha)$ of the trials. 
In contrast, the split conformal method, aimed at controlling the marginal coverage rate, allows the test coverage rate to fall below 0.9 in many of the trials, while providing a shorter prediction set. 
The second plot of Figure~\ref{fig:hist_pac} illustrates this tradeoff between the width of the prediction set and the strength of the target guarantee.

\begin{figure}[ht]
  \centering
  \includegraphics[width=0.9\linewidth]{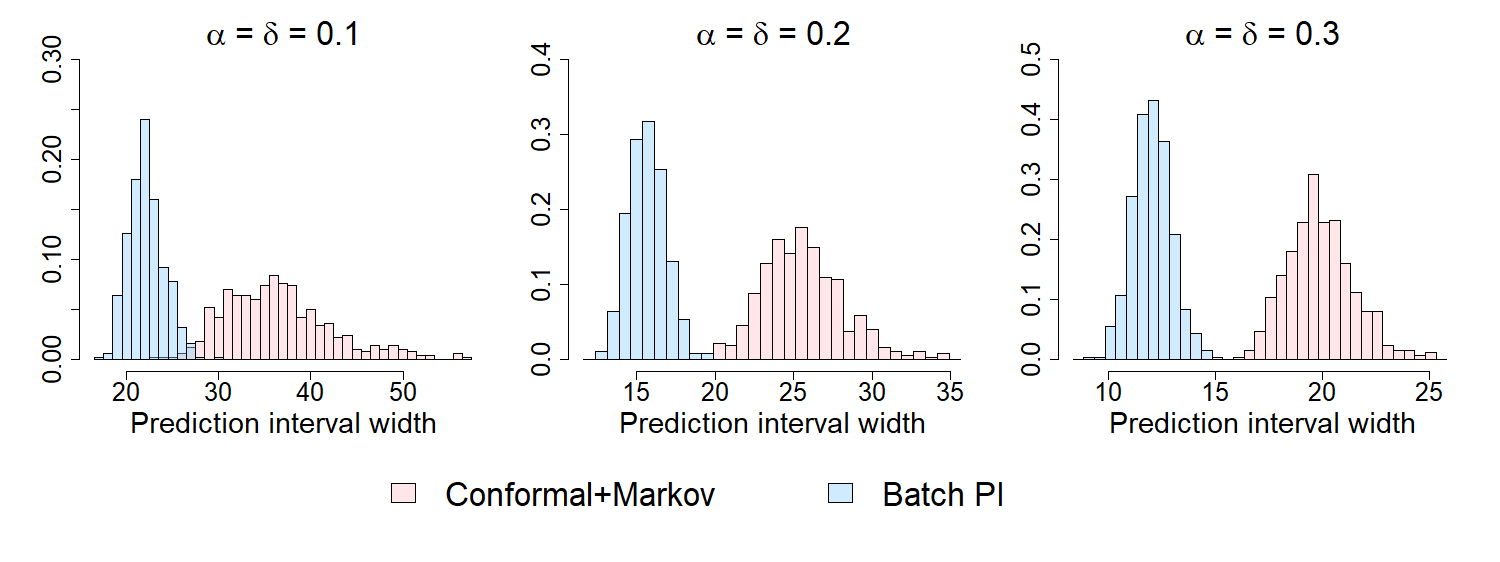}
  \vspace{-1em}
  \caption{\footnotesize Prediction interval widths from the batch PI-based method and the conformal prediction with Markov-based level adjustment at different levels. 
  }
  \label{fig:pac_comp}
    \vspace{-1em}
\end{figure}

Next, we compare the batch PI-based method with the 
baseline
Markov inequality-based
method discussed in Remark~\ref{rmk:pac_baseline}, 
which attains the same guarantee. We follow the same steps of the previous simulation, while additionally applying the baseline method, at three different pairs of levels: $(\alpha, \delta) = (0.1, 0.1)$, $(0.2, 0.2)$, and $(0.3, 0.3)$. Figure~\ref{fig:pac_comp} shows the widths of the prediction sets from the two methods across different trials, illustrating that the batch PI-based method provides significantly shorter prediction intervals.

\subsection{Selection with error control}\label{sec:exp_fwer}

Next, we illustrate the performance of batch PI procedure for the selection task described in~\ref{sec:selection}. 
We generate the data from the distribution
\[X \sim N_p(\mu_x,5 \cdot I_p), Y = \log(1+\exp(\beta^\top X + \sigma Z)), \text{ where } Z \sim \mathcal{N}(0,1).\]
The dimension is set to $p=20$, $\sigma=3$, and the mean vectors $\mu_x$ and $\beta$ are generated by drawing each component from uniform distributions over the unit interval. 
We consider the task of selecting individuals with $Y > 5$, while controlling the number of false claims, i.e., the number of individuals selected whose actual outcome is five or less.

\begin{figure}[ht]
  \centering
  \includegraphics[width=0.85\linewidth]{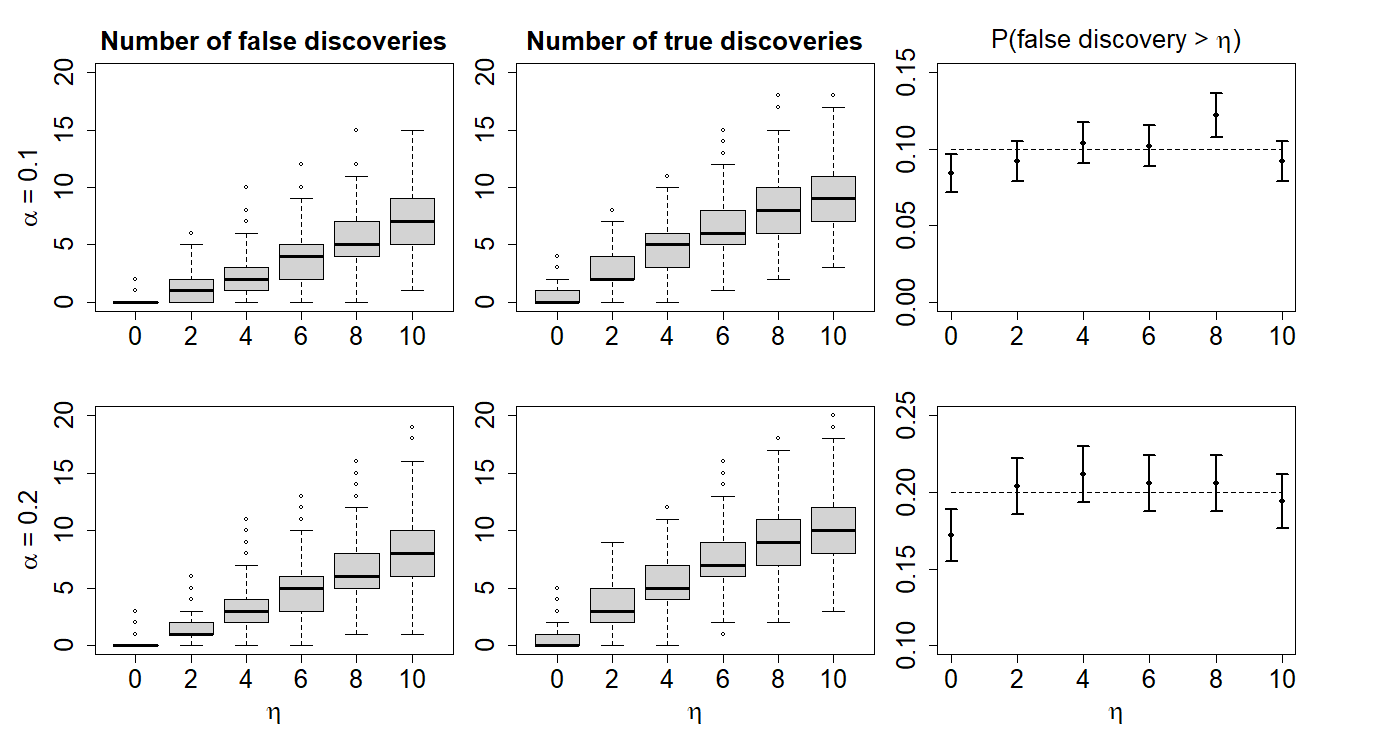}
    \vspace{-1em}
  \caption{\footnotesize Number of false claims, probability of the number of false claims being larger than the target level $\eta$, and the power of the batch PI-based selection procedure, for $\eta=0,2,4,6,8,10$ and $\alpha=0.1, 0.2$.}
  \label{fig:selection}
    \vspace{-1em}
\end{figure}

We first generate a training data of size $n_\text{train} = 500$, and then fit a random forest regression to construct the score function $s:(x,y) \mapsto \hat{\mu}(x) \One{y \le 5}$.
Next, we repeat the process of generating calibration data of size $n=1000$ and test data of size $m=100$, 500 times.
In each trial, we run the selection procedure~\eqref{eqn:selection_rule} at level $\alpha=0.1$ and $0.2$, with $\eta=0,2,4,6,8,10$. 
We record the number of false claims, as well as the number of true claims in each trial. The results are presented in Figure~\ref{fig:selection}, illustrating that the proposed procedure controls the number of false claims across various target levels $\eta$, satisfying the guarantee~\eqref{eqn:fd_control}.

\begin{figure}[ht]
  \centering
  \includegraphics[width=0.9\linewidth]{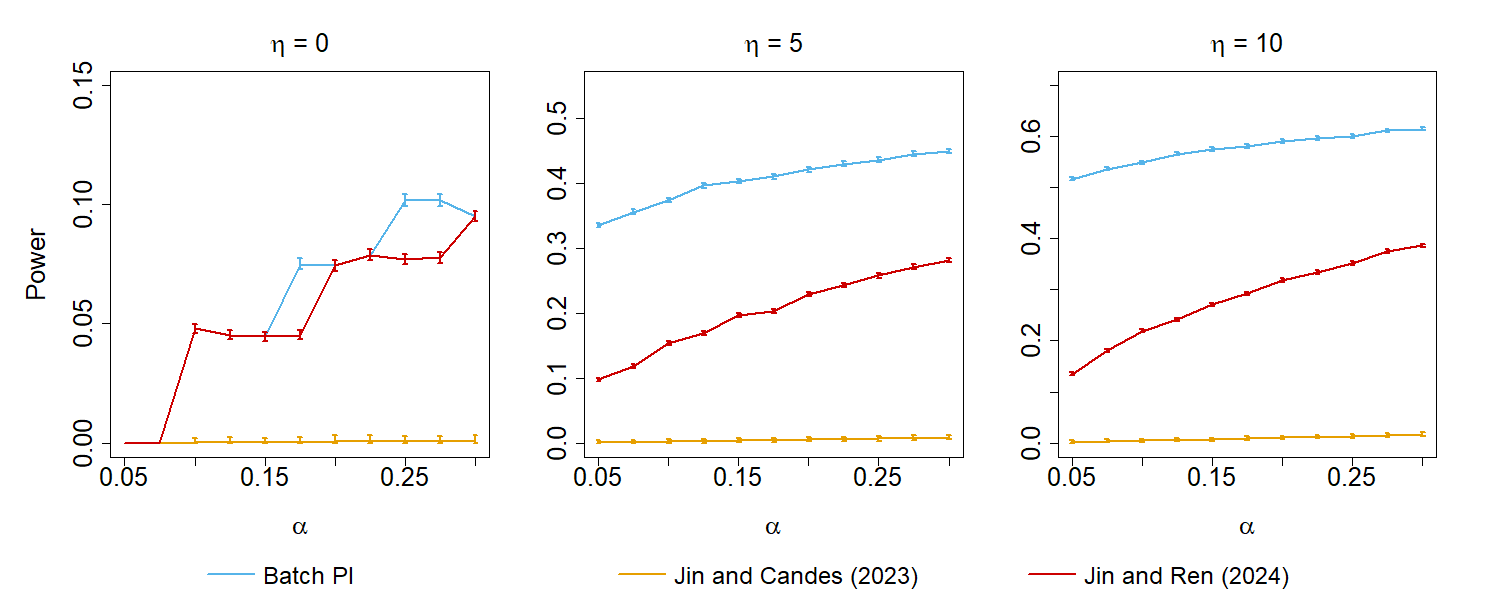}
    \vspace{-1em}
  \caption{\footnotesize Power of three procedures that control the $\eta$-FWER at different levels: (1) Batch PI-based procedure, (2) Procedure with conformal p-values by Jin\&Candes~\citep{jin2023selection}, (3) Procedure with conformal p-values by Jin\&Ren~\citep{jin2024confidence}.}
    \vspace{-1em}
  \label{fig:fwer_comparsion}
\end{figure}

Next, we compare the power of the proposed procedure and the methods based on \cite{jin2023selection,jin2024confidence}, discussed in Section~\ref{sec:fwer_baseline}.
We follow the same steps for the experiment as before but additionally run the procedures based on the conformal p-values~\eqref{eqn:cf_p_val_1} and~\eqref{eqn:cf_p_val_2}, at levels $\alpha = 0.05,0.075,0.1,\ldots,0.3$ and target false discovery bounds $\eta = 0,5,10$. The results are shown in Figure~\ref{fig:fwer_comparsion}, illustrating that the proposed method has significantly higher power than the conformal p-value-based methods in most settings.

\subsection{Inference on counterfactual variables}\label{sec:exp_counterfactual}

In this section, we provide experimental results for the predictive inference on counterfactual variables. 
We generate the data as $(X_i,A_i,Y_i^{a=0}, Y_i^{a=1}) \iidsim P_X \times P_{A \mid X} \times P_{Y^{a=0} \mid X} \times P_{Y^{a=1} \mid X}$, where $P_X$ is an entry-wise uniform distribution on $[0,1]^p$, and the treatment $A$ is assigned based on the logistic model
$\mathrm{logit}\, \PPst{A=1}{X=x} = \beta_A^\top x$ for all $x$,
where the parameter $\beta_A \in \R^p$ is generated randomly from a uniform distributions over $[0,1]^p$. 
The counterfactual distributions are set as
\begin{align*}
    Y^{a=0} \mid X \sim \textnormal{Beta}(1+X^\top \beta_Y,1-X^\top \beta_Y), \qquad Y^{a=1} \mid X \sim \textnormal{Beta}(1-X^\top \beta_Y,1+X^\top \beta_Y),
\end{align*}
where the parameter $\beta_Y$ is generated randomly from a uniform distribution $[0,1]^p$.

We first illustrate the performance of our procedure for inference on the quantiles of counterfactual variables.
We conduct experiments with a calibration (untreated group) size of $n=200$ and test (treated group) size of $m=40$---i.e., we investigate treatment-conditional inference where the treatment assignments are given.
We consider the following tasks:
\begin{enumerate}
    \item Inference on the median: Find $L, U$ such that
    $\PP{L \le Y_{(20)}^{a=0} \le U} \ge 1-\alpha$.
    \item Inference on quartiles: Find $L, U$ such that 
    $\PP{L \le Y_{(10)}^{a=0} \text{ and } Y_{(30)}^{a=0} \le U} \ge 1-\alpha$.
\end{enumerate}
Here, $Y_{(\zeta)}^{a=0}$ denotes the $\zeta$-th smallest value among $\{Y_{n+j}^{a=0} : j=1,2,\ldots,m\}$.

\begin{figure}[ht]
  \centering
  \includegraphics[width=0.8\linewidth]{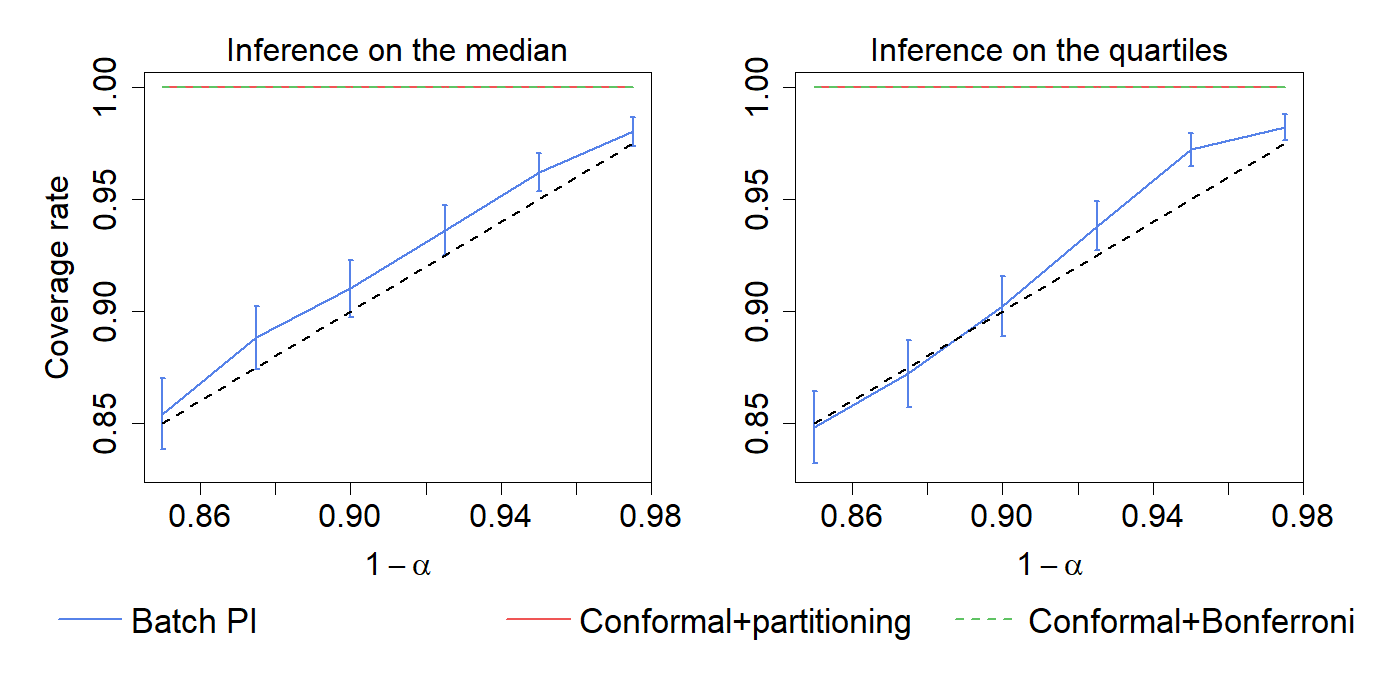}
    \vspace{-1em}
  \caption{\footnotesize Coverage rates of the batch PI prediction sets for the median and the quartiles of counterfactual variables at different levels. The dotted line corresponds to $y=x$ line. 
  Partitioning and the Bonferroni method both lead to trivial prediction sets that cover all possible outcomes, and have coverage equal to 100\% (their lines overlap).
  }
  \label{fig:counterfactual_quantiles}
    \vspace{-1em}
\end{figure}

We repeat the process of generating the calibration and test sets,
and then applying the procedures 500 times,
at levels $\alpha = 0.025,0.05,0.075,\ldots,0.15$.
Then we compute the coverage rates. For comparison, we also apply the baseline methods discussed in~\Cref{part} 
(conformal+partitioning)
and~\Cref{sec:cf_bonferroni} (conformal+Bonferroni). 
The results are summarized in Figure~\ref{fig:counterfactual_quantiles}.
They show that our procedure tightly attains the target coverage rate---while the alternative methods 
output uninformative prediction sets.

{\bf Inference on the mean of counterfactuals.}
Next, we investigate the task of inference on the mean of counterfactual variables, where we aim to construct a bound $B$ that satisfies
$\PP{\frac{1}{m}\sum_{j=1}^m Y_{n+j}^{a=0} \le B} \ge 1-\alpha$.
We perform the experiment with a calibration size of $n=100$ and the test sizes of $m=5$ and $m=10$. 
The calibration size 
after rejection sampling is smaller---around 40 in this experiment. 
Thus, neither the partitioning-based method (which requires a sufficiently large calibration-to-test ratio) nor the concentration-based method (which requires large calibration and test sizes) is useful. 
For illustration, we also provide results for
three baselines: 
conformal prediction with partitioning (see \Cref{part}), conformal prediction with Bonferroni correction (see \Cref{sec:cf_bonferroni}), and the concentration-based method (see \Cref{rmk:concentration}), and compare them with the batch PI-based procedure.

\begin{figure}[ht]
  \centering
  \includegraphics[width=0.7\linewidth]{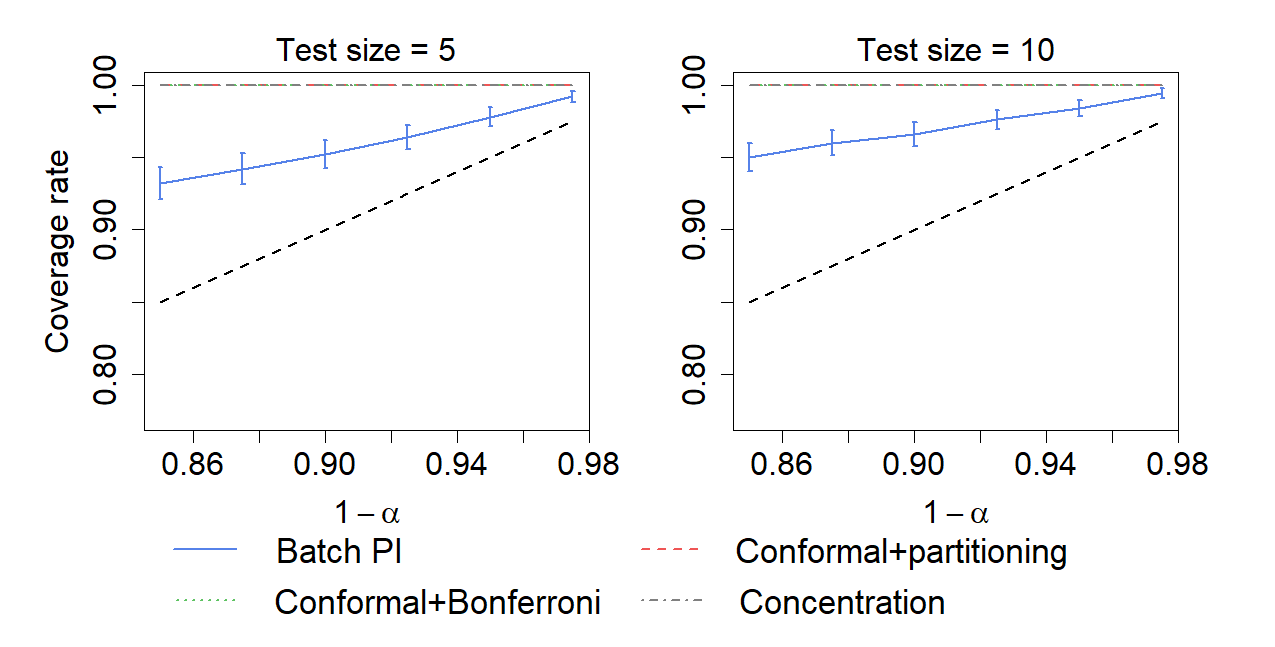}
    \vspace{-1em}
  \caption{\footnotesize Coverage rates of the prediction set for the mean of counterfactual variables obtained from batch PI and three baselines: conformal prediction with partitioning, conformal prediction with Bonferroni correction, and the concentration-based method, across different levels.
  The dotted line corresponds to $y=x$ line.   Partitioning, Bonferroni  and the concentration-based method all lead to trivial prediction sets that cover all possible outcomes, and have coverage equal to 100\% (their lines overlap).}
  \label{fig:counterfactual_mean}
    \vspace{-1em}
\end{figure}

We repeatedly generate the data and run the batch PI procedure with the dynamic programming approach from Section \ref{comp-struct} (which uses the rank-ordering function $\tilde{h}(r_1,\ldots,r_m) = \sum_{j=1}^m r_j$) along with two comparison methods, for 500 trials. We then compute the resulting coverage rates.
The results are shown in Figure~\ref{fig:counterfactual_mean}.

The results indicate that the batch PI prediction set satisfies the coverage guarantee, producing nontrivial prediction sets while 
the baseline method outputs nearly trivial prediction sets. 

{\bf Understanding over-coverage.}
The coverage of our method is here higher than the nominal level.
This reflects the inherent difficulty of the inference problem, rather than suggesting that the procedure is conservative. Observe that our inferential target is the following guarantee:
\[\inf_{\text{all distributions } P} \Pp{(X_i,Y_i)_{i \in [n+m]} \iidsim P}{\text{coverage event}} \geq 1-\alpha.\]
The batch PI procedure aims to attain the above distribution-free guarantee by ensuring that the coverage rate exceeds $1 - \alpha$ even in certain worst-case scenarios. 
As a result, in 
typical scenarios, the coverage may be higher than $1 - \alpha$. For certain targets---e.g., inference on quantiles---Corollary~\ref{cor:quantile} shows that we attain uniform tightness, i.e., 
\begin{multline*}
    1-\alpha \leq \inf_{\text{all distributions } P} \Pp{(X_i,Y_i)_{i \in [n+m]} \iidsim P}{\text{coverage event}}\\
\leq \sup_{\text{all distributions } P} \Pp{(X_i,Y_i)_{i \in [n+m]} \iidsim P}{\text{coverage event}} \leq 1-\alpha+O(\tfrac{1}{n}).
\end{multline*}
However, for general targets, the tightness typically varies with the underlying distribution.

\begin{figure}[ht]
  \centering
  \includegraphics[width=0.8\linewidth]{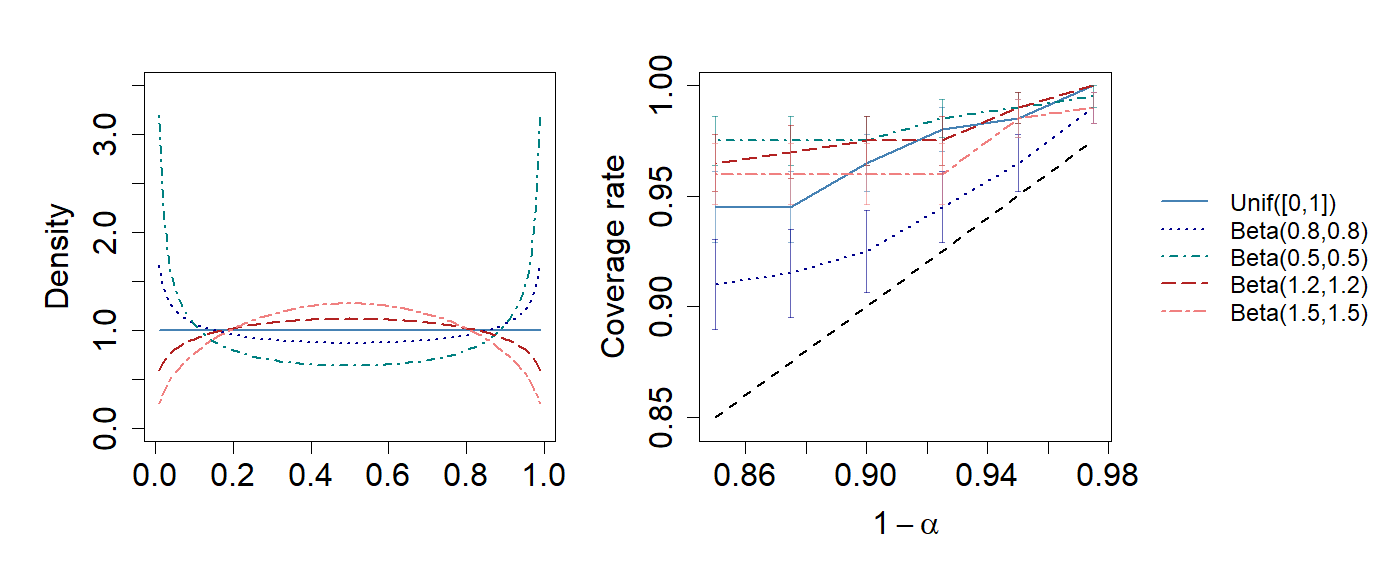}
    \vspace{-1em}
  \caption{\footnotesize Coverage rates of the prediction set for the mean of test scores under various score distributions. 
  The left plot visualizes the score distributions, while the right plot shows the coverage rates of the batch PI prediction sets. The dotted line represents the $y=x$ line.}
  \label{fig:batch_mean}
    \vspace{-1em}
\end{figure}

To further illustrate this, we empirically examine the coverage rates of the batch PI prediction sets for the mean of the test scores under various score distributions with bounded support, 
with calibration and test sizes set to $n=40$ and $m=10$, respectively. Figure~\ref{fig:batch_mean} demonstrates that the batch PI procedure achieves the target coverage guarantee across different distributions, though with varying levels of tightness. 
While the prediction set is designed to ensure a distribution-free guarantee---controlling for worst-case scenarios---it may be conservative in for particular data distributions. 
Nonetheless, these prediction sets remain useful and the only viable existing distribution-free method in this setting to our knowledge, as neither baseline methods nor concentration-based methods provide nontrivial prediction sets in this setting.

  \vspace{-1em}

In Section~\ref{sec:exp_cov_shift_est}, we provide simulation results in the setting where we do not have access to the true propensity score and instead use an estimate. 
These results demonstrate that our methodology similar results even when relying on the estimates.

\section{Empirical data illustration}

Next, we illustrate the performance of the batch PI procedure by applying it to a drug-target interaction (DTI) dataset to select high-scoring drug-target pairs. 
We use the dataset and the pre-trained model from the DeepPurpose library~\citep{huang2020deeppurpose}. 
The original dataset has 16,486 observations in both the calibration and the test sets.
The covariates consist of a pair of drug and target protein, and the response variable is the affinity score, which is a real-valued measure of the interaction between the drug and the target protein.

\begin{figure}[ht]
  \centering
  \includegraphics[width=0.8\linewidth]{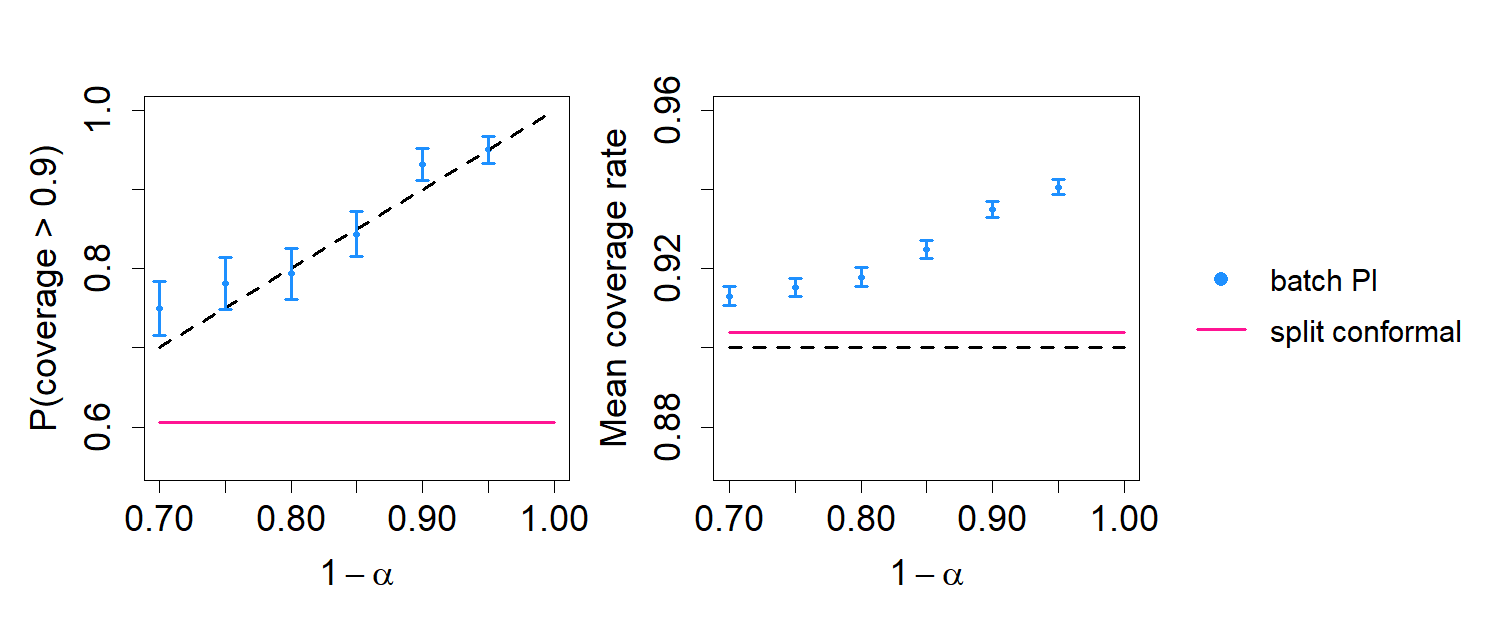}
    \vspace{-1em}
  \caption{\footnotesize The proportion of test sets whose test coverage exceeds 0.9, and the mean coverage rate of the batch PI and split conformal-based procedures at different levels. The dotted lines represent the $y=x$ line (left) and the $y=0.9$ line (right), respectively.}
  \label{fig:dti_pac}
    \vspace{-1em}
\end{figure}

We first consider the task of constructing prediction sets for each unobserved outcome variable---as discussed in Section~\ref{sec:pred_pac}. 
To illustrate performance under moderate sample sizes, we create a calibration set of size 500 randomly drawn from the original calibration data. 
We then construct 160 test sets, each of size 100, using a total of 16,000 observations from the test set. 
Denoting the pretrained estimator by $\hat{\mu}$, we run the batch PI-based procedure~\eqref{eqn:appl_pac} with the score
$s:(x,y) \mapsto |y-\hat{\mu}(x)|$ at levels $\delta=0.1$ and 
$\alpha=0.05,0.1,\ldots,0.3$. 
For comparison, we also run split conformal prediction at level $\delta=0.1$ for each of the test points. 
We compute the proportion of test sets (out of 160 total sets) where the coverage rate exceeds 0.9, as well as the mean coverage rate. 
The results are summarized in 
Figure~\ref{fig:dti_pac}. 


The results illustrate that both methods attain their respective target guarantees.
The batch PI-based procedure controls the probability of the test coverage exceeding 0.9 at different values of $\alpha$, whereas the split conformal method does not control this probability, and instead controls the mean coverage rate tightly.

\begin{figure}[ht]
  \centering
  \includegraphics[width=0.7\linewidth]{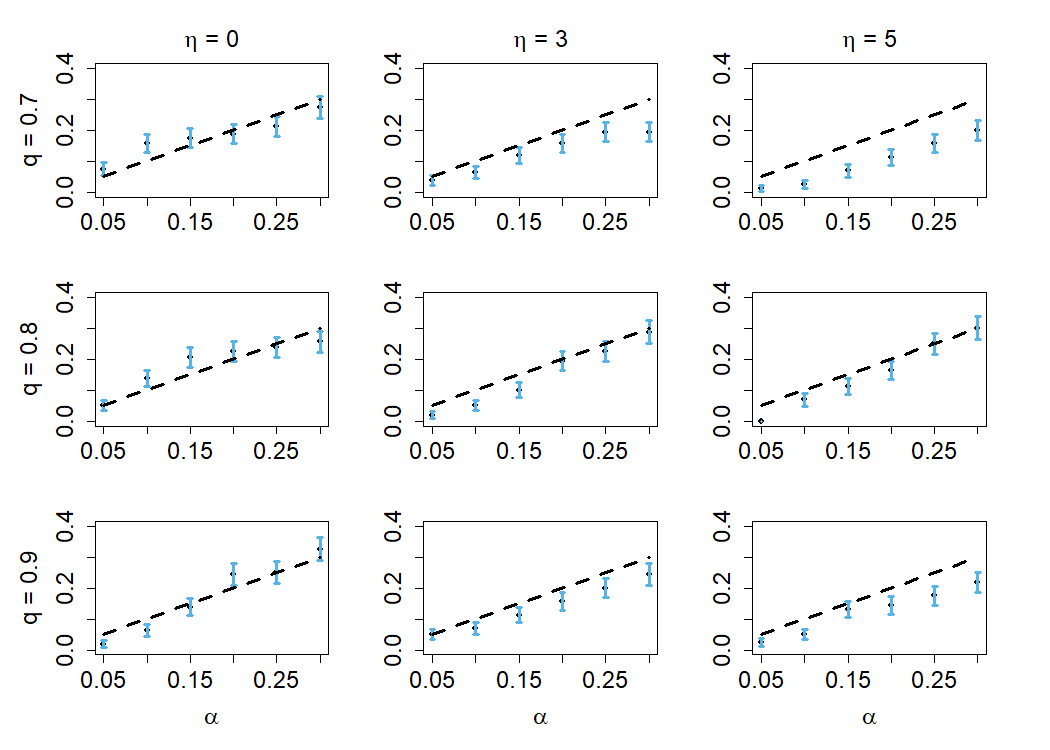}
    \vspace{-1em}
  \caption{\footnotesize The proportion of test sets whose number of false claims exceeds the target $\eta$, at levels $\alpha=0.05,\ldots,0.3$ and $\eta=0,3,5$, for three different cutoffs, with error bars. The dotted lines represent the $y=x$ line.}
    \vspace{-1em}
  \label{fig:dti_fd}
\end{figure}

Next, we examine the task of selecting drug-target pairs with high scores, following the discussion in Section~\ref{sec:selection}.
We construct a calibration set of size 2000, and 160 test sets of size 100. We aim to select drug–protein pairs whose corresponding scores exceed a certain cutoff. We experiment with three cutoffs, chosen as the $q$-th quantiles of the score values in the training data---the remaining points after sampling 2000 points for the calibration set---with $q = 0.7$, $0.8$, and $0.9$.\footnote{This experimental design follows that of~\cite{jin2023selection}.}
We run the procedure~\eqref{eqn:selection_rule} at levels $\alpha=0.05, 0.1, 0.15, 0.2, 0.25, 0.3$ and target numbers of false claims $\eta=0, 3, 5$ (recall that the procedure at $\eta=0$ controls a quantity analogous to the family-wise error rate (FWER)).
The results are shown in Figure~\ref{fig:dti_fd}, illustrating that the batch PI procedure achieves the target guarantee at various levels.

\section{Discussion}
This work introduces a distribution-free framework for joint predictive inference on a batch of multiple test points. The proposed batch PI method, provides procedures for various inference problems, such as constructing multiple prediction sets with PAC-type guarantees, constructing a selection procedure that controls the number of false claims, and inference on the mean or median of unobserved outcomes.

Many open questions remain.
For inference on one test point, several works have explored 
developing new distribution-free procedures that can achieve stronger targets or operate under more complex data structures. 
Examples include attaining training- or test-conditional coverage guarantees,
or developing methods that work with non-exchangeable data. 
Similar questions can be asked for joint inference on multiple objects. For example, can we achieve batch-conditional inference, and what kind of conditional coverage can be controlled? If we have a hierarchical structure in the data involving groups of observations, how can we perform inference for new groups? We leave these questions to future work.

\section*{Acknowledgements}
This work was supported in part by 
NIH R01-AG065276, R01-GM139926, NSF 2210662, P01-AG041710, R01-CA222147, 
ARO W911NF-23-1-0296,
NSF 2046874, ONR N00014-21-1-2843, and the Sloan Foundation. 
We thank Gilles Blanchard, Ulysse Gazin, and Etienne Roquain for helpful discussion.

\bibliographystyle{plainnat}
\bibliography{bib}

\appendix

\section{A simple example of our method and discussion of the choice of the rank ordering functions}\label{sec:ex_rank_ordering}

Our method follows the logical progression outlined below:
\begin{enumerate}
    \item For some random vector $R = (R_1, \ldots, R_m) \sim \text{Unif}(H)$, our target quantity is upper bounded by $h(S_{(R)}) := h(S_{(R_1)}, \ldots, S_{(R_m)})$.

    \item Once we construct a prediction set for $R$---i.e., a set $I \subset H$ such that $\PP{R \in I} \geq 1-\alpha$, it follows that $\PP{h(S_{(R)}) \leq \max_{r \in I} h(S_{(r)})} \geq 1-\alpha$, and thus the desired coverage guarantee also holds.
\end{enumerate}

For example, when $m = 1$ and $h$ is the identity function so that our method reduces to standard conformal prediction, this corresponds to setting $I = \{1, 2, \cdots, \lceil(1 - \alpha)(n + 1)\rceil\}$, and consequently, the upper bound for the test score is given by $\max_{r \in I} S_{(r)} = S_{(\lceil(1 - \alpha)(n + 1)\rceil)}$, which is exactly the bound provided by split conformal prediction. In this setting, the above $I$ is the one with the smallest $\max_{r \in I} h(S_{(r)})$, among all subsets of $H = [n+1]$ with coverage probability at least $1 - \alpha$.

More generally, to obtain a short or tight prediction set, we want $\max_{r \in I} h(S_{(r)})$ to be small. To achieve this, the set $I$ should consist of those elements $r$ in $H$ whose corresponding $h(S_{(r)})$ values are small---which becomes a nontrivial task when $m > 1$. For example, suppose $n = 10$, $m = 2$, and $h$ is the summation function, meaning that our target of inference is $S_6 + S_7$. Then we consider the following $\binom{10+2}{2} = 66$ sums:
\begin{equation}\label{eqn:score_sums}
    S_{(1)} + S_{(1)}, S_{(1)} + S_{(2)}, S_{(2)} + S_{(2)}, \cdots, S_{(10)} + S_{(10)}, S_{(10)} + \sup s, \sup s + \sup s.
\end{equation}
Mathematically, the set with the smallest $\max_{r \in I} h(S_{(r)})$ is the one containing the $\lceil 66 \cdot (1-\alpha) \rceil$ smallest elements from the above list:
\begin{equation}\label{eqn:I_opt}
    I = \left\{1 \leq r_1 \leq r_2 \leq n+1 : S_{(r_1)}+S_{(r_2)} \leq Q_{1-\alpha}\left(\left\{S_{(r_1')}+S_{(r_2')} : 1 \leq r_1' \leq r_2' \leq n+1\right\}\right)\right\}.
\end{equation}
However, this choice does not yield a prediction set with valid coverage, since it essentially selects $I$ in a calibration-data-dependent manner, thereby breaking the logic in the above Step 2---specifically, the condition $\PP{R \in I} \geq 1-\alpha$ does not hold for a data-dependent $I$.

Now, we require a set $I$ that does not depend on the data---for statistical validity---but still approximates the above ‘mathematically best’ $I$---to achieve a short and tight prediction set. The rank-ordering function $\tilde{h}$ was introduced to serve these two roles: it is independent of the data, but still tends to behave like $h$---since we want the resulting set $I$ to favor smaller elements in the list~\eqref{eqn:score_sums}, so that $\max_{r \in I} h(S_{(r)}) = \max_{(r_1, r_2) \in I} (S_{(r_1)} + S_{(r_2)})$ remains small. For example, in the paper, we discuss two strategies:
\begin{enumerate}
    \item Rank-ordering functionally identical to the batch score: we use
    \[I = \left\{1 \leq r_1 \leq r_2 \leq n+1 : r_1+r_2 \leq Q_{1-\alpha}\left(\left\{r_1'+r_2' : 1 \leq r_1' \leq r_2' \leq n+1\right\}\right)\right\}.\]

    \item Rank ordering based on independent split: we use
    \[I = \left\{1 \leq r_1 \leq r_2 \leq n+1 : \tilde{S}_{(r_1)}+\tilde{S}_{(r_2)} \leq Q_{1-\alpha}\left(\left\{\tilde{S}_{(r_1')}+\tilde{S}_{(r_2')} : 1 \leq r_1' \leq r_2' \leq n+1\right\}\right)\right\},\]
    where $\tilde{S}_i$ are scores from an independent data split (of the same size).
\end{enumerate}
In summary, the underlying intuition is to approximate the mathematically optimal---but statistically non-justified---prediction set $I$~\eqref{eqn:I_opt} for the ranks, using the function $\tilde{h}$ that mimics $h$.

\section{Naive method: extending weighted conformal prediction}
\label{nmwcp}
A simple approach one could consider 
for inference under covariate shift in \Cref{sec:cov_shift}
is to extend weighted conformal prediction. 
Specifically, suppose the propensity score $p_{A \mid X}$ (corresponding to some possibly unknown value of $\PP{A=1}$) is known. 
Then, for each subset $I \subset [n+m]$ of size $|I|=m$, define
\[p_{A \mid X}(I) = \frac{\prod_{i \in I} (1-p_{A \mid X}(X_i))/p_{A \mid X}(X_i)}{\sum_{I' \subset [n+m], |I'| = m}\prod_{i \in I'} (1-p_{A \mid X}(X_i))/p_{A \mid X}(X_i)}.\]
Also define, for each $I = \{i_1,i_2,\ldots,i_m\}$
with 
$1 \le i_1 < i_2 < \ldots <i_m \le n+m$,
the vectors
$\uS_I = (\uS_{i_1}, \uS_{i_2}, \ldots, \uS_{i_m})$, 
$\bS_I = (\bS_{i_1}, \bS_{i_2}, \ldots, \bS_{i_m})$,
where $\uS_i$ and $\bS_i$ follow the definition in~\eqref{eqn:uS_bS}.
Then we can construct the prediction set
\begin{equation}\label{eqn:wc_extended}
    \ch(\Dn) = \left[Q_{\beta}'\left(\sum_{I \subset [n+m], |I|=m} p_{A \mid X}(I) \cdot \delta_{h(\uS_I)}\right), Q_{1-\gamma}\left(\sum_{I \subset [n+m], |I|=m} p_{A \mid X}(I) \cdot \delta_{h(\bS_I)}\right)\right].
\end{equation}
This has the following property:
\begin{proposition}\label{prop:wc_extended}
    Suppose Condition~\ref{asm:target} holds and the data is generated by~\eqref{eqn:dist_mar}. Then the prediction set from~\eqref{eqn:wc_extended} satisfies
$\PP{g(\{Z_{n+1}, \ldots, Z_{n+m}\}) \in \ch(\Dn)} \ge 1-\alpha$,
where the probability is taken with respect to the model~\eqref{eqn:dist_mar}.
\end{proposition}

The prediction set~\eqref{eqn:wc_extended}, extending weighted split conformal prediction, suffers from a similar issue as the prediction set~\eqref{eqn:baseline_CP}, which extends split conformal prediction. Unless $n \gg m$, a substantial proportion of $\bS_i$s take the value $\sup s$ and $\uS_i$s take the value $\inf s$, likely resulting in a prediction set with a non-useful width.

\section{Additional details: One-sided batch PI}\label{1s}
\text{}

\begin{minipage}{\textwidth}
\begin{algorithm}[H]
\caption{One-sided Batch Predictive Inference (batch PI)}
{\bf Input} {Calibration data $\mathcal{D}_n= \{(X_1,Y_1),(X_2,Y_2), \ldots, (X_n,Y_n)\}$. 
Score function $s: \X \times \Y \rightarrow \R$. Test set size $m$. 
Batch score function $h : \R^m_{\uparrow} \rightarrow \R$.
Rank-ordering function $\tilde{h}: \N^m \rightarrow \R$. Target coverage level $1-\alpha\in[0,1]$.}

{\bf Goal:} Construct prediction set for $g(s(X_{n+1},Y_{n+1}), \ldots  s(X_{n+m},Y_{n+m})) = $ $ h( (s(X_{n+1},Y_{n+1}), \ldots  s(X_{n+m},Y_{n+m}))_{\uparrow})$.

{\bf Step 1:} With $H = \left\{ r_{1:m}:=(r_1,\ldots,r_m)^\top : 1 \le r_1 \le \ldots \le r_m \le n+1 \right\}$, compute the sample quantile induced by the rank-ordering function $\tilde h$:
$q = Q_{1-\alpha}\left(\sum_{r_{1:m} \in H} \delta_{\tilde{h}(r_{1:m})}/\binom{n+m}{m}\right)$.

{\bf Step 2:} Compute the scores $S_i = s(X_i,Y_i)$ for $i=1,2,\ldots,n$; and $S_{(n+1)} = \sup s$,.

{\bf Step 3:} Compute the upper bound
$B= \max\left\{h(S_{(r_1)}, \ldots, S_{(r_m)}) : r_{1:m} \in H,\, \tilde{h}(r_{1:m}) \le q\right\}$.

{\bf Return:} Prediction set $\ch(\mathcal{D}_n)=\big(-\infty, B \big]$.
\label{bpi1}
\end{algorithm}
\end{minipage}

\section{Batch predictive inference for general sparse functions}\label{sec:sparse}
Here, we describe the simplification of the batch PI procedure for general sparse function targets. 
As usual, we consider 
a target function $g$ that satisfies Condition~\ref{asm:target}, i.e., there exists a monotone function $h : \R^m_{\uparrow} \rightarrow \R$ such that
$g(\{z_1,\ldots,z_m\}) = h(s(z)_{\uparrow})$.
Further, 
we consider the case where
the function $h$ is sparse, meaning there exists a small subset $\{t_1, \ldots, t_l\} \subset [m]$, $t_1 < \ldots < t_l$,
such that $h(s_1, \ldots, s_m)$ depends only on $(s_{t_1}, \ldots, s_{t_l}$).
In other words, there exists
a function $h' : \R^l \rightarrow \R^{k_1}$ such that $h(s_1, \ldots, s_m) = h'(s_{t_1}, \ldots, s_{t_l})$ holds for all $(s_1, \ldots, s_m)$.
This is equivalent to $g$ depending only on $l$ order statistics of $s_1, \ldots, s_m$.

We first look into the computation of $q_L$ and $q_U$ in~\eqref{eqn:q_L_U}. Here we assume that the rank-ordering function $\tilde{h}$ is chosen ``reasonably", so that it also depends only on the $t_1,\ldots,t_l$-th components of the input. For instance, a natural choice would be
\begin{align*}
  \tilde{h}(r_1,\ldots,r_m) = \tilde{h}'(r_{t_1}, \ldots, r_{t_l}), \text{ where } \tilde{h}' = h' \big|_{H'}.
\end{align*}
Here,
\[H' = \left\{(r_1',r_2',\ldots,r_l') : 1 \le r_1' \le \ldots \le r_l' \le n+1\right\}.\]
The first step is to compute the sizes of the level sets of the function $(r_1, \ldots, r_m) \mapsto (r_{t_1}, \ldots, r_{t_l})$,
which equal
$L$ from \eqref{eqn:quantiles_level_set}. 
Then we compute
\[L_{\tilde{h}}(\tau) = \sum_{\substack{(\rho_1, \ldots, \rho_l) :\\ \tilde{h}'(\rho_1, \ldots, \rho_l)= \tau}}  L(\rho_1-1, \ldots, \rho_l-1)\,  \text{ and } U_{\tilde{h}}(\tau) = \sum_{\substack{(\rho_1, \ldots, \rho_l) :\\ \tilde{h}'(\rho_1, \ldots, \rho_l)= \tau}}  L(\rho_1, \ldots, \rho_l)\]
for each $\tau \in \textrm{Im}(\tilde{h}')$. Then,
$q_L$ and $q_U$ are given by
\[q_L = Q_\beta'\left(\sum_{\tau \in \textrm{Im}(\tilde{h}')} \frac{L_{\tilde{h}}(\tau)}{\binom{n+m}{m}} \delta_\tau\right) \text{ and } q_U = Q_{1-\alpha}\left(\sum_{\tau \in \textrm{Im}(\tilde{h}')} \frac{U_{\tilde{h}}(\tau)}{\binom{n+m}{m}} \delta_\tau\right).\]
The formula for $B_L$ and $B_U$ in can be written as
\begin{equation}
\begin{split}
    B_L&= \min\left\{h'(S_{(r_1'-1)}, \ldots, S_{(r_l'-1)}) : (r_1',\ldots, r_l') \in H', \tilde{h}'(r_1',\ldots, r_l') \ge q_L\right\},\\
    B_U&= \max\left\{h'(S_{(r_1')}, \ldots, S_{(r_l')}) : (r_1',\ldots, r_l') \in H', \tilde{h}'(r_1',\ldots, r_l') \le q_U\right\},
\end{split}
\end{equation}
and this requires the computation of the function values at $|H'|$ number of inputs, which scales as $n^l$. 
Therefore, we obtain a computationally feasible procedure for the case $h$ is sparse, i.e., $l$ is small.

\section{Simultaneous inference on multiple quantiles}

In this section, we extend the idea of batch PI 
to provide a simultaneous prediction set for multiple quantiles of the scores,
e.g., $h(s_1,\cdots,s_m) = (s_{\zeta_1}, s_{\zeta_2})^\top$.
This will allow us to provide fine-grained control of the test distribution, 
for instance by obtaining a prediction set for the interquartile range. 

Specifically, we examine the problem of constructing simultaneous bounds for multiple quantiles of test scores. Suppose the target function is given as
$h:(s_1,\cdots,s_m) \mapsto (s_{(t_1)}, \cdots, s_{(t_l)})^\top$,
where $1 \le t_1 \le \cdots \le t_l \le m$, and we aim to construct
vectors
$L = (L_1, \cdots, L_l)^\top$ 
and
$U = (U_1, \cdots, U_l)^\top$ 
serving as bounds
such that
\begin{equation}\label{eqn:multi_bound}
\PP{
L \preceq 
h(S_{(n+1)},\cdots,S_{(n+m)}) \preceq U} = \PP{
L_1 \le S_{(t_1)}^\textnormal{test} \le U_1,
\cdots,
L_n \le S_{(t_l)}^\textnormal{test} \le U_l} \ge 1-\alpha.
\end{equation}
To provide a procedure that attains the above guarantee, we first introduce some notation. 
For any $1 \le \rho_1 \le \ldots \le \rho_l \le n+1$,
we will need to compute the number of solutions
$r_{1:m} \in H$
of 
$r_{t_1} = \rho_1, \ldots, r_{t_l} = \rho_l$.
This equals
\begin{equation}\label{eqn:quantiles_level_set}
\begin{split}
    L(\rho_1, \ldots, \rho_l) &:= \left|\left\{(r_1, \ldots, r_m) \in H : r_{t_1} = \rho_1, \ldots, r_{t_l} = \rho_l\right\}\right|\\
    &= \scalebox{1}{$_{\rho_1}\mathrm{H}_{t_1-1} \cdot \left[\prod_{j=1}^n {}_{\rho_{j+1} - \rho_j+1}\mathrm{H}_{t_{j+1} - t_j -1}\right] \cdot _{n-\rho_l+2}\mathrm{H}_{m-t_l}.$}
\end{split}
\end{equation}
Next, 
define
for $(w_1,w_2,\cdots,w_l), (q_1,q_2,\cdots,q_l)$ satisfying $1 \le w_j \le q_j \le n+1$ for all $j \in [n+1]$,
\begin{multline}\label{eqn:multi_q_cdf}
    F_{n,m}(w_1,w_2,\cdots,w_l; q_1,q_2,\cdots,q_l) = \left|\left\{(r_1,r_2,\cdots,r_m)\in H, w_j \le r_{t_j} \le q_j,\;\forall\;j \in [l]\right\}\right|\nonumber\\
    = \scalebox{1}{$\sum_{\rho_1=w_1}^{q_1} \sum_{\rho_2=\max\{\rho_1,w_2\}}^{q_2} \cdots \sum_{\rho_m = \max\{\rho_{m-1},w_l\}}^{q_l} L(\rho_1,\cdots,\rho_l)$}.
\end{multline}
Applying the idea from the proof of batch PI, we can derive the following result.

\begin{theorem}\label{thm:multi_bound}
    Suppose that the data points $Z_1,\ldots,Z_n, Z_{n+1}, \ldots, Z_{n+m}$ are exchangeable, and that $(w_1,w_2,\cdots,w_l)$ and $(q_1,q_2,\cdots,q_l)$ satisfy $F_{n,m}(w_1,\cdots,w_l; q_1,\cdots,q_l) \ge (1-\alpha)\cdot\binom{n+m}{m}.$ Let $S_{(0)} = \inf s$ and $S_{(n+1)} = \sup s$. Then
    \[\PP{S_{(w_1-1)} \le S_{(t_1)}^\textnormal{test} \le S_{(q_1)}, S_{(w_2-1)} \le S_{(t_2)}^\textnormal{test} \le S_{(q_2)}, \cdots, S_{(w_l-1)} \le S_{(t_l)}^\textnormal{test} \le S_{(q_l)}} \ge 1-\alpha.\]
\end{theorem}

We also mention that \cite{gazin2024transductive} provided an approach that they refer to as "templates", which could also be used to derive joint prediction sets for the order statistics of the test scores.

Thus, it remains to determine vectors $(w_1,\cdots,w_l)$ and $(q_1,\cdots,q_l)$ that satisfy the condition of Theorem~\ref{thm:multi_bound}.
For instance, we can consider the following procedure.
Let $\tilde{t}_j = \text{round}(t_j \cdot n/m)$ for $j \in [l]$
represent---roughly speaking---the expected rank of the $j$-th largest test score among the $n$ calibration scores. 
Then our idea is to center the indices $w_j= \tilde{t}_j-a$,
$q_j=\tilde{t}_j+a$, $a\ge 0$,
around $\tilde{t}_j$, for $j \in [l]$.
Then, we find the smallest $a \in \mathbb{N}$ 
such that
\[
\scalebox{0.95}{$F_{n,m}\left((\tilde{t}_1 - a) \vee 1,\cdots, (\tilde{t}_l - a) \vee 1; (\tilde{t}_1 + a) \wedge (n+1), \cdots, (\tilde{t}_l + a) \wedge (n+1)\right)
\ge (1-\alpha)\binom{n+m}{m}$},
\]
and denote it by $t$.
Then define
\begin{equation}\label{eqn:multi_batch_PI_bound}
    L = (S_{((\tilde{t}_1-t-1)_+)}, \cdots, S_{((\tilde{t}_2-t-1)_+)}), \quad
    U = (S_{(\min\{\tilde{t}_l + t,n+1\})}, \cdots, S_{(\min\{\tilde{t}_l + t,n+1\})}).
\end{equation}
Applying Theorem~\ref{thm:multi_bound}, we have the following result.

\begin{corollary}\label{cor:multi_batch_PI}
    Suppose the data points $Z_1,\ldots,Z_n, Z_{n+1}, \ldots, Z_{n+m}$ are exchangeable. Then for $L$ and $U$ defined in~\eqref{eqn:multi_batch_PI_bound}, it holds that
    $\PP{L \preceq (S_{(t_1)}^\textnormal{test}, S_{(t_2)}^\textnormal{test}, \cdots, S_{(t_l)}^\textnormal{test}) \preceq U} \ge 1-\alpha$.
\end{corollary}

In Section~\ref{sec:exp_counterfactual}, we provide experimental results for 
the specific case of
inference on
\emph{quartiles}
$S_{(\text{round}(0.25 m))}^\textnormal{test},
S_{(\text{round}(0.75 m))}^\textnormal{test}$
with the following guarantee:
\[\PP{L \le S_{(\text{round}(0.25 m))}^\textnormal{test} \le S_{(\text{round}(0.75 m))}^\textnormal{test} \le U} \ge 1-\alpha.\]
For clarity, we include the specific procedure for this task below.

\begin{algorithm}
\caption{Batch Predictive Inference 
for quartiles}
\label{alg:multiple}
{\bf Input:} {Calibration data $\mathcal{D}_n= \{(X_1,Y_1),(X_2,Y_2), \ldots, (X_n,Y_n)\}$. 
Score function $s: \X \times \Y \rightarrow \R$. Test set size $m$. Target coverage level $1-\alpha \in [0,1]$.}

 \textbf{Step 1} Compute $t_1 = \text{round}(0.25 \cdot m)$, $t_2 = \text{round}(0.75 \cdot m)$, $\tilde{t}_1 = \text{round}(0.25 \cdot n)$ and $\tilde{t}_2 = \text{round}(0.75 \cdot n)$.

 \textbf{Step 2:} Compute 
 \[\scalebox{0.85}{$t = \min\left\{a \in \mathbb{N} : \sum\limits_{\rho_1 = \max\{\tilde{t}_1 - a,1\}}^{\min\{\tilde{t}_2+a,n+1\}} \sum\limits_{\rho_2 = \rho_1}^{\min\{\tilde{t}_2+a,n+1\}}  {}_{\rho_1}\mathrm{H}_{t_1-1} \cdot  {}_{\rho_2 - \rho_1+1}\mathrm{H}_{t_2 - t_1 -1} \cdot _{n-\rho_2+2}\mathrm{H}_{m-t_2} \ge (1-\alpha)\cdot\binom{n+m}{m}\right\}$}.\]

 \textbf{Step 3:} Compute the scores $S_i = s(X_i,Y_i)$ for $i=1,2,\ldots,n$; and let $S_{(0)} = \inf s$ and $S_{(n+1)} = \sup s$.

{\bf Return:} Bounds $L = S_{(\max\{\tilde{t}_1-t-1,0\})}$ and $U = S_{(\min\{\tilde{t}_2+t,n+1\})}$.
\end{algorithm}

This procedure also leads to
valid inference on the interquartile range $\text{IQR} = S_{(\text{round}(0.75 m))}^{\text{test}} - S_{(\text{round}(0.25 m))}^{\text{test}}$, with the guarantee
$\PP{\text{IQR} \le U-L} \ge 1-\alpha$.

\section{Algorithms for computation for compositional functions}
\label{alg}

\begin{algorithm}
\caption{\footnotesize Computation of $C_{m, n, k}$ for a compositional rank-ordering function $\tilde h$}
\begin{algorithmic}
\STATE Input: Rank-ordering function $\tilde h$ such that for any \(r \ge 1\), 
there is a strictly increasing function 
\(\tilde\Gamma(\cdot;r):\{0,1,\ldots\}\to \{0,1,\ldots\} \) 
such that 
for any \(\kappa \ge 1\),
$\tilde h(r_{1:\kappa}) = \tilde\Gamma(\tilde h(r_{1:(\kappa-1)});  r_\kappa)$.
Number of variables $m$, maximum variable $n$, target $k$
    \STATE Initialize $C_{1, \tilde{n}, \tilde{k}}=1$ if
$\tilde\Gamma(0;s)= \tilde k$
has a solution $s \in[n]$, and zero otherwise; 
    for $\tilde{n}\in[n]$, $\tilde{k}\in[k]$
    \FOR{$\tilde{m} = 2$ to $m$}
        \FOR{$\tilde k = 1$ to $k$}
            \FOR{$\tilde{n} = 1$ to $n$}
                \STATE $C_{\tilde{m}, \tilde{n},\tilde k} \gets \sum_{a=1}^{\tilde n} C_{\tilde m-1, a, \tilde\Gamma^{-1}(\tilde k;a)}$
            \ENDFOR
        \ENDFOR
    \ENDFOR
    \STATE Output: $C_{m, n, k}$, the number of 
\( 1 \le r_1 \le \ldots \le r_m \le n \) such that
$\tilde{h}(r_{1:m}) = k$.
\end{algorithmic}\label{dpgen}
\end{algorithm}

\begin{algorithm}
\caption{\footnotesize Computation of $M_{m, n, q}$ for the sum}
\begin{algorithmic}
\STATE Input: Scores $S_1, \ldots, S_n$, number of summands $m$, upper bound $q$ on sum of ranks 
    \STATE Initialize $M_{1, \tilde{n}, \tilde{k}} = S_{\min(\tilde n,\tilde q)}$ for $\tilde{n}\in[n], \tilde{q}\in[q]$
    \FOR{$\tilde{m} = 2$ to $m$}
        \FOR{$\tilde q = 1$ to $q$}
            \FOR{$\tilde{n} = 1$ to $n$}
                \STATE $M_{\tilde m,\tilde n,\tilde q} = \max \{ M_{\tilde m-1, a,\tilde  q-a} \mid 1 \le a \le \min(\tilde n,\tilde  q- \tilde m+1)\}$
            \ENDFOR
        \ENDFOR
    \ENDFOR
    \STATE Output: $M_{m,  n,  q}$, equal to $\max \{ S_{(r_1)} + S_{(r_2)} + \ldots + S_{(r_m)} \mid r_1 + \ldots + r_m \le q \}$
\end{algorithmic}\label{dp2}
\end{algorithm}

\noindent {\bf Computation of endpoints. }
The computation of the interval endpoints $B_L, B_U$ from \eqref{eqn:B_L_U} can be performed efficiently in a similar way. 
For concreteness, we consider $B_U$, and the reasoning for $B_L$ is entirely analogous.

For illustration, we will again first consider the case where 
\[h(S_{(r_1)}, \ldots, S_{(r_m)})=S_{(r_1)} + S_{(r_2)} + \ldots + S_{(r_m)} \text{ and } 
\tilde{h}(r_{1:m}) = r_1 + \ldots + r_m\]
for all $r_{1:m}$.
The problem becomes to compute
\[
M_{m,n,q}:=
M_{m,n,q}(S_1, \ldots, S_n) :=
\max \{ S_{(r_1)} + S_{(r_2)} + \ldots + S_{(r_m)} \mid r_1 + \ldots + r_m \le q \}.
\]

As above, we can obtain a recursion by considering the possible values of \( r_m \), to find that 
$M_{m,n,q} = \max \{ M_{m-1, a, q-a} \mid 1 \le a \le \min(n, q-m+1)\}$. 
This recursion can be initialized with
$M_{1,n,q} = S_{\min(n, q)}$,
leading to a similar dynamic programming algorithm.

More generally, 
consider the set
\[\mathcal{H} = \left\{h(S_{(r_1)}, \ldots, S_{(r_\kappa)}) : \kappa \in [m], 1\le r_1 \le \ldots \le r_\kappa \le n\right\}.\]
Suppose that
\eqref{htc} holds, and that similarly,
for all \(r \ge 1\), there is a strictly increasing  function  \(\Gamma(\cdot; r):\mathcal{H}\to \mathcal{H}\) 
such that
for any \(\kappa \ge 1\),
\beqs
h(S_{(r_{1})}, \ldots, S_{(r_{\kappa})}) = \Gamma( h(S_{(r_{1})}, \ldots, S_{(r_{\kappa-1})}); r_\kappa).
\eeqs
For instance, for $h(S_{(r_1)}, \ldots, S_{(r_m)})=S_{(r_1)} + S_{(r_2)} + \ldots + S_{(r_m)}$, 
we have 
$\Gamma( a; r) = a+S_{(r)}$.
Denote  $M_{m,n,q} = \max\left\{h(S_{(r_1)}, \ldots, S_{(r_m)}) : r_{1:m} \in H,\, \tilde{h}(r_{1:m}) \le q\right\}.$
Then, 
as above, we can obtain a recursion by considering the possible values of \( r_m \), to find that 
$M_{m,n,q} = \max \{ \Gamma( M_{m-1, a, \tilde\Gamma^{-1}(q;a)};a) \mid 1 \le a \le n\}$. 
By setting the initial conditions 
$
M_{1,n,q} = h (S_{(\tilde\Gamma^{-1}(q;n))}), 
$
we can obtain a dynamic programming algorithm similar to the ones presented above
for efficiently computing $M_{m,n,q}$.

\begin{algorithm}
\caption{\footnotesize Computation of $M_{m, n, q}$ for compositional functions $h,\tilde h$}
\begin{algorithmic}
\STATE \textbf{Input:} 
Scores $S_1, \ldots, S_n$,
number of summands $m$, constraint bound $q$; 
Rank-ordering function $\tilde h$ such that for any \(r \ge 1\), 
there is a strictly increasing function 
\(\tilde\Gamma(\cdot;r):\{0,1,\ldots\}\to \{0,1,\ldots\} \) 
such that 
for any \(\kappa \ge 1\),
$\tilde h(r_{1:\kappa}) = \tilde\Gamma(\tilde h(r_{1:(\kappa-1)});  r_\kappa)$; 
Batch score function $h$
such that
for all \(r \ge 1\), there is a strictly increasing  function  \(\Gamma(\cdot; r):\mathcal{H}\to \mathcal{H}\) such that
for any \(\kappa \ge 1\),
$h(S_{(r_{1})}, \ldots, S_{(r_{\kappa})}) = \Gamma( h(S_{(r_{1})}, \ldots, S_{(r_{\kappa-1})}); r_\kappa)$.

\STATE Initialize $M_{1, \tilde{n}, \tilde{k}} =h(S_{(\tilde{\Gamma}^{-1}(\tilde{q};\tilde{n}))})$ for $\tilde{n}\in[n], \tilde{q}\in[q]$
\FOR{$\tilde{m} = 2$ to $m$}
    \FOR{$\tilde{q} = 1$ to $q$}
        \FOR{$\tilde{n} = 1$ to $n$}
            \STATE $M_{\tilde{m},\tilde{n},\tilde{q}} = \max \left\{ \Gamma\left(M_{\tilde{m}-1, a, \tilde{\Gamma}^{-1}(\tilde{q}; a)}; a\right) : 1 \le a \le \tilde{n} \right\}$
        \ENDFOR
    \ENDFOR
\ENDFOR
\STATE \textbf{Output:} $M_{m, n, q}$, equal to $\max\left\{h(S_{(r_1)}, \ldots, S_{(r_m)}) : r_{1:m} \in H,\, \tilde{h}(r_{1:m}) \le q\right\}$
\end{algorithmic}\label{alg:generalized_dp}
\end{algorithm}

\section{Inference under covariate shift}\label{sec:cov_shift_appendix}

Here, we provide additional details for \Cref{sec:cov_shift}.

\subsection{Reformulation as a missing data problem}

To enable a concise argument, it helps to reformulate the problem as a missing data problem.
Let $A \in \{0,1\}$ be the binary variable that indicates
whether or not
the outcome $Y$ is observed. 
Then the set of all observed data points $(X_1,Y_1), \ldots, (X_n,Y_n),$ $X_{n+1}, \ldots, X_{n+m}$ can equivalently be viewed
as having $n+m$ tuples $(X_i,A_i,Y_iA_i)_{1 \le i \le n+m}$. 
The feature distributions $P_X$ and $Q_X$ in~\eqref{eqn:dist_cov_shift} correspond to the conditional distributions $P_{X \mid A=1}$ and $P_{X \mid A=0}$, respectively. Thus, we can rewrite the model~\eqref{eqn:dist_cov_shift} as
\begin{equation}\label{eqn:dist_mar}
\begin{split}
    &(X_1,Y_1), (X_2,Y_2)\ldots,(X_n,Y_n) \iidsim P_{X \mid A=1} \times P_{Y \mid X},\\
    &(X_{n+1},Y_{n+1}), (X_{n+2},Y_{n+2})\ldots,(X_{n+m},Y_{n+m}) \iidsim P_{X \mid A=0} \times P_{Y \mid X},
\end{split}
\end{equation}
and the target coverage guarantee~\eqref{eqn:guarantee_shift} can be written as
\[\PPst{g(\{Z_{n+1}, \ldots, Z_{n+m}\}) \in \ch(\Dn)}{A_1,\ldots,A_n=1, A_{n+1}, \ldots, A_{n+m}=0} \ge 1-\alpha.\]
Since the model~\eqref{eqn:dist_mar} and the target guarantee do not depend on the marginal distribution of $A$, we are free to assume any value for $\PP{A=1}$. Note that the tuple $(\PP{A=1}, P_{X \mid A=1}, P_{X \mid A=1})$ determines the joint distribution of $(X,A)$, and thus the distributions $P_X$ and $P_{A \mid X}$ are well-defined once $\PP{A=1}$ is fixed.

From this reframing, 
knowing
the likelihood ratio $dP_{X \mid A=1}/dP_{X \mid A=0}$ can equivalently be thought of as access to the propensity score 
$x\mapsto p_{A \mid X}(x) = \PPst{A = 1}{X=x}$ for some value of $\PP{A=1}$.
Indeed, for any $x$,
\[\frac{dP_{X \mid A=1}(x)}{dP_{X \mid A=0}(x)} = \frac{\PPst{A=1}{X}dP(x)}{\PPst{A=0}{X}dP(x)}\cdot\frac{\PP{A=0}}{\PP{A=1}} \propto \frac{1-p_{A \mid X}(x)}{p_{A \mid X}(x)}.\]
Based on this observation, we start by viewing 
propensity score as known.

A simple approach one could consider is to extend weighted split conformal prediction.
However, 
as we show in~\Cref{nmwcp}, 
this approach suffers from a similar issue as the standard extension of split conformal prediction. 
Unless $n \gg m$, it typically results in large prediction sets that can cover the entire range of the random variable of interest.

\subsection{Proposed method: batch PI with rejection sampling}\label{sec:rej_sam}

\begin{algorithm}
\caption{\footnotesize Batch Predictive Inference 
under Covariate Shift}
\label{bpi_cs}
{\bf Input:} {Calibration data $\mathcal{D}_n= \{(X_1,Y_1),(X_2,Y_2), \ldots, (X_n,Y_n)\}$. 
Propensity score $p_{A \mid X}$ with known pointwise lower bound $c>0$.
Score function $s: \X \times \Y \rightarrow \R$. Test set size $m$. 
Batch score
function $h : \R^m_{\uparrow} \rightarrow \R$.
Rank-ordering function $\tilde{h}: \N^m \rightarrow \R$. Target coverage level $1-\alpha\in[0,1]$. Lower and upper error levels $\beta, \gamma \in [0,1]$ satisfying $\beta+\gamma = \alpha$}

 \textbf{Step 1:} For  $i=1,2,\ldots,n$, draw 
$B_i \mid X_i \sim \textnormal{Bern}(p_{B \mid X}(X_i))$, where $p_{B \mid X}(x) $=$\frac{c}{1-c}\cdot\frac{1-p_{A \mid X}(x)}{p_{A \mid X}(x)}$.

 \textbf{Step 2:} Define the subset of the calibration data
$\tilde{\mathcal{D}}_n = \{(X_i, Y_i) : 1 \le i \le n, B_i = 1\}$.

{\bf Return:} Prediction set $\ch^{\textnormal{bPI-CovShift}}(\Dn):=\mathcal{C}^{\textnormal{bPI}}(\tilde{\mathcal{D}}_n)$, applying batch PI from Algorithm \ref{bpi} to $\tilde{\mathcal{D}}_n$
\end{algorithm}

As an alternative approach, we consider constructing an exchangeable dataset via rejection sampling, as it has been done for standard conformal prediction in \cite{park2021pac,qiu2023prediction},
and then applying the batch PI procedure. 

Suppose we have access to the conditional distribution $P_{A \mid X}$ (again, for some possibly unknown value of $\PP{A=1}$), with the following property:
\begin{condition}\label{asm:prop_bound}
    There exists a constant $c \in (0,1)$ such that $p_{A \mid X}(x) \ge c$ for all $x \in \X$.
\end{condition}
We draw a subset of the calibration data set as follows. For each $i=1,2,\ldots,n$, draw 
\begin{equation}\label{eqn:rej}
    B_i \mid X_i \sim \textnormal{Bern}(p_{B \mid X}(X_i)), \textnormal{ where } p_{B \mid X}(x) = \frac{c}{1-c} \cdot \frac{1-p_{A \mid X}(x)}{p_{A \mid X}(x)}.
\end{equation}
The Bernoulli distribution described above is well-defined for any value of $X_i$ 
if $p_{A \mid X}(x) >0$ for all $x \in \X$.

This sampling scheme was previously discussed in~\cite{park2021pac}, and intuitively, it constructs a subset of the calibration set that mimics the distribution of the test set through reweighting based on the propensity score. 
Let $\tDn$ be the subset of the calibration data defined as
\begin{equation}\label{eqn:tDn}
    \tDn = \{(X_i,Y_i) : 1 \le i \le n, B_i = 1\}.
\end{equation}
The subset $\tDn$ of the calibration data is exchangeable with the test data, and thus it follows that the batch PI prediction set $\ch^{\textnormal{bPI-CovShift}}(\Dn):=\ch(\tDn)$ from this subset achieves the target level of coverage:
\begin{corollary}\label{cor:bpi_rej}
    Under Conditions~\ref{asm:target} and~\ref{asm:prop_bound},
    with 
    $\tDn$ constructed by~\eqref{eqn:tDn},
    the batch PI prediction set $\ch^{\textnormal{bPI-CovShift}}(\Dn):=\ch(\tDn)$ based on~\eqref{eqn:batch_PI} satisfies
    \[\PPst{g(\{Z_{n+1}, \ldots, Z_{n+m}\}) \in \ch(\tDn)}{A_{1:n}, B_{1:n}} \ge 1-\alpha,\]
    where the probability is taken with respect to the model~\eqref{eqn:dist_mar}.
\end{corollary}

Similarly, we can conduct inference on multiple quantiles of test scores under covariate shift.
In general, rejection sampling translates any procedure designed for i.i.d. data to a procedure suitable for data with covariate shift.
The procedure $\ch(\tDn)$ is an application of this approach to batch PI.
Since rejection sampling reduces the sample size,
using 
naive procedures such as split conformal prediction 
may yield uninformative prediction sets after rejection sampling, even if the original calibration set is large.
The batch PI procedure addresses this issue as its usefulness does not depend heavily on the ratio of calibration to test sizes.

\section{Additional simulation results}\label{sec:exp_cov_shift_est}

\begin{figure}[ht]
  \centering
  \includegraphics[width=0.9\textwidth]{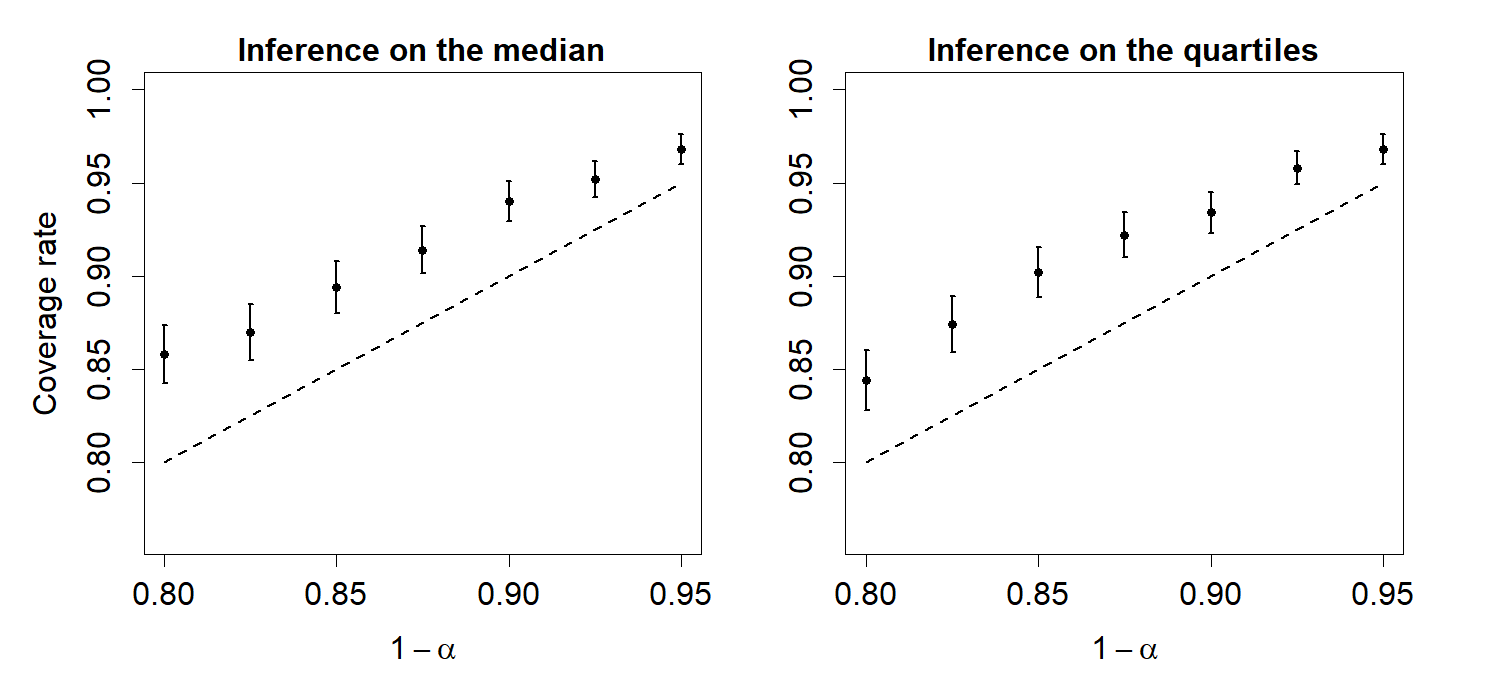}
  \caption{Coverage rates of the batch PI prediction sets for the median and quartiles of counterfactual variables using the estimated propensity score at different levels. The dotted line corresponds to the $y=x$ line.}
  \label{fig:counterfactual_quantiles_est}
\end{figure}

In this section, we reproduce the experimental results from Section~\ref{sec:exp_counterfactual} in the case where the true propensity score is unavailable, and instead, an estimate of the propensity score is used in the procedure. Specifically, we generate training data of size 200, fit a random forest classifier to construct an estimate $\hat{p}_{A \mid X}(\cdot)$ of the propensity score $p_{A \mid X}(\cdot)$, and then repeat the procedure with $p_{A \mid X}$ replaced by $\hat{p}_{A \mid X}$---i.e., we use the estimated propensity score in the rejection sampling step, and the following steps remain unchanged. The results for ths tasks of inference on the mean and quartiles are shown in Table~\ref{table:counterfactual_quantiles_est} and Figure~\ref{fig:counterfactual_quantiles_est}, illustrating that the prediction sets obtained with the estimated propensity score are similar to those from the true propensity score.

\begin{table}[ht]
\centering
\begin{tabular}{r|lllllll}
\hline
Target & $\alpha=0.05$ & $\alpha=0.075$ &$\alpha=0.1$ &$\alpha=0.125$ &$\alpha=0.15$ &$\alpha=0.175$ &$\alpha=0.2$\\
\hline
Median & \makecell{0.968\\(0.0079) } & \makecell{0.952\\(0.0096) } & \makecell{0.940\\(0.0106) } & \makecell{0.914\\(0.0126) } & \makecell{0.894\\(0.0138) } & \makecell{0.870\\(0.0151) } & \makecell{0.858\\(0.0156) }\\
\hline
Quartiles & \makecell{0.968\\(0.0079) } & \makecell{0.958\\(0.0090) } & \makecell{0.934\\(0.0111) } & \makecell{0.922\\(0.0120) } & \makecell{0.902\\(0.0133) } & \makecell{0.874\\(0.0149) } & \makecell{0.844\\(0.0162) }\\
\hline
\end{tabular}
\caption{Coverage rates of the batch PI prediction sets for counterfactual quartiles using the estimated propensity score (upper: median, lower: quartiles) at different levels, with standard errors.}
\label{table:counterfactual_quantiles_est}
\end{table}

\begin{figure}[ht]
  \centering
  \includegraphics[width=0.9\textwidth]{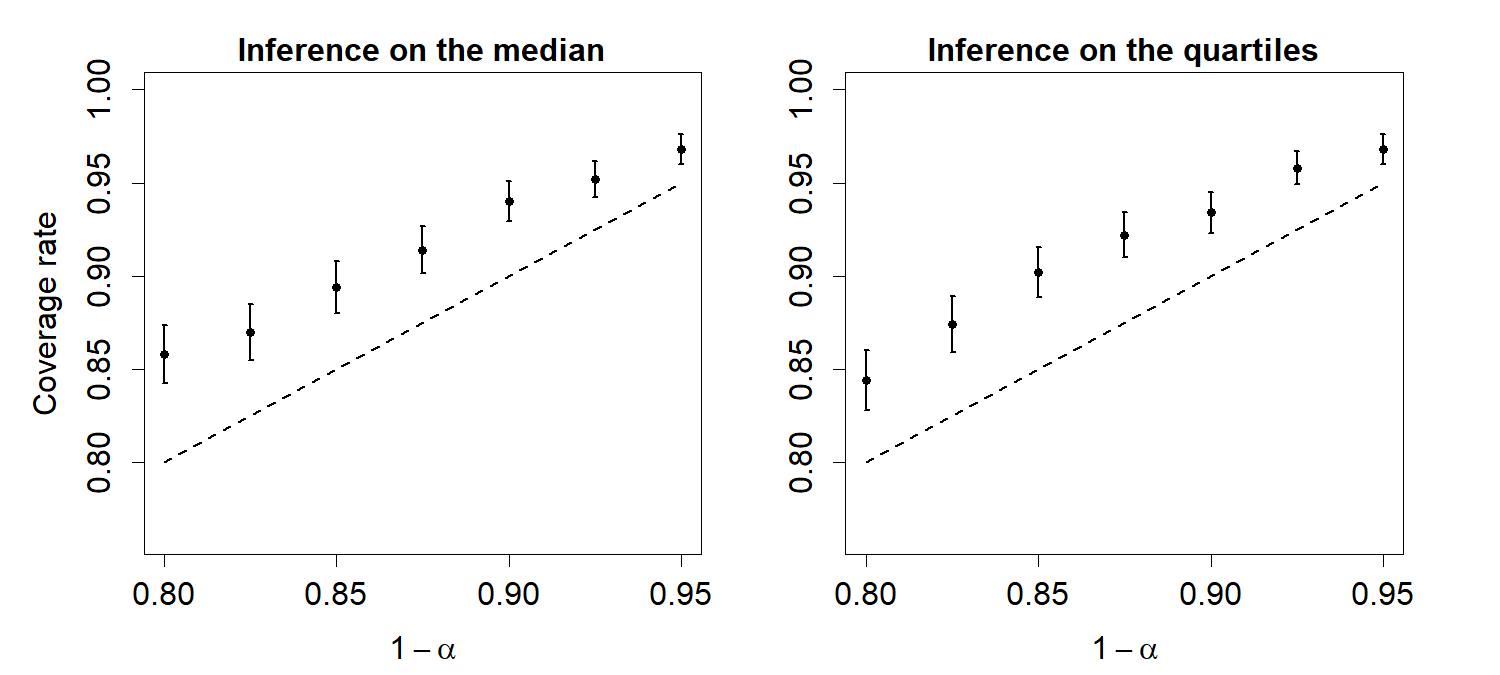}
  \caption{Coverage rates of the prediction set for the mean of counterfactual variables using the estimated propensity score for test sizes five and ten, at different levels. 
  The dotted line corresponds to $y=x$ line.}
  \label{fig:counterfactual_mean_est}
\end{figure}

Next, we present the results for inference on the mean using the estimated propensity score (Table~\ref{table:counterfactual_mean_est} and Figure~\ref{fig:counterfactual_mean_est}). The results illustrate that the prediction sets obtained from the estimate still achieve the coverage guarantee, although they are a bit more conservative.
\begin{table}[ht]
\centering
\begin{tabular}{c|lllllll}
\hline
Test size & $\alpha=0.05$ & $\alpha=0.075$ &$\alpha=0.1$ &$\alpha=0.125$ &$\alpha=0.15$ &$\alpha=0.175$ &$\alpha=0.2$\\
\hline
$m=5$ & \makecell{0.984\\(0.0056) } & \makecell{0.970\\(0.0076) } & \makecell{0.956\\(0.0092) } & \makecell{0.952\\(0.0096) } & \makecell{0.942\\(0.0105) } & \makecell{0.932\\(0.0113) } & \makecell{0.926\\(0.0117) }\\
\hline
$m=10$ & \makecell{0.998\\(0.0020) } & \makecell{0.992\\(0.0040) } & \makecell{0.986\\(0.0053) } & \makecell{0.980\\(0.0063) } & \makecell{0.968\\(0.0079) } & \makecell{0.962\\(0.0086) } & \makecell{0.950\\(0.0098) }\\
\hline
\end{tabular}
\caption{Coverage rates of the prediction sets for the mean of counterfactual variables using the estimated propensity score for test sizes of five and ten, at different levels, along with standard errors.}
\label{table:counterfactual_mean_est}
\end{table}

\section{Additional proofs}\label{sec:proofs}

\subsection{Proof of Theorem~\ref{thm:batch_PI}}
We first consider the case where the scores $S_1,\ldots S_n, S_{n+1}, \ldots, S_{n+m}$ are all distinct with probability one. 
By Condition~\ref{asm:target}, there exist functions $h : \R^m_{\uparrow} \rightarrow \R$ and $s : \X \times \Y \rightarrow \R$ such that $g(\{z_1,\ldots,z_m\}) = h(s(z)_{\uparrow})$ holds for any $z=(z_1,z_2,\ldots,z_m)$. 
Recall that $S_i = s(X_i,Y_i)$ for $i \in [n+m]$ and $S_{(1)}, S_{(2)}, \ldots, S_{(n)}$ are the order statistics of the observed scores $S_1,S_2,\ldots,S_n$. 

For $j=1,2,\ldots,m$, define
\begin{equation}\label{eqn:rank_1}
    R_{n+j} = \min\{r \in \{1,2,\ldots,n\} : S_{(r)} \ge S_{n+j}\},
\end{equation}
i.e., $R_{n+j}$ is the rank such that $S_{(R_{n+j})}$ is the smallest observed score that is larger than or equal to $S_{n+j}$. 
We define $R_{n+j} = n+1$ if $S_{(n)} < S_{n+j}$. Write $R^\textnormal{test} = (R_{n+1}, R_{n+2}, \ldots, R_{n+m})$. 
We also define $T_i$ as the rank (in increasing order) of $S_i$ among the set of all scores $\{S_1,\ldots,S_n,S_{n+1},\ldots, S_{n+m}\}$, for $i \in [n+m]$.

Now define the set
$C_{n+m} = \left\{r_{1:m} : 1 \le r_1 < r_2 < \ldots < r_m \le n+m\right\}$,
and let $T^{\text{test}} = (T_{n+1}, T_{n+2}, \ldots, T_{n+m})$ be the vector of ranks of the test scores. 
It is clear from the exchangeability of $S_1,\ldots,S_{n+m}$ that $T_{\uparrow}^{\text{test}}$ follows a uniform distribution over $C_{n+m}$---i.e., all the rank combinations appear with the same probability. 
Next, we construct a map $M$ from $C_{n+m}$ to $H$
such that for all $r_{1:m}\in C_{n+m}$,
\[M(r_{1:m}) = (r_1, r_2-1, \ldots, r_k-k+1, \ldots, r_m-m+1).\]
This is a well defined function, since for any $1 \le k \le m-1$, it holds that $r_{k+1} - (k+1) +1 \ge r_k+1 - (k+1) +1 = r_k - k + 1$.
Observe that $M$ is a bijection, since it has an inverse function
defined for all $r_{1:m} \in H$ by
\[M^{-1}(r_{1:m}) = (r_1,r_2+1, \ldots, r_k+k-1, \ldots, r_m+m-1).\]
Therefore,
$M(T_{\uparrow}^{\text{test}})$ follows a uniform distribution over $H$. 

The next step is to observe that $M(T_{\uparrow}^{\text{test}}) = R_{\uparrow}^{\text{test}}$.
To see this, assume $T_{n+1} < T_{n+2} < \ldots < T_{n+m}$, without loss of generality, and fix any $j \in [m]$. By the definition of $R_{n+j}$, we have
\begin{align*}
    R_{n+j} &= \sum_{i=1}^n \One{S_i < S_{n+j}} + 1 = \sum_{i=1}^{n+m} \One{S_i < S_{n+j}} - \sum_{i=n+1}^{n+m} \One{S_i < S_{n+j}} + 1\\
&= (T_{n+j}-1) - (j-1) + 1 = T_{n+j}-j+1.
\end{align*}
Putting everything together, we have shown that $R_{\uparrow}^{\text{test}} \sim \textnormal{Unif}(H)$. 
This implies that, for any fixed subset $I$ of $H$ with $|I| \ge (1-\gamma)|H|$, it holds that $\PP{R_{\uparrow}^{\text{test}} \in I} \ge 1-\gamma$. 
Let $S_{(n+1)}, \ldots, S_{(n+m)}$ represent the order statistics of $S_{n+1}, \ldots, S_{n+m}$, and $R_{(n+1)}, \ldots, R_{(n+m)}$ denote the order statistics of $R_{n+1}, \ldots, R_{n+m}$ (so that $R_{\uparrow}^{\text{test}} = (R_{(n+1)}, \ldots, R_{(n+m)})$). 
Now, $S_{n+j} \le S_{(R_{n+j})}$ holds for each $j \in [m]$ by the definition of $R_{n+j}$, 
and this implies that $S_{(n+j)} \le S_{(R_{(n+j)})}, j \in [m]$.
Therefore, we have
\begin{align*}
&\PP{h(S_{(n+1)}, \ldots, S_{(n+m)}) \le \max_{r_{1:m} \in I} h(S_{(r_1)}, \ldots, S_{(r_m)})} \\
&\ge \PP{h(S_{(R_{(n+1)})}, \ldots, S_{(R_{(n+m)})}) \le \max_{r_{1:m} \in I} h(S_{(r_1)}, \ldots, S_{(r_m)})}
\ge \PP{(R_{(n+1)}, \ldots, R_{(n+m)}) \in I} \ge 1-\gamma,
\end{align*}
where the first inequality applies the monotonicity assumption~\eqref{eqn:h_monotone} of $h$ and the definition of $R_{n+1}, \ldots, R_{n+m}$, 
and the second inequality uses the inclusion 
$\{f(x) \le \max_{y\in A} f(y)\}\supset \{x\in A\}$, valid for any function $f$ defined on a finite set $B$, for any $A\subset B$ and any $x\in B$.  
Further, $B_U = \max_{r_{1:m} \in I} h(S_{(r_1)}, \ldots, S_{(r_m)})$ 
where
$I := \left\{r_{1:m} \in H, \tilde{h}(r_{1:m}) \le q_U\right\}$.
Since $|I| \ge (1-\gamma)|H|$ by the definition of $q_U$, we have
$\PP{h(S_{(n+1)}, \ldots, S_{(n+m)}) \le B_U} \ge 1-\gamma.$

For the lower bound, we first observe that $S_{(R_{n+j}-1)} < S_{n+j}$ for each $j\in [m]$, by the definition of $R_{n+j}$. 
Then $S_{(R_{(n+j)}-1)} < S_{(n+j)}$ also holds, and thus
\[h(S_{(n+1)}, \ldots, S_{(n+m)}) \ge h(S_{(R_{(n+1)}-1)}, \ldots, S_{(R_{(n+m)}-1)})\]
holds deterministically.
Thus, following an argument similar to that for the upper bound, we can prove that
$\PP{h(S_{(n+1)}, \ldots, S_{(n+m)}) \ge B_L} \ge 1-\beta$
also holds, and this proves the desired inequality.

Now  consider the case where the scores can have ties. 
In such a case, we define $\tilde{T}_i$ as the rank of $S_i$ among $\{S_1,S_2,\ldots,S_{n+m}\}$, where we break the ties uniformly randomly. 
For example, if $S_2 < S_1 = S_3 < S_4$, then we have $T_2 = 1, T_4 = 4$ deterministically, and $(T_2,T_3) = (2,3)$ and $(T_2,T_3) = (3,2)$ each with probability $1/2$. 
Let $\tilde{T}_{(1)}^\text{cal} < \ldots < \tilde{T}_{(n)}^\text{cal}$ be the order statistics of $\{\tilde{T}_i : i \in [n]\}$. Then we let
\[\tilde{R}_{n+j} = \min\{r \in [n] : \tilde{T}_{(r)}^\text{cal} \ge T_{n+j}\}.\]
By the same argument as before, we have that $\tilde{R}_{\uparrow}^{\text{test}} = (\tilde{R}_{(n+1)}, \ldots, \tilde{R}_{(n+m)}) \sim \textnormal{Unif}(H)$. 
Also note that
$\tilde{R}_{n+j} \ge R_{n+j}$ holds for all $j \in [m]$, since $\tilde{T}_{(r)}^\text{cal} \ge T_{n+j}$ implies $S_{(r)} \ge S_{n+j}$ (i.e., $\tilde{T}_{(r)}^\text{cal} \ge T_{n+j}$ cannot happen if $S_{(r)} < S_{n+j}$). Therefore, we have $\tilde{R}_{\uparrow}^{\text{test}} \succeq R_{\uparrow}^{\text{test}}$, and thus it follows that
\begin{align*}
&\PP{h(S_{(n+1)}, \ldots, S_{(n+m)}) \le \max_{r_{1:m} \in I} h(S_{(r_1)}, \ldots, S_{(r_m)})} \\
&\ge \PP{h(S_{(R_{(n+1)})}, \ldots, S_{(R_{(n+m)})}) \le \max_{r_{1:m} \in I} h(S_{(r_1)}, \ldots, S_{(r_m)})}\\
&\ge \PP{h(S_{(\tilde{R}_{(n+1)})}, \ldots, S_{(\tilde{R}_{(n+m)})}) \le \max_{r_{1:m} \in I} h(S_{(r_1)}, \ldots, S_{(r_m)})}
\ge \PP{(\tilde{R}_{(n+1)}, \ldots, \tilde{R}_{(n+m)}) \in I} \ge 1-\gamma,
\end{align*}
proving the claim.

\subsection{Proof of Proposition~\ref{prop:np_hard}}
Given non-negative integers $\delta_1,\ldots, \delta_m$, define $r_i =\sum_{j\in[i]} \delta_j$
for all $i\in[m]$.
Further, for any $n\ge r_m$, 
recalling $\delta_{1:m} = (\delta_1,\ldots, \delta_m)$
define $g$ via $g(\delta_{1:m}) = h(S_{(r_1)}, \ldots, S_{(r_m)})$.
Clearly, the constraint $r_{1:m} \in H$ holds. Choosing $\tilde{h} \equiv 0$, $B_L$ from \eqref{eqn:B_L_U} becomes
\[ \min\left\{g(\delta_{1:m}) : \delta_i\in\{0,\ldots,n\}, i\in [m], \, \sum_{j\in[m]} \delta_j \le n \right\}.\]
By taking $g$ to take sufficiently large polynomial-sized values when 
any $\delta_i\ge 2$, $i\in [m]$, 
we can 
constrain
$\delta_i\in\{0,1\}, i\in [m]$.
Further, we can  take $n=m$.
Since $g$ can be arbitrary,
we now claim that
the above problem includes the vertex cover problem 
\citep[see e.g.,][]{garey1979computers}
as a special case. 

Indeed, given a graph $G = (V,E)$ and $\lambda\in \R$, we can 
take $g$ to be $g(\delta_{1:m}) =
\sum_{u\in V} \delta_u+
\lambda\sum_{(u,v)\in E} (1-\delta_u-\delta_v)_+$
for $\delta_{1:m}\in\{0,1\}^m$,
where $(\cdot)_+$ is the positive part.
Next, we claim that for  $\lambda\le |V|+1$, 
any minimizer $(\delta_{1:m})$ of $g$ must satisfy $\delta_u+\delta_v\ge 1$ for all $(u,v)\in E$.
Indeed, otherwise $\lambda\sum_{(u,v)\in E} (1-\delta_u-\delta_v)_+ \ge \lambda$; whereas setting $\tilde\delta_u = 1$ for all $u\in V$ leads to a value of 
$g(\tilde\delta_1, \ldots, \tilde\delta_m) = |V|<\lambda$; which is a contradiction with $(\delta_1, \ldots, \delta_m)$ being a minimizer.

Now, a minimizer of $\sum_{u\in V} \delta_u$ with $\delta_u \in\{0,1\}$ for all $u\in V$ 
and $\delta_u+\delta_v=1$  for all $(u,v)\in E$
exists and corresponds to a vertex cover; and all such minimizers are vertex covers. This shows that for this $\lambda$,  the minimizers of $g$ are precisely the vertex covers.
We conclude that our problem includes the vertex cover problem as a special case, and hence is NP-hard.

\subsection{Proof of Corollary~\ref{cor:quantile}}
The lower bound is a direct consequence of Theorem~\ref{thm:batch_PI}. To prove the upper bound, let us assume that the scores are all distinct almost surely. By the arguments in the proof of Theorem~\ref{thm:batch_PI} and the discussion in Section~\ref{sec:examples}, we have
\[R_{(n+\zeta)} \sim \sum_{k=1}^{n+1} p_{n,m,\zeta}(k)\cdot \delta_k,\]
and, by the definition of $q_U$, we have $\PP{R_{(n+\zeta)} \leq q_U - 1} \leq 1-\gamma$, and consequently $\PP{R_{(n+\zeta)} \leq q_U} \leq 1-\gamma + \PP{R_{(n+\zeta)} = q_U} \leq 1-\gamma+\eps_{n,m,\zeta}$. Since $\PP{R_{(n+\zeta)} < q_L} \leq \beta$ by the definition of $q_L$, it follows that
\[\PP{q_L \leq R_{(n+\zeta)} \leq q_U} = \PP{R_{(n+\zeta)} \leq q_U} - \PP{R_{(n+\zeta)} < q_L} \leq 1-\alpha+\eps_{n,m,\zeta}.\]
The proof is completed by observing that the event $\{q_L \leq R_{(n+\zeta)} \leq q_U\}$ is implied by $S_{(q_L-1)} \leq S_{(\zeta)}^\text{test} \leq S_{q_U}$.

To check $\eps_{n,m,\zeta} = O(1/n)$, we compute
\begin{align*}
    \eps_{n,m,\zeta} = \max_{k \in [n+1]} \frac{\tbinom{k+\zeta-2}{\zeta-1}\tbinom{n+m-k-\zeta+1}{m-\zeta}}{\tbinom{n+m}{m}} \leq \frac{\frac{(n+m)^{\zeta-1}}{(\zeta-1)!}\cdot \frac{(n+m)^{m-\zeta}}{(m-\zeta)!}}{\frac{n^m}{m!}} \leq \frac{(n+m)^{m-1}}{n^m} = \frac{1}{n}\cdot\left(1+\frac{m}{n}\right)^{m-1}.
\end{align*}
The term $\left(1+\frac{m}{n}\right)^{m-1}$ converges to one as $n$ grows, proving that $\eps_{n,m,\zeta} = O(1/n)$.

\subsection{Proof of Proposition~\ref{prop:k_fwer}}
By applying Markov's inequality, we have
\begin{multline*}
    \PP{\sum_{j=1}^m \One{p_j \leq \frac{k+1}{m}\alpha}\One{E_j} \geq k+1} \leq \frac{\sum_{j=1}^m \EE{\One{p_j \leq \frac{k+1}{m}\alpha}\One{E_j}}}{k+1}\\
    = \frac{\sum_{j=1}^m \PP{p_j \leq \frac{k+1}{m}\alpha \text{ and } E_j \text{ holds}}}{k+1} \leq \frac{\sum_{j=1}^m \frac{k+1}{m}\alpha}{k+1} = \alpha,
\end{multline*}
where the second inequality holds by the assumed property of $p_j$. Therefore, we have
\[\PP{\sum_{j=1}^m \One{p_j \leq \frac{k+1}{m}\alpha}\One{E_j} \leq k} \geq 1-\alpha.\]

\subsection{Proof of Corollary~\ref{cor:bpi_rej}}
It is sufficient to show that 
the random variables
 in the set
$\tDn \cup \{(X_i,Y_i) : n+1 \le i \le n+m\}$
are i.i.d. conditional on $B_{1:n}$. 
Since each outcome $Y_i$ depends only on $X_i$ (i.e., independent of every other random variable conditional on $X_i$) and is drawn from the same distribution $P_{Y\mid X}$, it is further enough to show that
$\{X_i : i \in [n], B_i = 1\} \cup \{X_i : n+1 \le i \le n+m\}$ are i.i.d. given $B_{1:n}$. 
The independence is clear under the model~\eqref{eqn:dist_mar}, and thus it remains to prove that the following two distributions are identical.
\begin{enumerate}
\item Conditional distribution of $X$ given $B=1$, where $X$ and $B$ are drawn by
    $X \sim P_{X \mid A=1}, B \mid X \sim \textnormal{Bern}(p_{B\mid X}(X))$.
    \item The distribution $P_{X \mid A=0}$.
\end{enumerate}
Take any measurable set $U \subset \X$. We have
\begin{align*}
    &\Ppst{X \sim P_{X \mid A=1}, B \mid X \sim \textnormal{Bern}(p_{B\mid X}(X))}{X \in U}{B=1}\\
    &= \Ppst{X \sim P_X, A \mid X \sim \textnormal{Bern}(p_{A\mid X}(X)), B \mid X \sim \textnormal{Bern}(p_{B\mid X}(X))}{X \in U}{B=1, A=1}\\
    &= \frac{\PPst{A=1, B=1}{X \in U} \cdot \PP{X \in U}}{\PP{A=1, B=1}}= \frac{\EEst{\PPst{A=1, B=1}{X}}{X \in U}\cdot\PP{X \in U}}{\EE{\PPst{A=1,B=1}{X}}}\\
    &= \frac{\EEst{p_{A \mid X}(X)\cdot\frac{c}{1-c} \cdot \frac{1-p_{A \mid X}(X)}{p_{A \mid X}(X)}}{X \in U} \cdot \PP{X \in U}}{\EE{p_{A \mid X}(X) \cdot \frac{c}{1-c} \cdot \frac{1-p_{A \mid X}(X)}{p_{A \mid X}(X)}}} = \frac{\EEst{1-p_{A \mid X}(X)}{X \in U} \cdot \PP{X \in U}}{\EE{1-p_{A \mid X}(X)}}\\
    &= \frac{\EEst{\PPst{A=0}{X}}{X \in U} \cdot \PP{X \in U}}{\EE{\PPst{A=0}{X}}} = \frac{\PPst{A=0}{X \in U}\cdot\PP{X \in U}}{\PP{A=0}} = \PPst{X \in U}{A=0}\\
    &= \Pp{X \sim P_{X \mid A=0}}{X \in U}.
\end{align*}
This shows that the above two distributions are identical, and thus the claim is proved.

\subsection{Proof of Corollary~\ref{cor:selection}}
By Theorem~\ref{thm:batch_PI} and the observations in Section~\ref{sec:examples} for inference on the quantile, we have
$\PP{S_{(m-\eta)}^\textnormal{test} \le \hat{T}} \ge 1-\alpha$.
Now, the event $\{S_{n+j} = \hat{\mu}(X_{n+j}) \One{Y_{n+j} \le c} > \hat{T}\}$ is equivalent to the event $\{\hat{\mu}(X_{n+j}) > \hat{T} \text{ and } Y_{n+j} \le c\}$, since $\hat{T} \ge 0$ holds almost surely. Therefore,

\[ \PP{\sum_{j=1}^m \One{\hat{\mu}(X_{n+j}) > \hat{T}, Y_{n+j} \le c} \le \eta} = \PP{\sum_{j=1}^m \One{S_{n+j} > \hat{T}} \le \eta} = \PP{S_{(m-\eta)}^\textnormal{test} \le \hat{T}} \ge 1-\alpha,\]
as desired.

\subsection{Proof of Proposition~\ref{prop:wc_extended}}
Fix any $z_1,\ldots,z_n,z_{n+1}, \ldots, z_{n+m}$, where each $z_i = (x_i,y_i)  \in \X \times \Y$, and let $\mathcal{E}_z$ denote the event that $\{Z_1, \ldots, Z_n, Z_{n+1}, \ldots, Z_{n+m}\} = \{z_1,\ldots, z_n,z_{n+1}, \ldots, z_{n+m}\}$, indicating that the data points are equal to these specified values as a (multi-)set. For simplicity, let us also write $\mathcal{E}_A$ to denote the event $A_1=\ldots=A_n = 1, A_{n+1}=\ldots=A_{n+m} = 0$.

Let 
$\mathcal{S}_{n+m}$ denote the set of all permutations of $[n+m]$. For $I = \{i_1,\ldots,i_m\}$ with $1 \le i_1 < \ldots < i_m \le n$, we compute
\begin{align*}
    &\PPst{\{Z_{n+1}, \ldots, Z_{n+m}\} = \{z_{i_1}, \ldots, z_{i_m}\}}{\mathcal{E}_z, \mathcal{E}_A}\\
    &=\frac{\PPst{\mathcal{E}_A}{\{Z_{n+1}, \ldots, Z_{n+m}\} = \{z_{i_1}, \ldots, z_{i_m}\}, \mathcal{E}_z}\cdot\PPst{\{Z_{n+1}, \ldots, Z_{n+m}\} = \{z_{i_1}, \ldots, z_{i_m}\}}{\mathcal{E}_z}}{\PPst{\mathcal{E}_A}{\mathcal{E}_z}}\\
    &=\frac{\prod_{k=1}^m (1-p_{A \mid X})(x_{i_k})\cdot\prod_{i \notin \{i_1,\ldots,i_m\}}p_{A \mid X}(x_i) \cdot \frac{n! m!}{(n+m)!}}{\sum_{\sigma \in S_{n+m}} \PPst{\mathcal{E}_A, Z_{n+1} = z_{\sigma(1)}, \ldots, Z_{n+m} = z_{\sigma(n+m)}}{\mathcal{E}_z}}\\
    &=\frac{\prod_{k=1}^m (1-p_{A \mid X}(x_{i_k}))\cdot\prod_{i \notin \{i_1,\ldots,i_m\}}p_{A \mid X}(x_i) \cdot \frac{n! m!}{(n+m)!}}{\sum_{\sigma \in \mathcal{S}_{n+m}} \frac{1}{(n+m)!}\prod_{i=1}^n p_{A \mid X}(x_{\sigma(i)}) \cdot \prod_{i=n+1}^{n+m} (1-p_{A \mid X}(x_{\sigma(i)}))}.
\end{align*}
By dividing both the numerator and the denominator by $\prod_{i=1}^{n+m} p_{A \mid X}(x_i)$, we find that this further equals
\begin{align*}
    &\frac{n! m!\prod_{k=1}^m \frac{1-p_{A \mid X}(x_{i_k})}{p_{A \mid X}(x_{i_k})}}{\sum_{\sigma \in \mathcal{S}_{n+m}} \prod_{k=1}^m \frac{1-p_{A \mid X}(x_{\sigma(i)})}{p_{A \mid X}(x_{\sigma(i)})}}
    =\frac{n! m!\prod_{k=1}^m \frac{1-p_{A \mid X}(x_{i_k})}{p_{A \mid X}(x_{i_k})}}{\sum_{I \subset [n+m] , |I|=m} \sum_{\sigma \in \mathcal{S}_{n+m} : \{\sigma(k) : k \in [m]\} = I} \prod_{i \in I} \frac{1-p_{A \mid X}(x_i)}{p_{A \mid X}(x_i)}}\\
    =& \frac{\prod_{k=1}^m \frac{1-p_{A \mid X}(x_{i_k})}{p_{A \mid X}(x_{i_k})}}{\sum_{I \subset [n+m] , |I|=m} \prod_{i \in I} \frac{1-p_{A \mid X}(x_i)}{p_{A \mid X}(x_i)}} ( =: p_{A \mid X}^z(I)).
\end{align*}
Therefore, we have
\[g(\{Z_{n+1}, \ldots, Z_{n+m}\}) \mid \mathcal{E}_z, \mathcal{E}_A \sim \sum_{I \subset [n+m], |I|=m} p_{A \mid X}^z(I) \cdot \delta_{h(S_I^z)},\]
where $S_I^z = (s(z_{i_1}), s(z_{i_2}), \ldots, s(z_{i_m}))$. It follows that
\[\PPst{g(\{Z_{n+1}, \ldots, Z_{n+m}\}) \le Q_{1-\gamma}\left(\sum_{I \subset [n+m], |I|=m} p_{A \mid X}^z(I) \cdot \delta_{h(S_I^z)}\right)}{\mathcal{E}_z, \mathcal{E}_A} \ge 1-\gamma,\]
and marginalizing with respect to $\mathcal{E}_z$ yields
\[\PPst{g(\{Z_{n+1}, \ldots, Z_{n+m}\}) \le Q_{1-\gamma}\left(\sum_{I \subset [n+m], |I|=m} p_{A \mid X}(I) \cdot \delta_{h(S_I^Z)}\right)}{\mathcal{E}_A} \ge 1-\gamma.\]
By the monotonicity assumption of $h$, $h(S_I^Z) \le h(\bS_I)$ holds deterministically, leading to
\[\PPst{g(\{Z_{n+1}, \ldots, Z_{n+m}\}) \le Q_{1-\gamma}\left(\sum_{I \subset [n+m], |I|=m} p_{A \mid X}(I) \cdot \delta_{h(\bS_I)}\right)}{\mathcal{E}_A} \ge 1-\gamma.\]
Similarly, we obtain
$\PPst{g(\{Z_{n+1}, \ldots, Z_{n+m}\}) \ge Q_{\beta}'\left(\sum_{I \subset [n+m], |I|=m} p_{A \mid X}(I) \cdot \delta_{h(\uS_I)}\right)}{\mathcal{E}_A} \ge 1-\beta$,
and the desired inequality follows.

\subsection{Proof of Theorem~\ref{thm:multi_bound}}

Let us define $R_{n+1}, R_{n+2}, \cdots, R_{n+m}$ as in~\eqref{eqn:rank_1}. Then, it holds that
\begin{multline*}
    \PP{S_{(w_1-1)} \le S_{(t_1)}^\textnormal{test} \le S_{(q_1)}, \cdots, S_{(w_l-1)} \le S_{(t_l)}^\textnormal{test} \le S_{(q_l)}}\\
    \ge \PP{S_{(w_1)} \le S_{(R_{n+t_1})} \le S_{(q_1)}, \cdots, S_{w_l} \le S_{(R_{n+t_l})} \le S_{(q_l)}}\\
    \ge \PP{w_1 \le R_{n+t_1} \le q_1, \cdots, w_l \le R_{n+t_l} \le q_l} \ge 1-\alpha,
\end{multline*}
where the last inequality holds by the condition $F_{n,m}(w_1,\cdots,w_l; q_1,\cdots,q_l) \ge (1-\alpha)\cdot |H|$ and the fact that $R_{\uparrow}^{\text{test}} \sim \textnormal{Unif}(H)$ holds by the result in the proof of Theorem~\ref{thm:batch_PI}.

\subsection{Proof of Corollary~\ref{cor:multi_batch_PI}}
The proof follows directly from the definition of $B$ in~\eqref{eqn:multi_batch_PI_bound} and Theorem~\ref{thm:multi_bound}.

\end{document}